\documentclass[prd,twocolumn,showpacs,preprintnumbers,nofootinbib,superscriptaddress]{revtex4}
\usepackage{graphicx}% Include figure files
\usepackage{enumerate}
\usepackage{amsmath,amssymb}
\usepackage{mathrsfs}
\usepackage{bm}

\usepackage{ulem}
\usepackage{cleveref}

\begin{document}
\preprint{CHIBA-EP-238, 2020.02.06}

\title{Reflection positivity and complex analysis of the Yang-Mills theory 
\\
from a viewpoint of gluon confinement
}

\author{Kei-Ichi Kondo}
\email{kondok@faculty.chiba-u.jp}
\affiliation{Department of Physics,  
Graduate School of Science, 
Chiba University, Chiba 263-8522, Japan
}
\affiliation{Department of Physics, 
Graduate School of Science and Engineering, 
Chiba University, Chiba 263-8522, Japan}

\author{Masaki Watanabe}
\affiliation{Department of Physics, 
Graduate School of Science and Engineering, 
Chiba University, Chiba 263-8522, Japan}

\author{Yui Hayashi}
\affiliation{Department of Physics, 
Graduate School of Science and Engineering, 
Chiba University, Chiba 263-8522, Japan}

\author{Ryutaro Matsudo}
%\email{afca3071@chiba-u.jp}
\affiliation{Department of Physics, 
Graduate School of Science and Engineering, 
Chiba University, Chiba 263-8522, Japan}

\author{Yutaro Suda}
\affiliation{Department of Physics,  
Graduate School of Science, 
Chiba University, Chiba 263-8522, Japan
}

\begin{abstract}
In order to understand the confining decoupling solution of the Yang-Mills theory in the Landau gauge, we consider the massive Yang-Mills model which is defined by just adding a gluon mass term to the Yang-Mills theory with the Lorentz-covariant gauge fixing term and the associated Faddeev-Popov ghost term.
First of all, we show that massive Yang-Mills  model is obtained as a gauge-fixed version of the gauge-invariantly extended theory which is identified with the gauge-scalar model with a single fixed-modulus scalar field in the fundamental representation of the gauge group.
This equivalence is obtained through the gauge-independent description of the Brout-Englert-Higgs mechanism proposed recently by one of the authors. 
Then, we reconfirm that the Euclidean gluon and ghost propagators  in the Landau gauge obtained by numerical simulations on the lattice are reproduced with good accuracy from the massive Yang-Mills  model  by taking into account one-loop quantum corrections. 
Moreover, we demonstrate in a numerical way that the Schwinger function calculated from the gluon propagator in the Euclidean region exhibits violation of the reflection positivity at the physical point of the parameters. 
In addition, we perform the analytic continuation of the gluon propagator from the Euclidean region to the complex momentum plane towards the Minkowski region.  
We give an analytical proof that the reflection positivity is violated for any choice of the parameters in the massive Yang-Mills  model, due  to the existence of a pair of complex conjugate poles and the negativity of the spectral function for the gluon propagator to one-loop order. 
The complex structure of the propagator enables us to explain why the gluon propagator in the Euclidean region is well described by the Gribov-Stingl form. 
We try to understand these results in light of the Fradkin-Shenker continuity between confinement-like and Higgs-like regions in a single confinement phase in the complementary gauge-scalar model. 

%In this seminar, we want to discuss how gluon and quark confinement/deconfinement are understood in the gauge-invariant way from a novel field theoretical point of view. 
%We focus on the decoupling solution of the Yang-Mills theory in the Landau gauge, since the decoupling solution is currently believed to be a true confining solution of the Yang-Mills theory as supported by the Schwinger-Dyson equation, functional renormalization group and lattice numerical simulations. 
%After a brief review of the development on the gluon and ghost propagators of the Yang-Mills theory, 

%Accordingly, we propose the positivity violation/restoration as a non-local parameter distinguishing gluon confinement and deconfinement phases, which are not discriminated by local order parameters due to 
 
%Then we derive gluon confinement/deconfinement from the reflection-positivity violation/restoration to give a phase structure in the phase diagram of the gauge-scalar model, which includes confinement phase in the pure Yang-Mills theory as a subregion. 

%This result is not restricted to the Landau gauge, rather it is a gauge-invariant result. 
%In fact, 
%This result is suggested from the Fradkin-Shenker continuity as an elucidation of the Osterwalder-Seiler theorem for the confinement-Higgs complementarity on the lattice. 
%Finally, we discuss how to understand quark confinement for gauge theories with matter fields in the  non-perturbative framework. 

\end{abstract}

\pacs{12.38.Aw, 21.65.Qr}

\maketitle

%%%%%%%%%%%%%%%%%%%%%%%%%%%%%%%%%%%%%%%%%%%%%%%%%%%%%%%%%%%%%
%%%%%%%%%%%%%%%%%%%%%%%%%%%%%%%%%%%%%%%%%%%%%%%%%%%%%%%%%%%%%
\section{Introduction}
%%%%%%%%%%%%%%%%%%%%%%%%%%%%%%%%%%%%%%%%%%%%%%%%%%%%%%%%%%%%%
%%%%%%%%%%%%%%%%%%%%%%%%%%%%%%%%%%%%%%%%%%%%%%%%%%%%%%%%%%%%%

It is still a challenging problem  in particle physics to explain \textit{quark and gluon confinement} in the framework of quantum gauge field theories \cite{YM54}. 
The very first question to this problem is to clarify what criterion should be adopted to understand confinement. 
For quark confinement, there is a well-established gauge-invariant criterion given by Wilson \cite{Wilson74}, namely, the area law falloff  of the Wilson loop average leading to the linear static quark potential with a non-vanishing string tension.
For gluon confinement, on the other hand, there is no known gauge-invariant criterion to the best of the authors' knowledge.  This is also the case for more general hypothesis of \textit{color confinement} including quark and gluon confinement as special cases. 
Once  the gauge is fixed, however, there are some proposals. 
For instance, the Kugo-Ojima criterion for color confinement is given for the Lorentz covariant Landau gauge \cite{KO79}.  
Indeed, it is rather difficult to prove the color confinement criterion even in a specific gauge, although there appeared an  announcement for a proof of the Kugo-Ojima criterion for color confinement in the covariant Landau gauge \cite{CF18}. 
Even if color confinement is successfully proved in a specific gauge, this does not automatically guarantee color confinement in the other gauges. 
Therefore the physical picture for confinement could change gauge by gauge.

The information on confinement is expected to be encoded in the gluon and ghost propagators which are obtained by fixing the gauge. 
Recent investigations have confirmed that in the Lorentz  covariant Landau gauge the \textit{decoupling solution}  \cite{decoupling-lattice,decoupling-analytical} is the confining solution of the Yang-Mills theory in the three- and  four-dimensional spacetime, while the scaling solution is realized in the two-dimensional spacetime. 
Therefore, it is quite important to understand the decoupling solution in the Lorentz covariant Landau gauge. 
Of course, there are so many approaches towards this goal.  
In this paper, we focus on the approach \cite{TW10,RSTW17,Machado16,Kondo15} which has been developed in recent several years and has succeeded to reproduce some features of the decoupling solution with good accuracy.  We call this approach the \textit{mass-deformed Yang-Mills theory} with the gauge fixing term or the \textit{massive Yang-Mills model} in the covariant gauge for short.

However, the reason why this approach is so successful is not fully  understood yet in our opinion. %, although some attempts in this direction exist.  
In the original works \cite{TW10} the massive Yang-Mills model in the Landau gauge  was identified with a special parameter limit of the Curci-Ferrari model \cite{CF76b}. 
However, the Curci-Ferrari model is not invariant under the usual Becchi-Rouet-Stora-Tyutin (BRST) transformation, but invariant just under the modified BRST transformation which does not respect the usual nilpotency. 

In this paper  we show based on the previous works \cite{Kondo16,Kondo18} that the mass-deformed Yang-Mills theory with the covariant gauge fixing term has the \textit{gauge-invariant extension} which is given by a gauge-scalar model with a single fixed-modulus scalar field in the fundamental representation of the gauge group, provided that a constraint called the \textit{reduction condition} is satisfied. 
We call such a model the \textit{complementary gauge-scalar model}. 
This equivalence is achieved based on the gauge-independent description \cite{Kondo16,Kondo18} of the \textit{Brout-Englert-Higgs (BEH) mechanism} \cite{Higgs1,Higgs2,Higgs3} which does not rely on the spontaneous breaking of gauge symmetry \cite{NJL61,Goldstone61}.  
This description enables one to give a \textit{gauge-invariant mass term of the gluon field} in the Yang-Mills theory which can be  identified with the gauge-invariant kinetic term of the scalar field in the complementary gauge-scalar model.

In this paper, we first confirm that the massive Yang-Mills model with one-loop quantum corrections being included in the Euclidean region reproduces with good accuracy the \textit{gluon and ghost propagators} of the decoupling solution of the Yang-Mills theory in the Landau gauge obtained by numerical simulations on the lattice.  In fact,  the resulting gluon and ghost propagators in the massive Yang-Mills model can be well fitted to those on the lattice by adjusting the parameters, namely, the coupling constant $g$ and the gluon mass parameter $M$. 

For gluon confinement, the violation of \textit{reflection positivity} is regarded as a necessary condition for confinement.
In fact, it is known that the gluon propagator in the Yang-Mills theory exhibits the violation of {reflection positivity}. 
This fact was directly shown by the numerical simulations on the lattice, e.g., in the covariant Landau gauge \cite{CMT05,Bowmanetal07}. 
In this paper, by using the relevant gluon propagator in the massive Yang-Mills model, we calculate the \textit{Schwinger function} in a numerical way to demonstrate that the reflection positivity is  violated at the physical point of parameters reproducing the Yang-Mills theory. 

In order to understand these facts and consider the meaning of gluon confinement, we perform the analytic continuation of the gluon and ghost propagators in the Euclidean region to those in the Minkowski region on the complex momentum squared plane. 
The consideration of the complex structure of the propagator enables us to give an analytical proof that the reflection positivity  is violated for any choice of the parameters without restricting to the physical point of the Yang-Mills theory in the massive Yang-Mills  model with one-loop quantum corrections being included.
For this proof, it is enough to show that the Schwinger function necessarily becomes negative in some region, which is achieved by calculating separately the contributions to the gluon  {Schwinger function} from the pole part and the continuous (branch cut) part of the gluon propagator based on the generalized spectral representation in the massive Yang-Mills model to one-loop order. 
It turns out that the violation of reflection positivity is an immediate consequence of the facts that the gluon propagator has a pair of complex conjugate poles and that the spectral function of the gluon propagator has negative value on the whole range, see \cite{HK18}. 
See e.g., \cite{DORS19,BT19} for the construction of the spectral function from the  Euclidean data of numerical simulations on the lattice. 
%due to  discontinuities across the branch cut on the positive real axis of time-like momentum %, see \cite{HK18} for more information.

The complex structure of the propagator enables us to explain why the gluon propagator in the Euclidean region is well described by the \textit{Gribov-Stingl form} \cite{Stingl86}, as demonstrated in the numerical simulations on the lattice \cite{DOS18}. 
Indeed, the pole part of the gluon propagator due to \textit{a pair of complex conjugate poles} exactly reproduces the Gribov-Stingl form which is fitted to the numerical simulations to very good accuracy, after subtracting the small contribution coming from the continuous part represented by the spectral function obtained from the discontinuity across the \textit{branch cut} on the positive real axis on the complex momentum plane.
See also \cite{Kondo11} for another explanation for the occurrence of the gluon propagator of the Gribov-Stingl form. 
 
The above result suggests that gluon confinement is not restricted to the confinement phase of the ordinary Yang-Mills theory, and can be  extended into more general situations, namely, anywhere represented by the massive Yang-Mills model, which includes the Higgs phase in the complementary gauge-scalar model.
In the lattice gauge theory, it is known that the confinement phase  in the pure Yang-Mills theory is analytically continued to the Higgs phase in the relevant gauge-scalar model, which is called the \textit{Fradkin-Shenker continuity} \cite{FS79} as a special realization of the Osterwalder-Seiler theorem \cite{OS78}.
There are no local order parameters which can distinguish the confinement and Higgs phases. 
There is no thermodynamic phase transition between confinement and Higgs phases \cite{lattice-gauge-scalar-fund}, in sharp contrast to the adjoint scalar case \cite{lattice-gauge-scalar-adj} where there is a clear phase transition between the two phases. 
Therefore, confinement and Higgs phases are just subregions of a single confinement-Higgs phase \cite{FMS80,tHooft80,Maas17}. 
Therefore, permanent violation of positivity can be understood in light of the {Fradkin-Shenker continuity} between confinement-like and Higgs-like regions in a single confinement phase in the gauge-scalar model.

This paper is organized as follows. 
In sec. II, we introduce the massive Yang-Mills model in the covariant gauge.  
In sec. III, we show that the massive Yang-Mills model with quantum corrections to one-loop order well reproduces the gluon and ghost propagators of the decoupling solution. 
In sec. IV, we show that the gluon propagator exhibits violation of reflection positivity through the calculation of the Schwinger function. 
In sect. V, we perform the analytic continuation of the propagator to the complex momentum to examine the complex structure.
% the model to the gauge-scalar model to  investigate the positivity violation/restoration transition. 
%The results are compared with the other approach based on the Gribov/Zwanziger theory. 
In the final section we draw the conclusion and discuss the future problems to be tackled.
In Appendix~A, we give a recursive construction of the transverse and gauge-invariant gluon field to show the gauge-invariant extension of the massive Yang-Mills model. 
In Appendix~B, we give another way for solving the reduction condition. 
%In Appendix~C, we give a detailed study of the running gauge coupling constant obtained in the infrared safe renormalization scheme. 
%In Appendix~C, we give the expression for the vacuum polarization tensor for gluons and the self-energy function for ghosts for  an arbitrary gauge fixing parameter in the Lorenz gauge. 

%\newpage
%%%%%%%%%%%%%%%%%%%%%%%%%%%%%%%%%%%%%%%%%%%%%%%%%%%%%%%%%%%%%
%%%%%%%%%%%%%%%%%%%%%%%%%%%%%%%%%%%%%%%%%%%%%%%%%%%%%%%%%%%%%
\section{Gauge-invariant extension of the mass-deformed Yang-Mills theory in the covariant Landau gauge
}
%%%%%%%%%%%%%%%%%%%%%%%%%%%%%%%%%%%%%%%%%%%%%%%%%%%%%%%%%%%%%
%%%%%%%%%%%%%%%%%%%%%%%%%%%%%%%%%%%%%%%%%%%%%%%%%%%%%%%%%%%%%

\subsection{
Mass deformation of the Yang-Mills theory in the covariant Landau gauge
}

We introduce the \textit{mass-deformed Yang-Mills theory in the covariant gauge} which is defined just by adding the naive mass term $\mathscr{L}_{\rm m}$ to the ordinary massless Yang-Mills theory in the (manifestly Lorentz) covariant gauge fixing.
The total Lagrangian density $\mathscr{L}_{\rm mYM}^{\rm tot}$ of the massive Yang-Mills model consists of the Yang-Mills Lagrangian $\mathscr{L}_{\rm YM}$, the gauge-fixing (GF) term $\mathscr{L}_{\rm GF}$, the associated Faddeev-Popov (FP) ghost term $\mathscr{L}_{\rm FP}$, and the mass term $\mathscr{L}_{\rm m}$,
%\begin{align}
%\mathscr{L}_{\rm mYM} 
%=  \mathscr{L}_{\rm YM}^{\rm tot} + \mathscr{L}_{\rm m}, \quad
%\mathscr{L}_{\rm m} = \frac12 M^2 \mathscr{A}^{\mu A}   \mathscr{A}_\mu^A ,
%= M^2 {\rm tr}(\mathscr{A}^\mu \mathscr{A}_\mu).
%\end{align}
%where 
%which we call the  \textit{massive Yang-Mills model} in the covariant gauge   for short. 
%The ordinary massless Yang-Mills theory in the (manifestly Lorentz covariant)  Lorenz gauge with a gauge-fixing parameter $\alpha$ is given by
\begin{align}
\mathscr{L}_{\rm  mYM}^{\rm tot} 
=&  \mathscr{L}_{\rm YM} + \mathscr{L}_{\rm GF} + \mathscr{L}_{\rm FP} + \mathscr{L}_{\rm m} , 
\nonumber\\
  \mathscr{L}_{\rm YM} =& -\frac{1}{4}\mathscr{F}^A_{\mu\nu}\mathscr{F}^{A\mu\nu}  ,  
\nonumber\\
\mathscr{L}_{\rm GF}  =& \mathscr{N}^A \partial^\mu \mathscr{A}^A_\mu + \frac{\alpha}{2}\mathscr{N}^A\mathscr{N}^A ,
%\to -\frac{1}{2}\alpha^{-1} (\partial^\mu \mathscr{A}^A_\mu)^2 ,
%\quad (\alpha \to 0) , 
\nonumber\\
  \mathscr{L}_{\rm FP} =& i\mathscr{\bar C}^A \partial^\mu \mathscr{D}_\mu[\mathscr{A}]^{AB} \mathscr{C}^B 
\nonumber\\
=&  i\mathscr{\bar{C}}^A\partial^\mu(\partial_\mu\mathscr{C}^A+gf_{ABC}\mathscr{A}^B_\mu \mathscr{C}^C) ,
\nonumber\\
\mathscr{L}_{\rm m}  =& \frac{1}{2}M^2\mathscr{A}^A_\mu\mathscr{A}^{\mu A} ,
\end{align}
where $\mathscr{A}_\mu^A$ denotes the Yang-Mills field, $\mathscr{N}^A$ the Nakanishi-Lautrup field,  and $\mathscr{C}^A, \mathscr{\bar{C}}^A$ the Faddeev-Popov ghost and antighost fields, which take their values in the Lie algebra $\mathscr{G}$ of a gauge group $G$ with the structure constants $f_{ABC}$ ($A,B,C=1,...,{\rm dim}G$). 
We call this theory the  \textit{massive Yang-Mills model} in the covariant gauge for short.

The expectation value of an operator $\mathscr{O}[\mathscr{A}]$ of $\mathscr{A}_\mu^A$ is given according to the path integral quantization using the total action $S_{\rm mYM}^{\rm tot}[\mathscr{A}, \mathscr{C} , \mathscr{\bar C} , \mathscr{N}]$ and the  integration measure $\mathcal{D}\mathscr{A} \mathcal{D}\mathscr{C} \mathcal{D}\mathscr{\bar C} \mathcal{D}\mathscr{N}$
\begin{align}
 \langle \mathscr{O}[\mathscr{A}] \rangle_{\rm mYM} 
:=&  \frac{\int \mathcal{D}\mathscr{A} \mathcal{D}\mathscr{C} \mathcal{D}\mathscr{\bar C} \mathcal{D}\mathscr{N} e^{iS_{\rm mYM}^{\rm tot}[\mathscr{A}, \mathscr{C} , \mathscr{\bar C} , \mathscr{N}]} \mathscr{O}[\mathscr{A}]}
{\int \mathcal{D}\mathscr{A} \mathcal{D}\mathscr{C} \mathcal{D}\mathscr{\bar C} \mathcal{D}\mathscr{N} e^{iS_{\rm mYM}^{\rm tot}[\mathscr{A}, \mathscr{C} , \mathscr{\bar C} , \mathscr{N}]} } .
\end{align}
In the Landau gauge $\alpha=0$, especially,  the average is cast into a simpler form by integrating the Nananishi-Lautrup field $\mathscr{N}^A$ and subsequently the ghost and antighost field $\mathscr{C}^A, \mathscr{\bar{C}}^A$ as
\begin{widetext}
\begin{align}
%\nonumber\\
  \langle \mathscr{O}[\mathscr{A}] \rangle_{\rm mYM} 
%\nonumber\\
  =& \frac{\int \mathcal{D}\mathscr{A} \mathcal{D}\mathscr{C} \mathcal{D}\mathscr{\bar C} \delta(\partial^\mu \mathscr{A}_{\mu}^A)  e^{ iS_{\rm YM}[\mathscr{A} ] +iS_{\rm FP}[\mathscr{A}, \mathscr{C} , \mathscr{\bar C}]+iS_{\rm m}[\mathscr{A} ]} \mathscr{O}[\mathscr{A}]}
{\int \mathcal{D}\mathscr{A} \mathcal{D}\mathscr{C} \mathcal{D}\mathscr{\bar C} \delta(\partial^\mu \mathscr{A}_{\mu}^A)  e^{iS_{\rm YM}[\mathscr{A} ] +iS_{\rm FP}[\mathscr{A}, \mathscr{C} , \mathscr{\bar C}]+iS_{\rm m}[\mathscr{A} ]}}
\nonumber\\
=& \frac{\int \mathcal{D}\mathscr{A}  \delta(\partial^\mu \mathscr{A}_{\mu}^A)  \Delta_{\rm FP}[\mathscr{A}] e^{iS_{\rm YM}[\mathscr{A} ] +iS_{\rm m}[\mathscr{A} ]} \mathscr{O}[\mathscr{A}]}
{\int \mathcal{D}\mathscr{A}  \delta(\partial^\mu \mathscr{A}_{\mu}^A)  \Delta_{\rm FP}[\mathscr{A}] e^{iS_{\rm YM}[\mathscr{A} ] +iS_{\rm m}[\mathscr{A} ]}} ,
\end{align}
\end{widetext}
with the Faddeev-Popov determinant, 
\begin{align}
\Delta_{\rm FP}[\mathscr{A}] := \det (\partial^\mu \mathscr{D}_\mu[\mathscr{A}]^{AB} ) 
 .  
\end{align}

%\footnote{
%In this paper, we ignore the Gribov problem. 
%We do not regard the inclusion of the gluon mass term with the reweighting to avoid the Neuberger 0/0 problem. 

In this paper we do not intend to take into account the Gribov problem. The reasons are as follows. 
In this paper we deal with the massive Yang-Mills model as a low-energy effective model of the Yang-Mills theory and perform the perturbative analysis based on this model.  

In the ultraviolet region the perturbative analysis of the Yang-Mills theory is valid due to the ultraviolet asymptotic freedom and is free from the Gribov problem, since the perturbative analysis can be done in the neighborhood of the origin of the configuration space of the gauge field within the first Gribov region and therefore does not reach the Gribov horizon where the Gribov problem becomes serious.  
This is also the case for the massive Yang-Mills model, since the effect of mass term can be ignored in the ultraviolet region.  

Of course, in the usual perturbative treatment of the Yang-Mills theory, we encounter the Landau pole at which the gauge coupling constant diverges and the perturbative analysis breaks down at an intermediate momentum scale before reaching the deep infrared region.  
For the massive Yang-Mills model, however, we can adopt the infrared safe renormalization scheme in which the perturbation theory does not break down and remains valid from the large momentum all the way down to the zero momentum, as can be seen from the fact that the gauge coupling constant remains finite without divergence in the whole momentum region, and even vanishes in the zero momentum limit, as reviewed in section III.  
Therefore, we think that the massive Yang-Mills model can be treated in the whole region without seriously worrying about the Gribov problem, although there is no rigorous proof on this claim.  

We regard the massive Yang-Mills model adopted in this paper as a low-energy effective model of the Yang-Mills theory where the mass term is generated in the dynamical way due to quantum corrections, for instance, according to the Wilsonian renormalization group.  
%The mass term plays also the role of an infrared regulator and the massive Yang-Mills model is valid even in the vanishing momentum limit. 
The mass term plays also the role of an infrared regulator and the massive Yang-Mills model is thereby  free from the infrared divergence even in the vanishing momentum limit. 
Of course, the generation of the gluon mass term originates from non-perturbative effects and should be investigated from the first principles, which is however beyond the scope of this paper.  Incidentally, we tried to show the existence of such mass term in  \cite{WMNSK18}.

%}

The massive Yang-Mills model just defined is a special case of a massive extension of the massless Yang-Mills theory in the most general renormalizable gauge having both BRST and anti-BRST symmetries given by \cite{Baulieu85}
\begin{subequations}
	\begin{align}
		\mathscr{L}^{\rm{tot}}_{m\rm{YM}} =& \mathscr{L}_{\rm{YM}} + \mathscr{L}_{\rm{GF+FP}} + \mathscr{L}_{m} , 
\\
		\mathscr{L}_{\rm{YM}}  =& - \frac{1}{4} \mathscr{F}_{\mu \nu} \cdot \mathscr{F}^{\mu \nu} , 
   \\
		\mathscr{L}_{\rm{GF+FP}}  =& 
%\frac{\alpha}{2} \mathscr{N} \cdot \mathscr{N}  + 
\frac{\beta}{2} \mathscr{N} \cdot \mathscr{N} 
\nonumber\\ & 
       + \mathscr{N} \cdot \partial^{\mu} \mathscr{A}_{\mu} 
		- \frac{\beta}{2} g \mathscr{N} \cdot (i \bar{\mathscr{C}} \times \mathscr{C}) 
  \nonumber\\
		& + i \bar{\mathscr{C}} \cdot \partial^{\mu} \mathscr{D}_{\mu}[\mathscr{A}] \mathscr{C}
%\nonumber\\&
		+ \frac{\beta}{4} g^2 (i \bar{\mathscr{C}} \times \mathscr{C}) \cdot (i \bar{\mathscr{C}} \times \mathscr{C}) 
\nonumber\\
		 =& \mathscr{N} \cdot \partial^{\mu} \mathscr{A}_{\mu} + i \bar{\mathscr{C}} \cdot \partial^{\mu} \mathscr{D}_{\mu}[\mathscr{A}] \mathscr{C}
\nonumber\\&
		 + \frac{\beta}{4} ( \bar{\mathscr{N}} \cdot \bar{\mathscr{N}} + \mathscr{N} \cdot \mathscr{N}) 
%+ \frac{\alpha}{2} \mathscr{N} \cdot \mathscr{N}
, 
   \\
		\mathscr{L}_{m}  =& \frac{1}{2} M^2 \mathscr{A}_{\mu} \cdot \mathscr{A}^{\mu} + \beta M^2 i \bar{\mathscr{C}} \cdot \mathscr{C} , 
	\end{align}
%		\label{mYM1}
\end{subequations}
where $\beta$ is a parameter which correspond to the gauge-fixing parameters in the $M \rightarrow 0$ limit, 
$
 \mathscr{D}_{\mu}[\mathscr{A}] \mathscr{C}(x) 
%:= \partial_{\mu} \mathscr{C}(x) - ig [\mathscr{A}(x), \mathscr{C}(x)] 
:=  \partial_{\mu}\mathscr{C}(x) + g \mathscr{A}(x) \times \mathscr{C}(x)  
$,
 and 
$ 
\bar{\mathscr{N}} :=-\mathscr{N}+gi\bar{\mathscr{C}} \times \mathscr{C} 
$. 
This model is called the Curci-Ferrari model \cite{CF76b} with the coupling constant $g$, the mass parameter $M$, and the parameter $\beta$.
[In the Abelian limit with  vanishing structure constants $f^{ABC}=0$, the FP ghosts decouple and the Curci-Ferrari model reduces to the Nakanishi model \cite{Nakanishi72}.]
For $M \not=0$, the physics depends on the parameter $\beta$.	
This result should be compared with the $M=0$ case, in which $\beta$ is a gauge fixing parameter and hence the physics should not depend on $\beta$. In the $M=0$ case, indeed, any choice of $\beta$ gives the same  physics. 
However, this is not the case for $M \not=0$.
See e.g., \cite{Kondo13} for more details.
%[K.-I. Kondo, PRD87, 025008 (2013)]; [K.-I. Kondo et al,, PRD87, 025017 (2013)]
%\cite{Esole04,DT86} 
The massive Yang-Mills model is regarded as a  $\beta=0$ case of the Curci-Ferrari model. 
This point of view taken in the preceding works \cite{TW10} is good from the viewpoint of renormalizability, since the Curci-Ferrari model is known to be renormalizable. However, the Curci-Ferrari model lacks the physical unitarity at least in the perturbation theory \cite{CF76b,Kondo13}. 
Indeed, the massive Yang-Mills model does not have the nilpotent BRST symmetry, although it has the modified BRST symmetry which does not respect the usual nilpotent property and reduces to the ordinary BRST symmetry only in the massless limit $M \to 0$. 
In this paper we try to find an extended theory with the ordinary nilpotent BRST symmetry, which reproduces the massive Yang-Mills model under an appropriate prescription.  As a candidate for such a theory we investigate a specific gauge-scalar model.
%This issue will be avoided in the gauge-invariant extension which retains the nilpotent BRST symmetry, as discussed later. 

In what follows we show that the massive Yang-Mills model in a covariant gauge has the \textit{gauge-invariant extension} which is given by the gauge-scalar model with a single radially fixed (or fixed modulus) scalar field in the fundamental representation of a gauge group if the theory is subject to an appropriate constraint which we call the \textit{reduction condition}. 
We call such a gauge-scalar model the \textit{complementary gauge-scalar model}. 
In other words, the complementary  gauge-scalar model with a single  {radially fixed scalar field}  in the fundamental representation   reduces to the mass-deformed Yang-Mills theory in a fixed gauge if an appropriate reduction condition is imposed. 
%We show that the the massive Yang-Mills model which is the (local) Yang-Mills theory in the Landau gauge with a mass term has the (non-local) gauge-invariant extension.  
%In other words, the massive Yang-Mills model is a local gauge non-invariant expression in a specific fixed gauge of the non-local gauge invariant extended theory. 
%This is performed by solving the reduction condition explicitly.

%\noindent
%[Theory 1] is the mass-deformed Yang-Mills theory in the covariant Landau gauge.
%[Theory 1] has no longer gauge symmetry, although it has the modified BRST symmetry. 
%However, [Theory 1] has a \textit{gauge-invariant extension} [Theory 2]. 
%
%\noindent
%[Theory 2] 
For $G=SU(2)$, the complementary gauge-scalar model is given by
%with a single radially fixed fundamental scalar %subject to the reduction condition.
\begin{align}
		\mathscr{L}_{\rm RF} =& \mathscr{L}_{\rm YM} + \mathscr{L}_{\rm kin} , 
 \nonumber\\ 
\mathscr{L}_{\rm YM} =& -\frac{1}{2} {\rm tr}[ \mathscr{F} _{\mu\nu} \mathscr{F}^{\mu\nu }  ], \
 \nonumber\\ 
\mathscr{L}_{\rm kin} :=&  ({D}_{\mu}[\mathscr{A}]\Phi )^{\dagger} \cdot ({D}^{\mu}[\mathscr{A}]\Phi ) , 
%\nonumber\\& 
%+ u(x) \left( \Phi ^{\dagger} \Phi  - \frac{1}{2}v^2   \right) ,
%\label{SU2-gauge-scalar-f1}
\end{align}
with a single fundamental scalar field $\Phi$ subject to the radially fixed condition, 
%with the holonomic constraint,  
%$u(x)$ is the \textit{Lagrange multiplier field} to incorporate the constraint:% that the radial degree of freedom or length of the scalar field is fixed $|\Phi(x)|=v/\sqrt{2}>0$: 
\begin{align}
		 f(\Phi(x)) :=  \Phi(x)^{\dagger} \cdot \Phi(x) - \frac{1}{2}v^2 = 0  ,
\label{SU2-s-constraint1}
\end{align}
where $v$ is a positive constant $(v>0)$ and $\Phi (x)$ is the \textit{$SU(2)$ doublet} 
%called the \textit{Higgs doublet} which is 
formed from two complex scalar  fields   $\bm{\phi}_1 (x), \bm{\phi}_2 (x)$,
% which are parameterized (by the reason clarified later) as
\begin{align}
		\Phi(x) =&
	 \begin{pmatrix}
			\bm{\phi}_1(x) \\ 
			\bm{\phi}_2(x)
	\end{pmatrix}   , \
 \bm{\phi}_1(x), \bm{\phi}_2(x) \in \mathbb{C} ,
%\nonumber\\  
%=& \frac{1}{\sqrt{2}}
%		 \begin{pmatrix}
%			\phi_2(x) + i \phi_1(x) \\ 
%			\phi_0(x) - i \phi_3(x)
%	\end{pmatrix}  ,
%\nonumber\\  &
%  \ \phi_0(x), \phi_A(x) \in \mathbb{R} \ (A=1,2,3) .
\end{align} 
where $D_\mu[\mathscr{A}]$ is the covariant derivative in the fundamental representation $D_\mu[\mathscr{A}]:=\partial_\mu -ig  \mathscr{A}_\mu $. 

% with $M=gv/2$. 
%Here $\Phi (x)$ is the \textit{$SU(2)$ doublet} 
%called the \textit{Higgs doublet} which is formed from two complex scalar  fields   $\bm{\phi}_1 (x), \bm{\phi}_2 (x)$.
This gauge-scalar model is invariant under the gauge transformation, 
\begin{align}
 {\mathscr{A}}_\mu(x) &\to {\mathscr{A}}_\mu^{U}(x)  := U(x) \mathscr{A}_{\mu}(x) U(x)^\dagger + ig^{-1} U(x) \partial_\mu U(x)^\dagger , 
 \nonumber\\
\Phi(x) &\to \Phi^{U}(x)  := U(x) \Phi(x)  , \quad U(x) \in G 
 .
\end{align}

%If we take the covariant Landau gauge and eliminate the scalar field, [Theory 2] reduces to [Theory 1].

%[Theory 1] well reproduces the decoupling solution in the Yang-Mills theory as a confining solution. 
%Based on this correspondence, [Theory 1] can describe also the Higgs phase by choosing the values of parameters.   

%If [Theory 1] describes the confining region of  the Yang-Mills theory, 
%This is regarded as a continuum realization of the \textit{Fradkin-Shenker continuity} connecting confinement and Higgs phase shown in the framework of lattice gauge theory. 

It is more convenient to convert the scalar field into the gauge group element. 
For this purpose, we introduce the \textit{matrix-valued scalar field} $\Theta$ by adding another $SU(2)$ doublet $\tilde\Phi:=\epsilon \Phi^*$ as
\begin{align}
		\Theta(x) :=&
		  \begin{pmatrix} 
		\tilde\Phi(x) &	\Phi(x) 
	\end{pmatrix} 
=
		  \begin{pmatrix} 
		\epsilon \Phi^*(x) &	\Phi(x) 
	\end{pmatrix} 
\nonumber\\
=&
		  \begin{pmatrix} 
		\bm{\phi}_2^*(x) &	\bm{\phi}_1(x) \\ 
		-\bm{\phi}_1^*(x) &	\bm{\phi}_2(x)
	\end{pmatrix} 
%\nonumber\\
%=&  \frac{1}{\sqrt{2}} (\phi_0 {\boldsymbol 1} + i \phi_A \sigma^A) 
%= \frac{1}{\sqrt{2}}
%	 \begin{pmatrix}
%			\phi_0 + i \phi_3 & \phi_2 + i \phi_1 \\ 
%			- \phi_2+i \phi_1  & \phi_0 - i \phi_3
%	\end{pmatrix}  
 , \
%\nonumber\\ 
  \epsilon =  
  \begin{pmatrix} 
		0 &	1 \\ 
		-1 & 0
	\end{pmatrix}
 .
	\label{Theta1}
\end{align} 
Then the complementary $SU(2)$ gauge-scalar model with a single  radially fixed scalar field in the fundamental representation is defined by   
\begin{align}
\mathscr{L}_{\rm RF} =&  \mathscr{L}_{\rm YM} + \mathscr{L}_{\rm kin} + \mathscr{L}_{\rm c} , 
\nonumber\\
 \mathscr{L}_{\rm YM}[\mathscr{A}] =&     \frac{-1}{2} {\rm tr}( \mathscr{F} _{\mu\nu} (x) \mathscr{F}^{\mu\nu } (x)) ,  
\nonumber\\
 \mathscr{L}_{\rm kin}[\mathscr{A},\Theta] =&   \frac{1}{2}{\rm tr}(  ( {D}_{\mu}[\mathscr{A}] \Theta (x))^\dagger 	{D}^{\mu}[\mathscr{A}] \Theta (x))  
 ,  
\nonumber\\
 \mathscr{L}_{\rm c}[u,\Theta] =&    u(x) f(\Theta(x)), 
\nonumber\\
& f(\Theta) := {\rm tr} \left(\Theta ^{\dagger}\Theta  - \frac{1}{2}v^2 \bm{1} \right) /{\rm tr}(\bm{1}) ,
\label{L-RF}
\end{align}
where $u$ is the Lagrange multiplier field to incorporate the holonomic constraint (\ref{SU2-s-constraint1}) written in the matrix form 
$f(\Theta)=0$.
The radially fixed gauge-scalar model with the Lagrangian density (\ref{L-RF}) is invariant under the gauge transformation, 
\begin{align}
 {\mathscr{A}}_\mu(x) &\to {\mathscr{A}}_\mu^{U}(x)  := U(x) \mathscr{A}_{\mu}(x) U(x)^\dagger + ig^{-1} U(x) \partial_\mu U(x)^\dagger , 
 \nonumber\\
\hat{\Theta}(x) &\to \hat{\Theta}^{U}(x)  := U(x) \hat{\Theta}(x)  , \ U(x) \in G  
 .
\end{align}

Then  we introduce the \textit{normalized matrix-valued scalar field} $\hat{\Theta}$ by
\begin{align}
 \hat{\Theta}(x) =  {\Theta} (x)/\left(\frac{v}{\sqrt{2}}\right), \ v > 0   
.
\end{align} 
The above constraint (\ref{SU2-s-constraint1})  implies that the normalized scalar field $\hat{\Theta}$ obeys the conditions:
%$\hat{\Theta}(x) \hat{\Theta}(x)^\dagger=\bm{1}$.
$
\hat{\Theta}(x)^\dagger \hat{\Theta}(x) = \hat{\Theta}(x) \hat{\Theta}(x)^\dagger =  \bm{1} 
,
$
and
$
 \det \hat{\Theta}(x) 
= 1
$.
Therefore,  $\hat{\Theta}$ is an element of $SU(2)$:
\begin{align}
 \hat{\Theta}(x) \in G=SU(2) 
.
\end{align}
This is an important property to provide a gauge-independent BEH mechanism. 

The massive vector boson field $\mathscr{W}_\mu \in \mathscr{G}=su(2)$ is defined in terms of the original gauge field $\mathscr{A}_\mu \in \mathscr{G}=su(2)$ and  the normalized scalar field $\hat{\Theta} \in G=SU(2)$ as shown in a previous paper \cite{Kondo18},
\begin{align}
	\mathscr{W}_\mu(x) 
 :=& ig^{-1} ( {D}_{\mu}[\mathscr{A}] \hat{\Theta}(x)) \hat{\Theta}(x)^\dagger 
\nonumber\\
=&  -ig^{-1} \hat{\Theta}(x)	({D}_{\mu}[\mathscr{A}] \hat{\Theta}(x) )^\dagger
\nonumber\\
=& \frac{1}{2} ig^{-1} [ ( {D}_{\mu}[\mathscr{A}] \hat{\Theta}(x)) \hat{\Theta}(x)^\dagger -  \hat{\Theta}(x)	({D}_{\mu}[\mathscr{A}] \hat{\Theta}(x) )^\dagger ]
 .
\label{W1-SU2}
\end{align}
According to the gauge-independent BEH mechanism \cite{Kondo16,Kondo18}, the kinetic term of the scalar field $\Theta$ is identical to the mass term of $\mathscr{W}_\mu$,
%the mass term of $W$ is  
\begin{align}
 \mathscr{L}_{\rm kin}[\mathscr{A},\hat{\Theta}] =&    \frac{1}{2}{\rm tr}(  ( {D}_{\mu}[\mathscr{A}] \hat{\Theta} (x))^\dagger 	{D}^{\mu}[\mathscr{A}] \hat{\Theta}(x) )  
\nonumber\\
=& M^2{\rm tr}(\mathscr{W}_\mu(x) \mathscr{W}^\mu(x) ) , \ M=g\frac{v}{2} .
\label{kin=mass}
\end{align}

The massive vector field $\mathscr{W}_\mu$ is rewritten using $\hat{\Theta}(x)  \hat{\Theta}(x)^\dagger=\bm{1}$ into  
\begin{align}
 \mathscr{W}_\mu(x)  
=  \mathscr{A}_{\mu}(x)  -ig^{-1} \hat{\Theta}(x) \partial_{\mu} \hat{\Theta}(x)^\dagger .
\label{W1b-SU2}
\end{align}
Then it is shown that the massive vector boson field $\mathscr{W}_\mu$ has the expression,
\footnote{
In \cite{Kondo18} $\mathscr{A}_{\mu}^{\hat{\Theta}^\dagger}(x) $ was written as $\tilde{\mathscr{W}}_\mu(x)$ .
}
\begin{align}
 {\mathscr{W}}_\mu(x)  = \hat{\Theta}(x)  \mathscr{A}_{\mu}^{\hat{\Theta}^\dagger}(x)   \hat{\Theta}(x)^\dagger	   
 ,
\end{align}
where 
$\mathscr{A}_{\mu}^{\hat{\Theta}^\dagger}$ denotes the gauge transform of $\mathscr{A}_{\mu}$ by $\hat{\Theta} \in G$.
%$\mathscr{A}_{\mu}^{U}$ denotes the gauge transform of $\mathscr{A}_{\mu}$ by $U \in G$,
%\begin{align}
% {\mathscr{A}}_\mu^{U}(x)  = U(x) \mathscr{A}_{\mu}(x) U(x)^\dagger + ig^{-1} U(x) \partial_\mu U(x)^\dagger     .
%\end{align}
Notice that $\mathscr{W}_{\mu}$ transforms according to the adjoint representation under the gauge transformation, 
\begin{align}
 \mathscr{W}_\mu(x)   \to \mathscr{W}_\mu^{U}(x) = U(x) \mathscr{W}_{\mu}(x) U(x)^\dagger
 ,
\end{align}
whereas $\mathscr{A}_{\mu}^{\hat{\Theta}^\dagger}$ is gauge invariant,
\begin{align}
 \mathscr{A}_{\mu}^{\hat{\Theta}^\dagger}(x)   \to (\mathscr{A}_{\mu}^{\hat{\Theta}^\dagger})^{U}(x) = \mathscr{A}_{\mu}^{\hat{\Theta}^\dagger}(x)
 .
\end{align}
Therefore, the mass term can be written in terms of the gauge-invariant field $\mathscr{A}_{\mu}^{\hat{\Theta}^\dagger}$ as
\begin{align}
 \mathscr{L}_{\rm kin}[\mathscr{A},\hat{\Theta}^\dagger] =  M^2{\rm tr}(\mathscr{A}_{\mu}^{\hat{\Theta}^\dagger}(x) \mathscr{A}^{\mu}{}^{\hat{\Theta}^\dagger}(x)) , \ M=g\frac{v}{2} .
\label{kin=mass2}
\end{align}

This theory is supposed to obey the \textit{reduction condition} for the {massive vector field mode} $\mathscr{W}_\mu(x)$.
%$\mathscr{W}_\mu(x) = \mathscr{W}_\mu[\mathscr{A}(x), \Phi(x)]$. % defined in terms of $\mathscr{A}_\mu$ and $\Phi$. 
The stationary form of the \textit{reduction condition} is given by 
\begin{equation}
 \chi(x) := \mathscr{D}^\mu[\mathscr{A}]  \mathscr{W}_\mu(x) = 0 ,
\end{equation}
where $\mathscr{D}_\mu[\mathscr{A}]$ is the covariant derivative in the adjoint representation $\mathscr{D}_\mu[\mathscr{A}]:=\partial_\mu -ig [\mathscr{A}_\mu , \cdot ]$. 
%which follows from the \textit{gauge-independent BEH mechanism. 
The stationary {reduction condition} is cast into 
\begin{align}
 \chi(x)   :=&  \mathscr{D}^{\mu}[\mathscr{A}]  \mathscr{W}_\mu(x)   
\nonumber\\ 
=& (\hat{\Theta}(x)  \mathscr{D}^{\mu}[\mathscr{A}^{\hat{\Theta}^\dagger}]   \hat{\Theta}(x)^\dagger)(\hat{\Theta}(x)  \mathscr{A}_{\mu}^{\hat{\Theta}^\dagger}(x)   \hat{\Theta}(x)^\dagger)
\nonumber\\ 
=&  \hat{\Theta}(x)  \mathscr{D}^{\mu}[\mathscr{A}^{\hat{\Theta}^\dagger}]   \mathscr{A}_{\mu}^{\hat{\Theta}^\dagger}(x)   \hat{\Theta}(x)^\dagger 
\nonumber\\ 
=&  \hat{\Theta}(x)  \partial^{\mu}   \mathscr{A}_{\mu}^{\hat{\Theta}^\dagger}(x)   \hat{\Theta}(x)^\dagger 
 .
 \label{red-cond2}
\end{align}
This implies that imposing the reduction condition $\chi(x)   :=\mathscr{D}^{\mu}[\mathscr{A}]  \mathscr{W}_\mu(x)=0$ is equivalent to imposing the ``Landau gauge condition''  $\partial^{\mu}\mathscr{A}_{\mu}^{\hat{\Theta}^\dagger}(x) = 0$ or transverse condition for the gauge-invariant field  $\mathscr{A}_{\mu}^{\hat{\Theta}^\dagger}(x)$.
Therefore, we can use the (gauge-transformed) reduction condition  $\chi^{\hat{\Theta}^\dagger}$ written as 
\begin{align}
\chi^{\hat{\Theta}^\dagger}(x) 
:= \hat{\Theta}(x)^\dagger \chi(x) \hat{\Theta}(x) 
= \partial^{\mu}  \mathscr{A}_{\mu}^{\hat{\Theta}^\dagger}(x)  = 0 ,
 \label{transverse-A} 
\end{align}
and the associated Faddeev-Popov determinant $\Delta_{\rm FP}^{\rm red}$ reads
\begin{align}
 \Delta_{\rm FP}^{\rm red}[\mathscr{A}^{\hat{\Theta}^\dagger}]
:= \det \left[ \frac{\delta \chi^{\hat{\Theta}^\dagger} }{\delta \hat{\Theta}^\dagger } \right]
= \det \left[ \frac{\delta \partial^\mu \mathscr{A}_{\mu}^{\hat{\Theta}^\dagger} }{\delta \hat{\Theta}^\dagger} \right] .
\label{det-red}
\end{align}
Notice that the reduction condition $\chi^{\hat{\Theta}^\dagger}$ and the associated FP determinant $\Delta^{\rm red}$ are written in terms of $\mathscr{A}_{\mu}^{\hat{\Theta}^\dagger}$ alone, $\chi^{\hat{\Theta}^\dagger}=\chi[\mathscr{A} ^{\hat{\Theta}^\dagger}]$ and $\Delta^{\rm red}=\Delta^{\rm red}[\mathscr{A} ^{\hat{\Theta}^\dagger}]$,
and hence they are gauge invariant.

\begin{widetext}

We show that the massive Yang-Mills (mYM) model in the Landau gauge can be converted to the complementary gauge-scalar (CGS) model, namely, radially fixed  gauge-scalar model subject to  the reduction condition.
In fact, the vacuum expectation value of a gauge-invariant operator $\mathscr{O}[\mathscr{A}]$ of $\mathscr{A}_\mu^A$  reads 
\begin{align}
  \langle \mathscr{O}[\mathscr{A}] \rangle_{\rm mYM} 
%\nonumber\\
:=& \frac{\int \mathcal{D}\mathscr{A}    \Delta_{\rm FP}[\mathscr{A}] \delta(\partial^\mu \mathscr{A}_{\mu} ) e^{iS_{\rm YM}[\mathscr{A} ] +iS_{\rm m}[\mathscr{A} ]} \mathscr{O}[\mathscr{A}]}
{\int \mathcal{D}\mathscr{A}    \Delta_{\rm FP}[\mathscr{A}] \delta(\partial^\mu \mathscr{A}_{\mu} ) e^{iS_{\rm YM}[\mathscr{A} ] +iS_{\rm m}[\mathscr{A} ]}} 
\nonumber\\
=&  
\frac{\int \mathcal{D} \hat{\Theta}^\dagger \int \mathcal{D}\mathscr{A}   \Delta_{\rm FP}[\mathscr{A}]  \delta(\partial^\mu \mathscr{A}_{\mu} ) e^{iS_{\rm YM}[\mathscr{A} ] +iS_{\rm m}[\mathscr{A} ]} \mathscr{O}[\mathscr{A}]}
{\int \mathcal{D} \hat{\Theta}^\dagger \int \mathcal{D}\mathscr{A}    \Delta_{\rm FP}[\mathscr{A}] \delta(\partial^\mu \mathscr{A}_{\mu} ) e^{iS_{\rm YM}[\mathscr{A} ] +iS_{\rm m}[\mathscr{A} ]}} 
\nonumber\\
=&  
\frac{\int \mathcal{D} \hat{\Theta}^\dagger \int \mathcal{D}\mathscr{A}^{\hat{\Theta}^\dagger}   \Delta_{\rm FP}[\mathscr{A}^{\hat{\Theta}^\dagger}] \delta(\partial^\mu \mathscr{A}_{\mu}^{\hat{\Theta}^\dagger} )  e^{iS_{\rm YM}[\mathscr{A}^{\hat{\Theta}^\dagger} ] +iS_{\rm m}[\mathscr{A}^{\hat{\Theta}^\dagger} ]} \mathscr{O}[\mathscr{A}^{\hat{\Theta}^\dagger}]}
{\int \mathcal{D} \hat{\Theta}^\dagger \int \mathcal{D}\mathscr{A}^{\hat{\Theta}^\dagger}   \Delta_{\rm FP}[\mathscr{A}^{\hat{\Theta}^\dagger}] \delta(\partial^\mu \mathscr{A}_{\mu}^{\hat{\Theta}^\dagger} )  e^{iS_{\rm YM}[\mathscr{A}^{\hat{\Theta}^\dagger} ] +iS_{\rm m}[\mathscr{A}^{\hat{\Theta}^\dagger} ]}} 
\nonumber\\
=&  
\frac{\int \mathcal{D} \hat{\Theta}^\dagger \int \mathcal{D}\mathscr{A}  \Delta_{\rm FP}[\mathscr{A}^{\hat{\Theta}^\dagger}] \delta(\partial^\mu \mathscr{A}_{\mu}^{\hat{\Theta}^\dagger} )  e^{iS_{\rm YM}[\mathscr{A} ] +iS_{\rm m}[\mathscr{A}^{\hat{\Theta}^\dagger} ]} \mathscr{O}[\mathscr{A} ]}
{\int \mathcal{D} \hat{\Theta}^\dagger \int \mathcal{D}\mathscr{A}   \Delta_{\rm FP}[\mathscr{A}^{\hat{\Theta}^\dagger}] \delta(\partial^\mu \mathscr{A}_{\mu}^{\hat{\Theta}^\dagger} ) e^{iS_{\rm YM}[\mathscr{A} ] +iS_{\rm m}[\mathscr{A}^{\hat{\Theta}^\dagger} ]}} 
\nonumber\\
=&  
\frac{\int \mathcal{D}\mathscr{A}  \{ \int \mathcal{D} \hat{\Theta}^\dagger  \Delta_{\rm FP}^{\rm red}[\mathscr{A}^{\hat{\Theta}^\dagger}]  \delta(\partial^\mu \mathscr{A}_{\mu}^{\hat{\Theta}^\dagger} )  \} e^{iS_{\rm YM}[\mathscr{A} ] +iS_{\rm kin}[\mathscr{A},\hat{\Theta}^\dagger]} \mathscr{O}[\mathscr{A} ]}
{\int \mathcal{D}\mathscr{A} \{ \int \mathcal{D} \hat{\Theta}^\dagger    \Delta_{\rm FP}^{\rm red}[\mathscr{A}^{\hat{\Theta}^\dagger}]  \delta(\partial^\mu \mathscr{A}_{\mu}^{\hat{\Theta}^\dagger} ) \} e^{iS_{\rm YM}[\mathscr{A} ] +iS_{\rm kin}[\mathscr{A},\hat{\Theta}^\dagger]}}  
\nonumber\\
=&  
\frac{\int \mathcal{D}\mathscr{A} e^{iS_{\rm YM}[\mathscr{A} ] +iS_{\rm kin}[\mathscr{A},\hat{\Theta}^\dagger]} \mathscr{O}[\mathscr{A} ]}
{\int \mathcal{D}\mathscr{A}  e^{iS_{\rm YM}[\mathscr{A} ] +iS_{\rm kin}[\mathscr{A},\hat{\Theta}^\dagger}]}
=:  \langle \mathscr{O}[\mathscr{A}] \rangle_{\rm CGS} 
 ,
\end{align}  
where the normalized matrix-scalar field $\hat{\Theta}$ is introduced and the integration over the gauge volume $\int \mathcal{D} \hat{\Theta}^\dagger$ is inserted in the second equality, 
  the integration variable $\mathscr{A}$ is renamed to $\mathscr{A}^{\hat{\Theta}^\dagger}$ in the third equality, 
 the gauge invariance of the Yang-Mills action $S_{\rm YM}[\mathscr{A}^{\hat{\Theta}^\dagger}]=S_{\rm YM}[\mathscr{A} ]$, the integration measure $\mathcal{D}\mathscr{A}^{\hat{\Theta}^\dagger}=\mathcal{D}\mathscr{A}$ and the operator $\mathscr{O}[\mathscr{A}^{\hat{\Theta}^\dagger}]=\mathscr{O}[\mathscr{A} ]$ is used in the fourth equality, 
 and the FP determinant $\Delta_{\rm FP}[\mathscr{A}] $ for the Landau gauge $ \partial^\mu \mathscr{A}_{\mu}=0$ in the massive Yang-Mills model is identified with the FP determinant (\ref{det-red}) for the reduction condition  (\ref{transverse-A})  in the fifth equality.
In the last step, the delta function $\delta(\hat{\Theta}^\dagger, h[\mathscr{A}])$ on the group $G$ satisfying  
$\int \mathcal{D} \hat{\Theta}^\dagger \delta(\hat{\Theta}^\dagger, h[\mathscr{A}])=1$  
 is used to rewrite  
\begin{align}
 \delta(\hat{\Theta}^\dagger, h[\mathscr{A}]) 
 = \det \left[ \frac{\delta \partial^\mu \mathscr{A}_{\mu}^{\hat{\Theta}^\dagger} }{\delta \hat{\Theta}^\dagger} \right] \Biggr|_{\hat{\Theta}^\dagger= h[\mathscr{A}]} 
 \delta( \partial^\mu \mathscr{A}_{\mu}^{\hat{\Theta}^\dagger} ) 
= \Delta_{\rm FP}^{\rm red}[\mathscr{A}^{\hat{\Theta}^\dagger}] \delta( \partial^\mu \mathscr{A}_{\mu}^{\hat{\Theta}^\dagger} )  ,
 \label{FP-red-proc1}
\end{align}
%where the equality holds 
which is valid when the following equation for a given $\mathscr{A}_{\mu}$ has a unique solution of $h=h[\mathscr{A}] \in G$,
\begin{align}
%\partial_{\mu}  \mathscr{A}_{\mu}^{\hat{\Theta}^\dagger}(x) =
 \partial_{\mu}  \mathscr{A}_{\mu}^{h[\mathscr{A}]}(x)   = 0 .
\end{align}
\end{widetext}
This uniqueness of the solution corresponds to assuming that there are no Gribov copies if $\partial_{\mu}  \mathscr{A}_{\mu}^{h[\mathscr{A}]}(x) = 0$ is regarded as the gauge-fixing condition.  
Notice that we have taken into account the radially fixed constraint (\ref{SU2-s-constraint1}) in replacing the scalar field ${\Theta}^\dagger$ by the normalized matrix-valued (or group-valued) scalar field $\hat{\Theta}^\dagger$ in the last step.

We have assumed that the solution is unique in showing the equivalence in the above. 
Therefore, the equivalence is valid up to the Gribov copies. As mentioned already, however, we do not intend to seriously consider the Gribov problem in this paper, since we take the same standpoint as before explained in the above.  

%Incidentally, by taking the absolute Landau gauge, 
Incidentally, by adopting the absolute Landau gauge for $\mathscr{A}_{\mu}^{\hat{\Theta}^\dagger}$ as the reduction condition, 
we can extract the gauge field configuration as the unique solution without Gribov copies.  Then we can show the exact equivalence between the massive Yang-Mills model and a specific gauge-scalar model.  Consequently, the resulting theory inevitably becomes nonloal as expected from the effective theory, which however does not affect the perturbative analysis done in this paper.

\subsection{
Solving the reduction condition
}

In the complementary gauge-scalar model, the scalar field $\Phi$ and the gauge field $\mathscr{A}$ are not independent field variables, because we intend to obtain the massive pure Yang-Mills theory which does not contain the scalar field $\Phi$.
Therefore, the scalar field $\Phi$  is to be eliminated in favor of the gauge field $\mathscr{A}$.  
This is in principle achieved by solving the reduction condition as an off-shell equation, which is different from solving the field equation for the scalar field $\Phi$ as adopted in the preceding studies \cite{Stueckelberg38,KG67,SF70,Cornwall74,Cornwall82,DT86}. 
\footnote{
See e.g., \cite{DTT88,RRA04} for reviews of the St\"uckelberg field. 
Notice that the reduction condition is an off-shell condition. 
Therefore, solving the reduction condition is different from solving the field equation for the St\"uckelberg field as done in the preceding works \cite{Capri-etal05}.
This means that the solution of the reduction condition does not necessarily satisfy the field equation, while the solution of the field equation of the complementary gauge-scalar model automatically satisfies the reduction condition \cite{Kondo18}.
}
%However, the resulting expression for the scalar field $\Phi$ would be given by a complicated form, e.g., an infinite series in perturbation theory. 
%In particular, the scalar field $\Phi$ becomes trivial when the gauge fields is transverse, namely, in the Landau gauge $\partial^\mu \mathscr{A}_\mu=0$.  
Consequently, the resulting massive Yang-Mills model with the covariant gauge-fixing term and the associated Faddeev-Popov ghost term becomes power-counting renormalizable in the perturbative framework, as demonstrated to  one-loop order in the next section. 

Moreover, the entire theory is invariant under the usual Becchi-Rouet-Stora-Tyutin (BRST) transformation $\delta_{BRST}$.  The nilpotency $\delta_{BRST}\delta_{BRST}=0$ of the usual BRST transformations ensures the unitarity of the theory in the physical subspace of the total state vector space  as the  BRST-invariant sector according to Kugo and Ojima  \cite{KO79}. 
This situation should be compared with the Curci-Ferrari model \cite{CF76b} which is not invariant under the ordinary BRST transformation, but instead  can be made invariant under the modified BRST transformation $\delta_{BRST}^\prime$. Nevertheless, this fact does not guarantee the unitarity of the  Curci-Ferrari model  due to the lack of usual nilpotency   of the modified BRST transformation satisfying $\delta_{BRST}^\prime\delta_{BRST}^\prime\delta_{BRST}^\prime=0$, see e.g., \cite{Kondo13}.

We proceed to eliminate the scalar field $\Phi$ or $\Theta$ by solving the reduction condition to obtain the massive Yang-Mills model from the complementary gauge-scalar model  
\begin{align}
  \langle \mathscr{O}[\mathscr{A}] \rangle_{\rm CGS} 
 = \langle \mathscr{O}[\mathscr{A}] \rangle_{\rm mYM} 
 .  
\end{align}
Notice that introducing the reduction condition does not break the original gauge symmetry. 
The general form of the transverse and gauge-invariant Yang-Mills gauge field $\mathscr{A}_{\mu}^{h[\mathscr{A}]}$ satisfying (\ref{transverse-A})  can be obtained explicitly by order by order expansion in powers of the gauge field $\mathscr{A}$ up to the \textit{Gribov copies}.
Indeed, $\mathscr{A}_{\mu}^{h[\mathscr{A}]}$ satisfying the transverse condition,  
\begin{align}
\partial_{\mu} \mathscr{A}_{\mu}^{h[\mathscr{A}]} = 0 ,
\end{align}
is obtained as a power series in $\mathscr{A}$, %see e.g. \cite{Capri-etal05}:
\begin{align}
\mathscr{A}_{\mu}^{h[\mathscr{A}]} = &
\mathscr{A}_{\mu}^T
- i g \frac{\partial_{\mu}}{\partial^{2}} \biggl[ \mathscr{A}_{\nu} , \partial_{\nu} \frac{\partial \cdot \mathscr{A}}{\partial^{2}} \biggr]
\nonumber\\
&- \frac{i}{2} g \frac{\partial_{\mu}}{\partial^{2}} \biggl[ \partial \cdot \mathscr{A} , \frac{1}{\partial^{2}} \partial \cdot \mathscr{A} \biggr]
  + i g \biggl[ \mathscr{A}_{\mu} , \frac{1}{\partial^{2}} \partial \cdot \mathscr{A} \biggr]
\nonumber\\
&+ \frac{i}{2} g \biggl[ \frac{1}{\partial^{2}} \partial \cdot \mathscr{A} , \frac{\partial_{\mu}}{\partial^{2}} \partial \cdot \mathscr{A} \biggr] 
+ \mathcal{O} (\mathscr{A}^{3})
,
\label{solution_A}
\end{align}
where we have defined the transverse  field $\mathscr{A}_{\mu}^T$ in the lowest order term linear in $\mathscr{A}$ as 
\begin{align}
\mathscr{A}_{\mu}^T := \mathscr{A}_{\mu} -\partial_{\mu} \frac{\partial \cdot \mathscr{A}}{\partial^2} .
\end{align}
Then we find that the transverse  field $\mathscr{A}_{\mu}^{h[\mathscr{A}]}$ is rewritten into 
\begin{align}
\mathscr{A}_{\mu}^{h[\mathscr{A}]} = & \left( \delta_{\mu \nu} - \frac{\partial_{\mu} \partial_{\nu}}{\partial^{2}} \right) \Psi_{\nu} ,  
\nonumber\\
\Psi_{\nu} = & \mathscr{A}_{\nu} - i g \biggl[ \frac{1}{\partial^{2}} \partial \cdot \mathscr{A} , \mathscr{A}_{\nu} \biggr]
\nonumber\\
&+ \frac{i}{2} g \biggl[ \frac{1}{\partial^{2}} \partial \cdot \mathscr{A} , \partial_{\nu} \frac{1}{\partial^{2}} \partial \cdot \mathscr{A} \biggr]
+ \mathcal{O} (\mathscr{A}^{3})
.
\label{solution_A2}
\end{align}
Under an infinitesimal gauge transformation $\delta_\Lambda$ defined by  
%\begin{align}
%  \delta_\Lambda \mathscr{A}_{\mu}^{\hat{\Theta}^\dagger}(x) = 0 ,
% \label{invariant-A}
%\end{align}
$%\begin{equation}
\delta_\Lambda \mathscr{A}_{\mu} = \mathscr{D}_{\mu} [\mathscr{A}] \Lambda
:= \partial_{\mu} \Lambda - i g [ \mathscr{A}_{\mu} , \Lambda ]
 ,
%\label{ing-gt}
$ %\end{equation}
 $\Psi_{\nu}$ transforms as
\begin{equation}
\delta_\Lambda \Psi_{\nu} = \partial_{\nu} \left( \Lambda - \frac{i}{2} g  \biggl[ \frac{\partial \cdot \mathscr{A}}{\partial^{2}} , \Lambda \biggr] \right)
+ \mathcal{O} (g^{2})
.
\label{gauge_transform_psi}
\end{equation}
Therefore, $\mathscr{A}_{\mu}^{h}$ given by (\ref{solution_A2}) is left invariant by infinitesimal gauge transformations order by order of the expansion,% in the coupling constant $g$:
\begin{equation}
\delta_\Lambda \mathscr{A}_{\mu}^{h[\mathscr{A}]} (x) = 0
.
\end{equation}
In Appendix~A, we give a recursive construction of the transverse field $\mathscr{A}_{\mu}^{h[\mathscr{A}]}$ and the proof of gauge invariance  of the resulting $\mathscr{A}_{\mu}^{h[\mathscr{A}]}$.

The mass term of $\mathscr{W}_\mu$ is equal to that of $\mathscr{A}_{\mu}^{\hat{\Theta}^\dagger}$, 
\begin{align}
 \mathscr{L}_{\rm kin} 
%=&    \frac{1}{2}{\rm tr}(  ( {D}_{\mu}[\mathscr{A}] \Theta (x))^\dagger 	{D}^{\mu}[\mathscr{A}] \Theta(x) )  
%\nonumber\\
=&    M^2{\rm tr}(\mathscr{W}_\mu(x) \mathscr{W}^\mu(x) )  
\nonumber\\
=&    M^2{\rm tr}(\mathscr{A}_{\mu}^{\hat{\Theta}^\dagger}(x)\mathscr{A} ^{\mu \hat{\Theta}^\dagger}(x))  .
\label{kin=mass3}
\end{align}
Therefore, the ``mass term'' of gauge-invariant field $\mathscr{A}_{\mu}^{h}$   is used to rewrite the kinetic term of the scalar field:
%The resulting mass term for $W$ agrees with the gauge-fixing functional given by
\begin{align}
& S_{\rm kin}^*[\mathscr{A}] 
\nonumber\\
=&   \int d^Dx \ M^2 {\rm tr}(\mathscr{A}_\mu^{h[\mathscr{A}]} \mathscr{A}^{\mu h[\mathscr{A}]} ) .
\nonumber\\  
=&  \int d^Dx \ M^2 {\rm tr}\Big\{  \mathscr{A}_{\mu}^T  \mathscr{A}^{\mu T}      
 -i   g \mathscr{A}_{\mu}^T \left[\frac{\partial \cdot \mathscr{A}}{\partial^2} ,\partial_{\mu} \frac{\partial \cdot \mathscr{A}}{\partial^2} \right] 
 \Big\} 
\nonumber\\ &
 + \mathcal{O}(\mathscr{A}^4) ,
 \label{Sh}
\end{align}
In this way, we have eliminated the scalar field by solving the reduction condition. 

Only when we adopt the covariant Landau gauge  $\partial \cdot \mathscr{A}=0$ as the gauge-fixing condition, the infinite number of nonlocal terms disappear so that $S_{\rm kin}^*$ reduces to the naive mass term of $\mathscr{A}$, 
\begin{align}
 S_m[\mathscr{A}] =  \int d^Dx \  M^2 {\rm tr} (\mathscr{A}_{\mu}(x)  \mathscr{A}_{\mu}(x) )  .
\end{align}
In the Landau gauge, thus, the complementary gauge-scalar model with the reduction condition reduces to the massive Yang-Mills model with the naive mass term.
% which is renormalizable. 

The explicit expression of the massive vector field $\mathscr{W}_\mu$ in terms of $\mathscr{A}_\mu$ is given in  Appendix~B. 
Notice that $\mathscr{W}_\mu$ agrees with $\mathscr{A}_{\mu}^T=\mathscr{A}_{\mu}$ in the Landau gauge $\partial \cdot \mathscr{A}=0$.

%\newpage
%%%%%%%%%%%%%%%%%%%%%%%%%%%%%%%%%%%%%%%%%%%%%%%%%%%%%%%%%%%%%
%%%%%%%%%%%%%%%%%%%%%%%%%%%%%%%%%%%%%%%%%%%%%%%%%%%%%%%%%%%%%
\section{
Massive Yang-Mills model and decoupling solutions
}
%%%%%%%%%%%%%%%%%%%%%%%%%%%%%%%%%%%%%%%%%%%%%%%%%%%%%%%%%%%%%
%%%%%%%%%%%%%%%%%%%%%%%%%%%%%%%%%%%%%%%%%%%%%%%%%%%%%%%%%%%%%

%%%%%%%%%%%%%%%%%%%%%%%%%%%%%%%%%%%%%%%%%%%%%%%%%%%%%%%%%%%%%
%\subsection{
%Quantum correction to the massive Yang-Mills model and decoupling solution
%}
%%%%%%%%%%%%%%%%%%%%%%%%%%%%%%%%%%%%%%%%%%%%%%%%%%%%%%%%%%%%%

%%%%%%%%%%%%%%%%%%%%%%%%%%%%%%%%%%%%%%%%%%%%%%%%%%%%%%%%%%%%%

In this section we give a review of the pertubative results \cite{TW10,RSTW17} obtained for the massive Yang-Mills model and reconfirm them from our viewpoint for later convenience. 

In order to reproduce the decoupling solution of the Yang-Mills theory in the covariant Landau gauge, we calculate one-loop quantum corrections to the gluon and ghost propagators in the massive Yang-Mills model.
The Nakanishi-Lautrup field $\mathscr{N}^A$ can be eliminated  so that the gauge-fixing term reduces to
\begin{align}
 \mathscr{L}_{GF} \to -\frac{1}{2}\alpha^{-1} (\partial^\mu \mathscr{A}^A_\mu)^2 .
\end{align} 
The results in the Landau gauge is obtained by taking the limit $\alpha \to 0$ in the final step of the calculations. 
Only in the Landau gauge $\alpha=0$ the massive Yang-Mills model with a mass term $\mathscr{L}_m$ has the gauge-invariant extension.
In order to obtain the gauge-independent results in the other gauges with $\alpha \not=0$, we need to include an infinite number of non-local terms in addition to the naive mass term $\mathscr{L}_m$ for gluons, as shown in the previous section.   

%Therefore, the results obtained in the Feynman gauge $\alpha=1$ from this theory must be different from those in the Landau gauge $\alpha=0$. 
%We confirm that this is indeed the case by examining the positivity violation/restoration transition. 

\subsection{Feynman rules for the massive Yang-Mills model}

\begin{figure}[t]
\centering
\includegraphics[height=4.0cm]{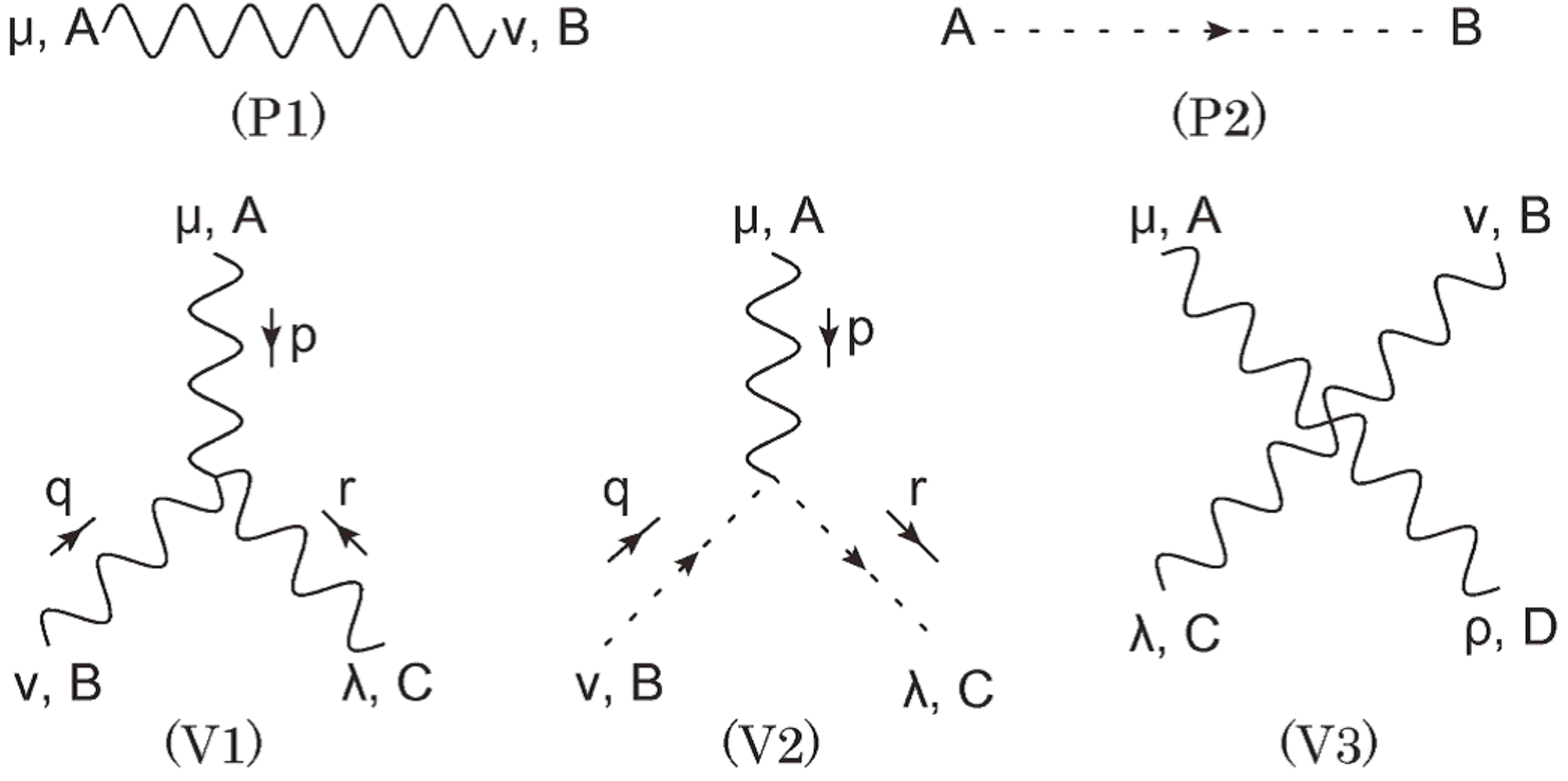}
\caption{\small
Feynman rules for the massive Yang-Mills model in the covariant gauge.
}
\label{feynrule1}
\end{figure}

The Feynman rules for the massive Yang-Mills model are given as follows. 
The diagrammatic representations of the Feynman rules are given in Fig.~\ref{feynrule1}.

\begin{description}
\item[(P1) gluon propagator $\langle\mathscr{AA}\rangle$]
\begin{subequations}
\begin{flalign}
 & \tilde D^{AB}_{\mu\nu}\left(k\right)
\nonumber\\
&:=\frac{-\delta^{AB}}{k^2-M^2}\left[g_{\mu\nu}-\left(1-\alpha\right)\frac{k_\mu k_\nu}{k^2-\alpha M^2}\right] 
\\
&=\delta^{AB}\left[\frac{-1}{k^2-M^2}\left(g_{\mu\nu}-\frac{k_\mu k_\nu}{M^2}\right)-\frac{k_\mu k_\nu}{M^2}\frac{1}{k^2-\alpha M^2}\right] 
\label{propM}
\\
&=\delta^{AB}\left[\frac{-1}{k^2-M^2}\left(g_{\mu\nu}-\frac{k_\mu k_\nu}{k^2}\right)-\frac{\alpha}{k^2-\alpha M^2}\frac{k_\mu k_\nu}{k^2}\right] ,
\end{flalign}
\label{propTL}
\end{subequations}
\item[(P2) ghost propagator $\langle\mathscr{C\bar{C}}\rangle$]
\begin{flalign}
\Delta_{gh}^{AB}\left(k\right):=\frac{-i\delta^{AB}}{k^2+i\epsilon} ,
\end{flalign}
\item[(V1) three-gluon vertex function $\langle\mathscr{AAA}\rangle$]
\begin{flalign}
\Gamma^{ABC}_{\mu\nu\lambda}\left(p,q,r\right) 
&=gf^{ABC} [(q-r)_\mu g_{\nu\lambda}+(r-p)_\nu g_{\mu\lambda}\nonumber\\ &
\hspace{35pt}+(p-q)_\lambda g_{\mu\nu} ] ,
%\nonumber\\ &
%:= gf^{ABC}V_{\mu\nu\lambda}\left(p,q,r\right) ,
\end{flalign}
\item[(V2) gluon-ghost-antighost vertex function $\langle\mathscr{AC\bar{C}}\rangle$]
\begin{flalign}
\Gamma^{ABC}_\mu\left(p,q,r\right):= igf^{ABC}r_\mu ,
\end{flalign}
\item[(V3) four-gluon vertex function $\langle\mathscr{AAAA}\rangle$] 
\begin{flalign}
\Gamma^{ABCD}_{\mu\nu\lambda\rho}\left(p,q,r,k\right)
 =&-ig^2 [f^{ABE}f^{ECD}(g_{\mu\lambda}g_{\nu\rho}-g_{\mu\rho}g_{\nu\lambda})
\nonumber\\ &
\hspace{15pt} +f^{ADE}f^{EBC}(g_{\mu\nu}g_{\lambda\rho}-g_{\mu\lambda}g_{\nu\rho}) 
\nonumber\\ &
\hspace{15pt}+f^{ACE}f^{EBD}(g_{\mu\nu}g_{\lambda\rho}-g_{\mu\rho}g_{\nu\lambda}) ]  .
%\notag \\
%:=&-ig^2W^{ABCD}_{\mu\nu\lambda\rho} .
\end{flalign}
\end{description}
Here the momentum conservation is omitted and the momentum flow at each vertex is regarded as incoming, while the momentum of antighost as outgoing. 
Notice that the Feynman rules are the same as those of the ordinary Yang-Mills theory in the Lorenz gauge except for the gluon propagator which was replaced by the massive propagator (\ref{propTL}). 

\begin{figure}[t]
\centering\includegraphics[height=2.0cm]{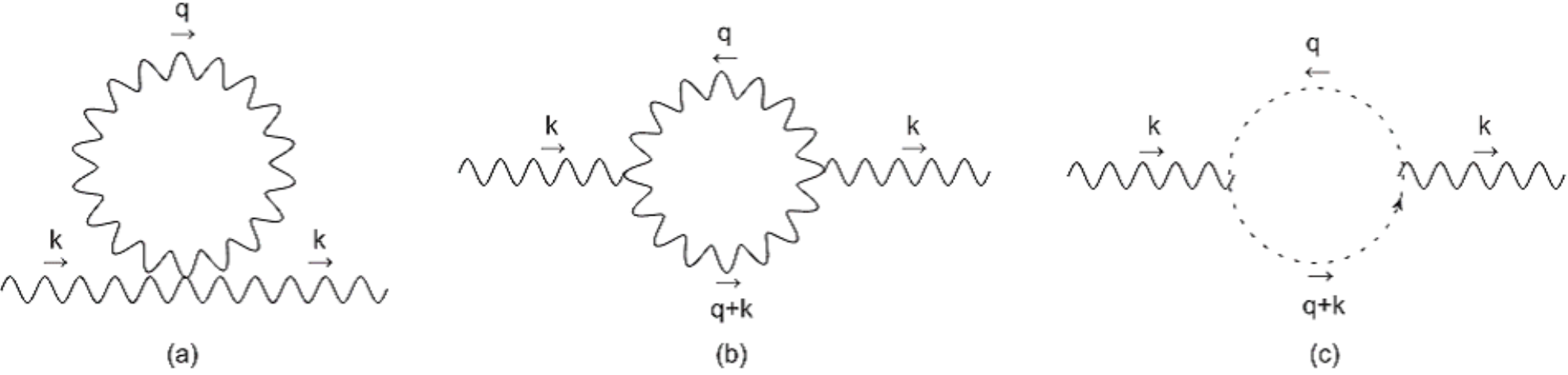}
\quad\quad\quad\quad
\centering\includegraphics[height=2cm]{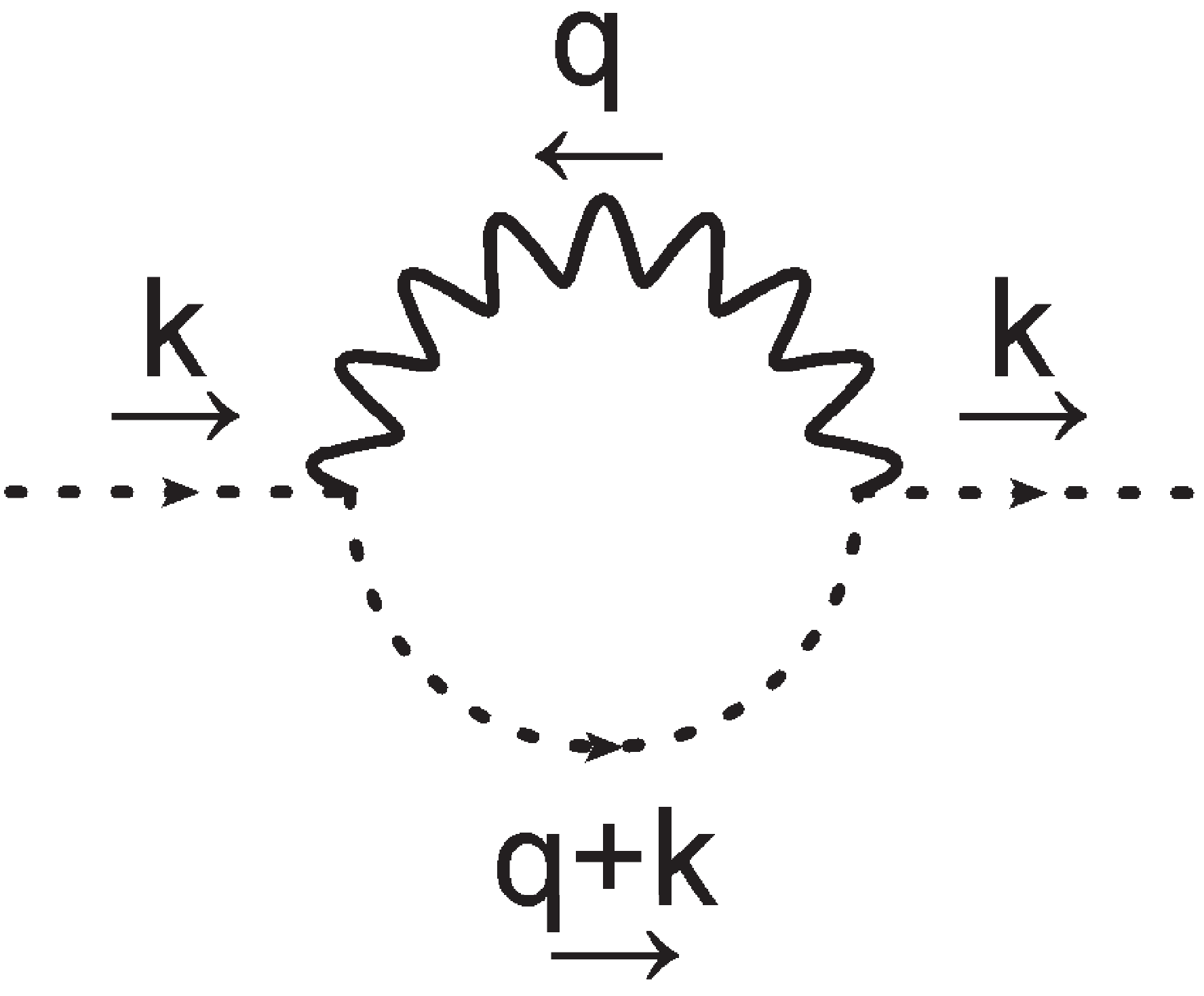}
\caption{\small
(top) gluon vacuum polarization diagrams (a),(b),(c) to one-loop order, 
(bottom) ghost self-energy diagram to one-loop order.
}
\label{1-loop}
\end{figure}

The gluon propagator (\ref{propTL}) has the same form as that in the renormalizable $R_\xi$ gauge where unitarity is not manifest. For any finite values of $\alpha$, the gluon propagator has good high-energy behavior, namely, the asymptotic behavior $O(1/k^{2})$ as $k \to \infty$, and hence the theory is renormalizable by power counting. 
%\footnote{ 
%This theory is multiplicatively renormalizable to all orders of perturbation theory. See e.g.. Dudal et al.\cite{Dudal-etal03}.
%}
For example, the choice $\alpha=1$ leads to the propagator $\frac{-1}{k^2-M^2} g_{\mu\nu}$. 
In the limit $\alpha \to \infty$, the gluon propagator reduces to the standard form for a massive spin-one particle, as can be easily seen in the second form. In the unitary gauge particle content is manifest, since there are no unphysical fields, and hence unitarity is transparent, while renormalizability is not transparent. 

For any finite values of $\alpha$, the gluon propagator has an extra unphysical pole at $k^2=\alpha M^2$ besides the physical pole (massive gauge bosons) at $k^2=M^2$, as can be seen in the second form of (\ref{propTL}). In order to preserve unitarity, the unphysical poles must be eliminated or mutually cancel in the $S$-matrix element involving only physical particles. In the spontaneously broken gauge theory, the would-be Nambu-Goldstone boson field has the propagator with the unphysical pole at $k^2=\alpha M^2$, and this unphysical pole of the would-be Nambu-Goldstone particle cancels one of the gauge boson in order to preserve unitarity. This is not the case in our model, since there are no Nambu-Goldstone particles without spontaneous symmetry breaking. The above type of cancellation of unphysical poles can be proven to all orders in perturbation theory by using the generalized Ward-Takahashi identities which are a consequence of the gauge invariance of the theory.   

In the limit $\alpha \to 0$, however, the gluon propagator reduces to the simple form for a massive spin-one particle with the transverse projector $ \frac{-1}{k^2-M^2}\left(g_{\mu\nu}-\frac{k_\mu k_\nu}{k^2}\right)$, as can be seen in the third form of  (\ref{propTL}), and the contribution from the unphysical pole at $k^2=0$ disappears in this limit. Therefore, the Landau gauge is the very special gauge which guarantees renormalizability and 
 allows the existence of the gauge-invariant extension as demonstrated for the massive Yang-Mills model in the previous section.
%unitarity in agreement with the gauge invariance of the theory. This result is consistent with our point of view that the massive Yang-Mills model has the gauge-invariant extension only in the Landau gauge. 

\subsection{One-loop quantum corrections and renormalization }

We now take into account quantum corrections to the gluon and ghost propagators to one-loop order.
In Fig.~\ref{1-loop}, we enumerate the one-loop diagrams which contribute to the gluon and ghost propagators to one-loop order.

%
%\begin{figure}[h]
%\centering\includegraphics[height=2cm]{suda/fig3.pdf}
%\caption{\small
%ghost self-energy diagram to one loop}
%\label{fig3}
%\end{figure}
%

In the \textit{massive Yang-Mills model}  we introduce the renormalization factors $Z_{\mathscr{A}}, Z_{\mathscr{C}}=Z_{\mathscr{\bar C}}, Z_{g}, Z_{M^2}, \tilde{Z}_\alpha$ to connect the bare unrenormalized fields (gluon $\mathscr{A}_B$, ghost $\mathscr{C}_B$ and antighost $\mathscr{\bar C}_B$) and bare parameters (the coupling constant $g_B$, the mass parameter $M_B$ and the gauge-fixing parameter $\alpha_B$) to the renormalized fields ($\mathscr{A}_R$, $\mathscr{C}_R$ and  $\mathscr{\bar C}_R$) and renormalized parameters ($g_R$, $M_R$ and $\alpha_R$) respectively \cite{Taylor71,BSNW96,Wschebor08}:
\begin{align}
 & \mathscr{A}_B = Z_{\mathscr{A}}^{1/2} \mathscr{A}_R , \
 \mathscr{C}_B = Z_{\mathscr{C}}^{1/2} \mathscr{C}_R , \
 \mathscr{\bar C}_B = Z_{\mathscr{C}}^{1/2} \mathscr{\bar C}_R , \
\nonumber\\ 
& g_B = Z_{g} g_R , \
 M_B^2 = Z_{M^2} M_R^2 , \
 \alpha_B = \tilde{Z}_\alpha^{-1} Z_{\mathscr{A}} \alpha_R .
\end{align}
For comparison with the lattice data, we move to the Euclidean region and use $k_{E}$ to denote the Euclidean momentum so that $k^2=-k_{E}^2$.

For gluons, we introduce the two-point vertex function $\Gamma_{\mathscr{A}}^{(2)}$ as the inverse of the transverse part $\mathscr{D}_{\rm T}$ of the propagator 
\footnote{
In this paper we focus on the Landau gauge.  For the gluon propagator, therefore, we discuss the transverse part alone. 
}
and the vacuum polarization function $\Pi_T$ as
\begin{align}
\Gamma_{\mathscr{A}}^{(2)}(k_{E}) &:= [ \mathscr{D}_{\rm T}(k_{E}^2) ]^{-1} 
\nonumber\\ &
 = k_{E}^2 + M^2 + \Pi_T(k_{E}^2) + k_{E}^2 \delta_Z + M^2 \delta_{M^2} 
\nonumber\\ &
 = k_{E}^2 + M^2 + \Pi_T^{\rm fin}(k_{E}^2)  ,
%\nonumber\\
%&:= M^2  [ \hat{\mathscr{D}}_{\rm T}(s) ]^{-1}
%\nonumber\\
%&= M^2 \left[ s + 1 + \hat{\Pi}_T(s) + s \delta_Z + \delta_{M^2} \right]  
%\nonumber\\
%&= M^2 \left[ s + 1 + \hat{\Pi}_{\rm fin}(s) \right] ,
%\nonumber\\
%&:= M^2 \hat{\Gamma}(s) ,
%\nonumber\\ &
%s:=\frac{k^2}{M^2} ,
\end{align}
where
$\delta_Z$ and $\delta_{M^2}$ are counterterms to cancel the divergence coming from the vacuum polarization function $\Pi_T$ to obtain the finite renormalized one ${\Pi}_T^{\rm fin}$ 
\begin{align}
 \Pi_T^{\rm fin}(k_{E}^2) := \Pi_T(k_{E}^2) + k_{E}^2 \delta_Z + M^2 \delta_{M^2} ,
\end{align}
under the suitable renormalization conditions to be discussed shortly, 
and they are related to the renormalization factors as 
\begin{align}
\delta_Z = Z_{\mathscr{A}} -1 ,
\quad
 \delta_{M^2} = Z_{M^2} Z_{\mathscr{A}} -1 .
 \label{delta-Z1}
\end{align}
We define the dimensionless versions $\hat{\mathscr{D}}_{\rm T}(s)$ and $\hat{\Pi}(s)$ 
% and $\hat{\Gamma}(s)$ 
of $\mathscr{D}_{\rm T}(k_{E})$ and $\Pi_T(k_{E}^2)$ 
% and $\Gamma_{\mathscr{A}}^{(2)}(k_{E})$
with the hat respectively
\begin{align}
\Gamma_{\mathscr{A}}^{(2)}(k_{E})/M^2  
&:=  [\hat{\mathscr{D}}_{\rm T}(s) ]^{-1}
\nonumber\\
&=  s + 1 + \hat{\Pi}_T(s) + s \delta_Z + \delta_{M^2}  
\nonumber\\
&=  s + 1 + \hat{\Pi}_T^{\rm fin}(s) ,
%\nonumber\\
%&:= M^2 \hat{\Gamma}(s) ,
%\nonumber\\ &
%s:=\frac{k^2}{M^2} ,
\end{align}
with
the dimensionless squared momentum
\begin{align}
 s :=  \frac{k_{E}^2}{M^2}, %, \ \nu:=\frac{\mu^2}{M^2} .
\end{align}
and 
\begin{align}
 \hat{\Pi}_T^{\rm fin}(s) := \hat{\Pi}_T(s) + s \delta_Z + \delta_{M^2} .
 \label{Pi_T^fin}
\end{align}

The gluon vacuum polarization function in the covariant Landau gauge $\alpha=0$ calculated using the dimensional regularization  in Euclidean space is given to one-loop order as the power-series Laurent expansion in $\epsilon :=2-\frac{D}{2}$ 
\footnote{
These expressions are obtained by taking the limit $\alpha \to 0$ of those with an arbitrary $\alpha$ given in \cite{Suda18}.
}
\begin{align}
\hat{\Pi}_T (s) &= \frac{g^2 C_2 (G)}{16 \pi^2} \frac{1}{12}   
\notag \\ & \times 
 s \Biggl[ \left( \frac{9}{s} - 26 \right) \left\{  \epsilon^{-1}  - \gamma + \ln  \left( 4 \pi \right) + \ln \eta \right\} 
\nonumber\\ & \hspace{10pt}
+ \frac{63}{s} - \frac{121}{3}     
+ h(s) \Biggr] ,
%\notag \\ &
%\hspace{10pt} - \frac{1}{s^2}  + \left( 1 - \frac{s^2}{2} \right) \ln  \left( s \right) 
%\nonumber\\ &
%\hspace{10pt} + \left( 1 + \frac{1}{s} \right)^3 \left(s^2 - 10s + 1 \right) \ln  \left( s + 1 \right) 
%\notag \\ &
%\hspace{10pt} + \frac{1}{2} \left( 1 + \frac{4}{s} \right)^\frac{3}{2} \left(s^2 - 20s + 12 \right) \ln \left( \frac{\sqrt{4 + s} - \sqrt{s}}{\sqrt{4 + s} + \sqrt{s}} \right) \Biggr] , 
%\nonumber\\ &
%\ s := \frac{k^2}{M^2} , \ \eta:=\frac{\mu^2}{M^2} .
\label{VP1}
\end{align}
where $C_2 (G)$ is the quadratic Casimir operator of a gauge group $G$, $\gamma$ is the Euler constant, 
and $\eta$ is the value of $s$ at the scale $\tilde\mu$ introduced through the dimensional regularization for dimensional reasons
\begin{align}
\eta:=\frac{\tilde\mu^2}{M^2} .
\end{align}
Here we have defined the functions of $s$, 
\begin{align}
h(s) :=& K_1(s) + K_2(s) + K_3(s) ,
\nonumber\\  
K_1(s) :=& - \frac{1}{s^2}  
 + \left( 1 - \frac{s^2}{2} \right) \ln s   ,
\nonumber\\  
K_2(s) :=& \left( 1 + \frac{1}{s} \right)^3 \left(s^2 - 10s + 1\right) \ln  \left( s + 1 \right) ,
\nonumber\\  
K_3(s)  :=& \frac{1}{2} \left( 1 + \frac{4}{s} \right)^\frac{3}{2} \left(s^2 - 20s + 12 \right) \ln \left( \frac{\sqrt{4 + s} - \sqrt{s}}{\sqrt{4 + s} + \sqrt{s}} \right) .
\end{align}
Notice that there are no singular term in the finite part $\hat{\Pi}_T^{\rm fin}(s)$ even at $s=0$, because there does not exist $O(s^{-2})$ term in the bracket [...] of (\ref{VP1}), since the expansion of $h(s)$ around $s=0$ reads 
\begin{align}
h(s) %:=& K_1(s) + K_2(s) + K_3(s) 
%\nonumber\\  
=& -\frac{111}{2 s}+\left( \ln s +\frac{251}{6}\right)+\frac{389 s}{60}
\nonumber\\  &
+s^2
   \left(-\frac{1}{2}\ln s -\frac{141}{140}\right)+\frac{269
   s^3}{420}+O\left(s^4\right) , 
\end{align}
which follows from 
\begin{align}
K_2(s) =& \frac{1}{s^2}-\frac{15}{2 s}-\frac{133}{6}-\frac{187
   s}{12}-\frac{43 s^2}{60}+\frac{23
   s^3}{30}+O\left(s^4\right) , 
\nonumber\\  
K_3(s) =& -\frac{48}{s}+64+\frac{331 s}{15}-\frac{61 s^2}{210}-\frac{53s^3}{420}+O\left(s^4\right) . 
\end{align}
Thus, we have the finite part of the gluon vacuum polarization to one-loop 
\begin{align}
\hat{\Pi}_T^{\rm fin}{}^\prime(s) =  \frac{g^2 C_2 (G)}{16 \pi^2} \frac{1}{12} s \left[ \frac{63}{s} - \frac{121}{3}    + h(s) \right] ,
\end{align}
which has  the $s = 0$ limit, 
%\begin{align}
%\hat{\Pi}(s = 0) 
%&= g^2 C_2 (G) \frac{1}{48 \pi^2} \frac{1}{4} \Big[ 9 \left\{\epsilon^{-1} - \gamma - \ln  \left( 4 \pi \right) + \ln  \left( \nu \right) \right\} 
%\nonumber\\ &
%+ 63 - \frac{15}{2} - 48  \Big] 
%\notag \\
%=  \frac{g^2 C_2 (G)}{16 \pi^2} \frac{1}{12} \Big[ 9 \left\{\epsilon^{-1} - \gamma - \ln  \left( 4 \pi \right) + \ln  \left( \nu \right) \right\} 
%\nonumber\\ &
%+ \frac{15}{2} \Big] ,
%\end{align}
\begin{align}
\hat{\Pi}_T^{\rm fin}{}^\prime(s = 0) 
=  \frac{g^2 C_2 (G)}{16 \pi^2} \frac{1}{12} \frac{15}{2}  .
\end{align}
%In this original expression, we calculate the value at $s = 0$ as
%\begin{align}
%\tilde{\Pi}(s = 0) &= g^2 C_2 (G) \frac{1}{48 \pi^2} \frac{1}{4} \left[ 9 \left\{\epsilon^{-1} - \gamma - \ln  \left( 4 \pi \right) + \ln  \left( \nu \right) \right\} + 63 - \frac{15}{2} - 48 \right] \notag \\
%&= g^2 C_2 (G) \frac{1}{48 \pi^2} \frac{1}{4} \left[ 9 \left\{\epsilon^{-1} - \gamma - \ln  \left( 4 \pi \right) + \ln  \left( \nu \right) \right\} + \frac{15}{2} \right].
%\end{align}
%Then we do (i) and (ii) procedure.

%The divergences coming from the gluon vacuum polarization function is removed by adding the counterterms $k^2 \delta_Z + M^2 \delta_{M^2}$. 

\subsection{Naive (zero-momentum) renormalization conditions}

For gluons, we can take a naive vanishing-momentum  renormalization condition such that 
\begin{align}
 \ \Gamma_{\mathscr{A}}^{(2)}(k_{E} = 0) = M^2 \Longleftrightarrow
\hat{\Pi}_T^{\rm fin}(s = 0)=0  .
\label{ren-cond-0}
\end{align}
%This is achieved by redefining $\hat{\Pi}_T^{\rm fin}(s) \to \hat{\Pi}_T^{\rm fin}(s)-\hat{\Pi}_T^{\rm fin}(s = 0)$ so that we obtain 
%\begin{align}
%\hat{\Pi}_T^{\rm fin}(s) =  \frac{g^2 C_2 (G)}{16 \pi^2} \frac{1}{12} s \left[ \frac{111}{2s} - \frac{121}{3}    + h(s) \right] .
%\ (\Longrightarrow \hat{\Pi}_{\rm fin}(s = 0)=0) .
%\end{align}
%The following renormalization conditions for gluons are adopted in the preceding investigations. 
The first renormalization condition adopted by  Tissier and Wschebor \cite{TW10} is the vanishing-momentum renormalization condition which is written in terms of $\Gamma_{\mathscr{A}}^{(2)}$ or equivalently $\hat\Pi_T^{\rm fin}$ as%\cite{bibTW}:
%\footnote{
%If we impose a renormalization condition, 
%\begin{align}
%\label{KK}
% \Gamma_{\mathscr{A}}^{(2)}(k = \mu) = \mu^2 + M^2 \ Longleftrightarrow   
% \hat{\Pi}(s = \nu) = 0 ,
% ( \text{at} \ \mu = 1 \ \text{GeV} ). 
%\tag{\text{TW}}
%\end{align} 
%then the renormalized gluon vacuum polarization reads
%\begin{align}
%\hat{\Pi}_{\rm K}(s) &= \frac{g^2 C_2 (G)}{16 \pi^2} \frac{1}{12}  
%\notag \\ & \times 
% s \left[  \frac{63}{s} + h(s)     - ( s \rightarrow \nu )   \right] . 
%\end{align}
%Then we find 
%\begin{align}
%\hat{\Pi}(s = 0) 
%=  \frac{g^2 C_2 (G)}{16 \pi^2} \frac{5}{8} 
%\Longleftrightarrow   
%\Gamma_{\mathscr{A}}^{(2)}(k = 0) = M^2 \left(1+\frac{g^2 C_2 (G)}{16 \pi^2} \frac{5}{8}\right) .
%\end{align}
%}
\begin{align}
\label{TW}
\text{[TW1]}
%\text{[TW]}
\begin{cases}
\Gamma_{\mathscr{A}}^{(2)}(k_{E} = 0) = M^2 
\\
\Gamma_{\mathscr{A}}^{(2)}(k_{E} = \mu) = \mu^2 + M^2
\end{cases}
  \Longleftrightarrow  &
\begin{cases}
\hat{\Pi}_T^{\rm fin}(s = 0) = 0 
\\
\hat{\Pi}_T^{\rm fin}(s = \nu) = 0
\end{cases}
\nonumber\\  &
 ( \text{at} \ \mu = 1 \ \text{GeV} ) , 
%\tag{\text{TW}}
\end{align}
where we have introduced the dimensionless ratio of the renormalization scale $\mu$ to the mass defined by
\begin{align}
 \nu:=\frac{\mu^2}{M^2} .
\end{align}

%where $\mu$ is the renormalization scale.
%\begin{flalign}
%s:=\frac{k^2}{M^2}, \quad \nu:=\frac{\mu^2}{M^2} .
%\end{flalign}
Adopting the renormalization condition  [TW1], we obtain the renormalized gluon vacuum polarization function,
\begin{align}
\hat{\Pi}_{\rm TW1}^{\rm fin}(s) &= \frac{g^2 C_2 (G)}{16 \pi^2} \frac{1}{12}   
%\notag \\& \times
  s \Biggl[  \frac{111}{2 s} + h(s)    
%- \frac{1}{s^2} + \left( 1 - \frac{s^2}{2} \right) \ln  \left( s \right) 
%\nonumber\\&
%\hspace{10pt} + \left( 1 + \frac{1}{s} \right)^3 \left(s^2 - 10s + 1\right) \ln  \left( s + 1 \right) \notag \\
%& \hspace{10pt} + \frac{1}{2} \left( 1 + \frac{4}{s} \right)^\frac{3}{2} \left(s^2 - 20s + 12 \right) \ln \left( \frac{\sqrt{4 + s} - \sqrt{s}}{\sqrt{4 + s} + \sqrt{s}} \right) 
%\notag \\
%& \hspace{20pt} 
%\nonumber\\&
%\hspace{10pt} 
 - ( s \rightarrow \nu )   \Biggr] .%, \ \nu := \frac{\mu^2}{M^2} .
\label{exCL}
\end{align}
Note that constant terms in $[ ... ]$ are canceled by the subtraction: $- ( s \rightarrow \nu )$.

However, it has been shown \cite{TW10} that the vanishing-momentum renormalization condition (\ref{ren-cond-0}):
$\Gamma_{\mathscr{A}}^{(2)}(k_{E} = 0) = M^2$ or  
$\hat{\Pi}^{\rm fin}(s = 0) = 0$
yields the infrared  Landau pole, namely, the coupling constant diverging at a certain momentum in the infrared region. 
% due to another non-renormalization theorem.
Therefore, we use another renormalization condition given in the next subsection.

\subsection{Infrared safe renormalization condition}

For ghost, we introduce the two-point vertex function $\Gamma_{gh}^{(2)}$, the propagator $\Delta_{gh}$ and the self-energy function $\Pi_{gh}$,
 % together with the dimensionless versions denoted by the hat as 
\begin{align}
\Gamma_{gh}^{(2)}(k_{E}) &:= -[ \Delta_{gh}(k_{E}^2) ]^{-1} 
\nonumber\\ &
 = k_{E}^2  + \Pi_{gh}(k_{E}^2) + k_{E}^2 \delta_C   
\nonumber\\ &
 = k_{E}^2  + \Pi_{gh}^{\rm fin}(k_{E}^2)  ,   
%\nonumber\\
%&:= M^2  [ \hat{\Delta}_{gh}(s) ]^{-1}
%\nonumber\\
%&= M^2 \left[ s + \hat{\Pi}_{gh}(s) + s \delta_C \right]  
%\nonumber\\
%&= M^2 \left[ s + \hat{\Pi}_{gh}^{\rm fin}(s) \right] ,
%\nonumber\\
%&:= M^2 \hat{\Gamma}(s) ,
%\nonumber\\ &
%s:=\frac{k^2}{M^2} ,
\end{align}
where $\delta_C$ is a counterterm to cancel the divergence coming from the ghost self-energy function $\Pi_{gh}$ to obtain the finite one $\Pi_{gh}^{\rm fin}$.  
\begin{align}
 \Pi_{gh}^{\rm fin}(k_{E}^2) := \Pi_{gh}(k_{E}^2) + k_{E}^2 \delta_C , 
\end{align}
and is related to the renormalization factor as
\begin{align}
\delta_C = Z_{\mathscr{C}} -1 .
\label{delta-Z2}
\end{align}
We also define the dimensionless versions $\hat{\Delta}_{gh}(s)$ and $\hat{\Pi}_{gh}(s)$
% and $\hat{\Gamma}(s)$ 
of $\Delta_{gh}(k_{E}^2)$ and $\Pi_{gh}(k_{E}^2)$  
% and $\Gamma_{\mathscr{A}}^{(2)}(k_{E})$
as
\begin{align}
\Gamma_{gh}^{(2)}(k_{E})/M^2 &:= [ \hat{\Delta}_{gh}(s) ]^{-1}
\nonumber\\ &
=  s + \hat{\Pi}_{gh}(s) + s \delta_C   
\nonumber\\ &
=  s + \hat{\Pi}_{gh}^{\rm fin}(s) ,
%\nonumber\\
%&:= M^2 \hat{\Gamma}(s) ,
%\nonumber\\ &
%s:=\frac{k^2}{M^2} ,
\end{align}
with 
\begin{align}
 \hat{\Pi}_{gh}^{\rm fin}(s  ) =& \hat{\Pi}_{gh}(s )  + s \delta_C  .
\end{align}

The ghost self-energy function $\Pi_{gh}(k)$ in the covariant Landau gauge $\alpha = 0$ is also calculated using the dimensional regularization  and the dimensionless version $\hat{\Pi}_{gh}(s)$ is given to one-loop order by
\begin{align}
\hat{\Pi}_{gh}(s) % :=& \Pi_{gh}(k_{E}^2) / M^2 
%\notag \\
 =&  \frac{g^2 C_2 (G)}{16 \pi^2} \frac{1}{4}   s 
%\notag \\ & \times 
\Big[ - 3 \left\{ \epsilon^{-1} - \gamma + \ln(4 \pi) + \ln \eta \right\} 
\nonumber\\& \hspace{60pt}
- 5 + f(s) \Big] ,
\nonumber\\
  f(s) :=& - \frac{1}{s} - s \ln s  
%\nonumber\\ & \hspace{10pt} 
+ \frac{(1 + s)^3}{s^2} \ln (1 + s) 
%\end{align}
%which has the expansion around $s=0$
%\begin{align}
\nonumber\\
=& \frac{5}{2}+\left(\frac{11}{6}- \ln s \right)
   s+\frac{s^2}{4}+O\left(s^3\right) .
\end{align}
%The ghost propagator is related to the ghost self-energy function as 
%\begin{align}
%\Delta_{gh}(k)^{AB} &= \delta^{AB} \frac{1}{k^2 + \Pi_{gh}(k) + k^2 \delta_C} .
%[\Delta_{gh}(k)^{AB}]^{-1} &= \delta^{AB} [ {k^2 + \Pi_{gh}(k^2) + k^2 \delta_C} ].
%\end{align}
%The divergence coming from the ghost self-energy function is removed by adding the counterterm $k^2 \delta_C$. 

For ghosts, we impose the renormalization condition% which is common to both [TW] and [OS]:
\begin{align}
%\Delta_{gh}(k_{E} = \mu ) = \frac{1}{\mu^2} \Longleftrightarrow 
\Gamma_{gh}^{(2)}(k_{E} = \mu ) = \mu^2 
%\nonumber\\  
\Longleftrightarrow
\hat{\Pi}_{gh}^{\rm fin}(s = \nu) = 0 .
\label{renor-cond-ghost}
\end{align}
The renormalization condition (\ref{renor-cond-ghost}) determines the counterterm $\delta_C$ as  
%\footnote{
%In the MS scheme, the counterterm is  determined to one loop as 
%\begin{align}
% \delta_C^{(1)} = \frac{g^2C_2\left(G\right)}{16\pi^2}\frac{3}{4} \epsilon^{-1}  .
%\end{align}
%}
\begin{align}
%\hat{\Pi}_{gh}^{\rm fin}(s = \nu) =& \hat{\Pi}_{gh}(s = \nu) + \nu \delta_C  = 0   \Longrightarrow  
\delta_C^{(1)} =& - \hat{\Pi}_{gh}(s = \nu)/\nu  
\nonumber\\
% \delta_C^{(1)}   
=& - \frac{g^2 C_2 (G)}{16 \pi^2} \frac{1}{4}  
%\notag \\ & \times 
\Big[ - 3 \left\{ \epsilon^{-1} - \gamma + \ln(4 \pi) + \ln \eta \right\} 
\nonumber\\& \hspace{60pt}
- 5 + f(\nu) \Big]  .
\label{delta_C}
\end{align}
Then we obtain the renormalized ghost self-energy function under the renormalization condition (\ref{renor-cond-ghost})
\begin{align}
\hat{\Pi}_{gh}^{\rm fin}(s) &= \frac{g^2 C_2 (G)}{16 \pi^2} \frac{1}{4}   s 
%\notag \\& \hspace{10pt} \times 
\left[ f(s)  - f(\nu)  \right] .
\end{align}

We now return to the gluon renormalization. 
To avoid the infrared Landau pole for the coupling, we replace the vanishing-momentum renormalization condition (\ref{ren-cond-0}) by the second one:
% (\ref{renormalization-non}) following from (\ref{non-renor1a}). 
\begin{align}
\text{[TW2]}
%\text{[TW]}
\begin{cases}
Z_{M^2} Z_{\mathscr{A}} Z_{\mathscr{C}} = 1   
\\
\Gamma_{\mathscr{A}}^{(2)}(k_{E} = \mu) = \mu^2 + M^2
\end{cases}
  \Longleftrightarrow  &
\begin{cases}
Z_{M^2} Z_{\mathscr{A}} Z_{\mathscr{C}} = 1   
% \delta_{M^2} + \delta_C = 0  
\\
\hat{\Pi}_T^{\rm fin}(s = \nu) = 0
\end{cases}
\nonumber\\  &
 ( \text{at} \ \mu = 1 \ \text{GeV} ). 
%\tag{\text{TW}}
\label{TW2}
\end{align} 
There is a well-known non-renormalization for the coupling in the Taylor scheme \cite{Taylor71} which also holds in the massive Yang-Mills model in the Landau gauge:
The identity 
\begin{align}
 Z_{g} Z_{\mathscr{A}}^{1/2} Z_{\mathscr{C}} = \tilde{Z}_{\alpha}^2 ,
  \label{non-renor1b}
\end{align}
 implies in the Landau gauge
\begin{align}
 Z_{g} Z_{\mathscr{A}}^{1/2} Z_{\mathscr{C}} = 1 ,
  \label{non-renor1b0}
\end{align}
since in the Landau gauge, 
\begin{align}
  \tilde{Z}_{\alpha} =1 \ \text{for} \ \alpha = 0 .
  \label{non-renor1c}
\end{align}

The implication of the first renormalization condition of (\ref{TW2}) is explained as follows. 
For the massive Yang-Mills model in the Landau gauge $\alpha=0$ as a special limit of the Curci-Ferrari model, the \textit{non-renormalization theorem} holds in the sense that a combination of renormalization factors is finite to all orders in the loop expansions \cite{BSNW96,Wschebor08}:
The identity 
\begin{align}
 Z_{M^2} Z_{\mathscr{A}} Z_{\mathscr{C}} = \tilde{Z}_{\alpha}^2  ,
  \label{non-renor1a}
\end{align}
  implies in the Landau gauge
\begin{align}
 Z_{M^2} Z_{\mathscr{A}} Z_{\mathscr{C}} = 1 .
  \label{non-renor1a0}
\end{align}
%By using these relations, $Z_{\mathscr{A}}$ and $Z_{\mathscr{C}}$ are expressed by $Z_{g}$ and $Z_{M^2}$,
%\begin{align}
%  Z_{\mathscr{A}} = Z_{g}^2 Z_{M^2}^{-2}, \
%  Z_{\mathscr{C}} = Z_{g}^{-2} Z_{M^2} ,
%  \label{non-renor2}
%\end{align}
%and vice versa,
%\begin{align}
%  Z_{g}^{-1} = Z_{\mathscr{A}}^{1/2} Z_{\mathscr{C}}, \
%  Z_{M^2}^{-1} = Z_{\mathscr{A}} Z_{\mathscr{C}}  .
%  \label{non-renor3}
%\end{align}
As $Z_{M^2} Z_{\mathscr{A}}=1+\delta_{M^2}$ from (\ref{delta-Z1})  and $Z_{\mathscr{C}}=1+\delta_C$ from (\ref{delta-Z2}), the non-renormalization theorem (\ref{non-renor1a}) in the Landau gauge reduces to the relation between the counterterms 
\begin{align}
 \delta_{M^2} 
=& Z_{M^2} Z_{\mathscr{A}} -1
= Z_{\mathscr{C}}^{-1} -1
=  (1+\delta_C)^{-1} -1
 ,
\end{align}
%\begin{align}
% 1 =& Z_{M^2} Z_{\mathscr{A}} Z_{\mathscr{C}}  = (1+\delta_{M^2})(1+\delta_C) 
%\nonumber\\
%=& 1+\delta_{M^2} +\delta_C + \delta_{M^2} \delta_C
%\nonumber\\
%=&   1+\delta_{M^2}^{(1)} + \delta_C^{(1)} + ...   ,
%\end{align}
which means in the one-loop level 
\begin{align}
  \delta_{M^2}^{(1)} = -  \delta_C^{(1)} .% = - (\ref{delta_C})    .
  \label{delta_M}
\end{align}
This is the result of the first renormalization condition of (\ref{TW2}).
%\begin{align}
% \hat{\Pi}_T^{\rm fin}(s)  
%= \hat{\Pi}_T(s) + s \delta_Z + \delta_{M^2}  
%=  [\hat{\Pi}_T(s) - \delta_C] + s \delta_Z .
%\ \text{at} \ s = \nu ,
%\end{align}

Then the remaining $\delta_Z$ is determined from the second renormalization condition of (\ref{TW2}): $\hat{\Pi}_T^{\rm fin}(s = \nu) =  \hat{\Pi}_T(s= \nu)  + \nu \delta_Z^{(1)} - \delta_C^{(1)} = 0$  
by using $\delta_C^{(1)}  = (\ref{delta_C})$ as 
%\footnote{
%In the MS scheme, the counterterms are determined to one loop as 
%\begin{align}
% \delta_Z^{(1)} =&  \frac{g^2C_2\left(G\right)}{16\pi^2}\frac{13}{6} \epsilon^{-1}, \ 
%\delta_{M^2}^{(1)} = -\frac{g^2C_2\left(G\right)}{12\pi^2}\frac{3}{4}  \epsilon^{-1} .
%\end{align}
%} 
\begin{align}
\delta_Z^{(1)} =& - [\hat{\Pi}_T(s=\nu) - \delta_C^{(1)}]/\nu
\nonumber\\
% \delta_C^{(1)}   
=& - \frac{g^2 C_2 (G)}{16 \pi^2} \frac{1}{12}   
%\notag \\ &  \times 
\Biggl[  - 26 \left\{  \epsilon^{-1}  - \gamma + \ln  \left( 4 \pi \right) + \ln \eta \right\} 
\nonumber\\ & \hspace{10pt}
+ \frac{48 }{\nu} - \frac{121}{3}     
+ h(\nu) + \frac{ 3}{\nu} f(\nu)\Biggr] .
\label{delta_Z}
\end{align}
Then, by substituting (\ref{delta_Z}) and (\ref{delta_M}) into (\ref{Pi_T^fin}): $\hat{\Pi}_T^{\rm fin}(s) = \hat{\Pi}_T(s) + s \delta_Z^{(1)} + \delta_{M^2}^{(1)}$, the renormalized gluon vacuum polarization function is modified into \cite{HK18}
\begin{align}
\hat{\Pi}_T^{\rm fin}(s) =  \frac{g^2 C_2 (G)}{16 \pi^2} \frac{1}{12} s \left[ \frac{48+3f(\nu)}{s}  + h(s) - ( s \rightarrow \nu ) \right] .
%\ (\Longrightarrow \hat{\Pi}_{\rm fin}(s = 0)=0) .
\end{align}
The gluon vacuum polarization at $s=0$ has a positive value
\begin{align}
\hat{\Pi}_T^{\rm fin}(s=0) =  \frac{g^2 C_2 (G)}{16 \pi^2} \frac{1}{12}  \left[   3f(\nu) - \frac{15}{2}  \right] > 0,
\end{align}
where we have used the fact that $f(s)$ is a monotonically increasing function of $s$ with $f(0)=\frac52$.

We enumerate the obtained renormalization factors as functions of $g^2$ and $\nu$ 
\begin{align}
Z_{\mathscr{A}}^{(1)} =&  \delta_Z^{(1)} 
\nonumber\\
=&  %- \frac{g^2 C_2 (G)}{16 \pi^2} \frac{1}{12}   
%\notag \\   \times 
   \frac{C_2 (G)g^2 }{16 \pi^2} \frac{1}{12} 
\Biggl[    26 \left\{  \epsilon^{-1}  - \gamma + \ln  \left( 4 \pi \right) + \ln \eta \right\} 
\nonumber\\ & \hspace{10pt}
- \frac{48 }{\nu} + \frac{121}{3}     
- h(\nu) - \frac{ 3}{\nu} f(\nu)\Biggr] ,
\\
%\hat{\Pi}_{gh}^{\rm fin}(s = \nu) =& \hat{\Pi}_{gh}(s = \nu) + \nu \delta_C  = 0   \Longrightarrow  
Z_{\mathscr{C}}^{(1)} =&  \delta_C^{(1)} 
\nonumber\\
=&  %- \frac{g^2 C_2 (G)}{16 \pi^2} \frac{1}{4}  
%\notag \\   \times 
 \frac{C_2 (G)g^2 }{16 \pi^2} \frac{1}{4}  
\Big[  3 \left\{ \epsilon^{-1} - \gamma + \ln(4 \pi) + \ln \eta \right\} 
\nonumber\\& %\hspace{60pt}
+ 5 - f(\nu) \Big]  ,
\\
Z_{g}^{(1)} =& - \frac12 Z_{\mathscr{A}}^{(1)} - Z_{\mathscr{C}}^{(1)} 
= - \frac12 \delta_Z^{(1)} - \delta_C^{(1)} 
\nonumber\\  
=& %  - \frac{g^2 C_2 (G)}{16 \pi^2} \frac{1}{12}   
%\notag \\ &  \times 
  \frac{C_2 (G)g^2 }{16 \pi^2} \frac{1}{12} 
\Biggl[  - 22 \left\{  \epsilon^{-1}  - \gamma + \ln  \left( 4 \pi \right) + \ln \eta \right\} 
\nonumber\\ & \hspace{10pt}
+ \frac{24}{\nu} - \frac{211}{6}  
+ \frac12 h(\nu)  + 3f(\nu) + \frac{ 3}{2\nu} f(\nu)\Biggr] ,\end{align}
and
\begin{align}
Z_{M^2}^{(1)} =& - Z_{\mathscr{A}}^{(1)} - Z_{\mathscr{C}}^{(1)} 
= - \delta_Z^{(1)} - \delta_C^{(1)}
\nonumber\\  
=& %  - \frac{g^2 C_2 (G)}{16 \pi^2} \frac{1}{12}   
%\notag \\ &  \times 
  \frac{C_2 (G)g^2 }{16 \pi^2} \frac{1}{12} 
\Biggl[  - 35 \left\{  \epsilon^{-1}  - \gamma + \ln  \left( 4 \pi \right) + \ln \eta \right\} 
\nonumber\\ & \hspace{10pt}
+ \frac{48 }{\nu} - \frac{166}{3}     
+ h(\nu) + 3f(\nu)+ \frac{ 3}{\nu} f(\nu)\Biggr] .
\end{align}

We can obtain the renormalization group functions using these renormalization factors. 
For instance, the anomalous dimension of the  field $\Phi$ is obtained from the renormalization factor $Z_{\Phi}=1+Z_{\Phi}^{(1)}+ \cdots$ according to  
\begin{align}
  \gamma_{\Phi}(g^2 ,M^2) 
:=& \frac{\partial \ln Z_{\Phi}}{\partial \ln \mu} \Big|_{g_B,M_B}
=  \frac{\partial \ln [1+Z_{\Phi}^{(1)}+ \cdots]}{\partial \ln \mu} \Big|_{g_B,M_B}
\nonumber\\ 
=&  2\nu  \frac{\partial  Z_{\Phi}^{(1)}}{\partial \nu} \Big|_{g_B,M_B} + \cdots ,
\end{align}
where the replacement of the derivative with respect to $\mu^2$ by $\nu=\mu^2/M^2$ is valid to one-loop order, since $M$ is the renormalized mass which depends on the renormalization scale $\mu$. 
Therefore, the ghost field has the anomalous dimension to one-loop order 
\begin{align}
   \gamma_{\mathscr{C}} (g^2 ,M^2) 
:=&  \frac{\partial \ln Z_{\mathscr{C}}}{\partial \ln \mu} \Big|_{g_B,M_B}
%=& 2\nu  \frac{\partial  Z_{\mathscr{C}}^{(1)}}{\partial \nu} \Big|_{g_B,M_B}
%\nonumber\\
%=& 
%\frac{C_2 (G)g^2 }{16 \pi^2}  \frac{1}{4} \mu^{-2\epsilon}  2\nu  \frac{\partial }{\partial \nu}
%\Big[  3 \left\{ \epsilon^{-1} - \gamma + \ln(4 \pi) + \ln \nu %\right\} 
%\nonumber\\& %\hspace{60pt}
%+ 5 - f(\nu) \Big]  
%\nonumber\\&
%+ \frac{C_2 (G)g^2 }{16 \pi^2}  \frac{1}{4}  \frac{\partial \mu^{-2\epsilon}}{\partial \ln \mu} 
%\Big[  3 \left\{ \epsilon^{-1} - \gamma + \ln(4 \pi) + \ln \nu \right\} 
%\nonumber\\
%=& 
% \frac{C_2 (G)g^2 }{16 \pi^2}  \frac{1}{4}   
%\Big[  3 \left\{  2 - 2  \right\}    - 2\nu f^\prime (\nu) \Big]  
\nonumber\\
=& - 
  \frac{C_2 (G)g^2 }{16 \pi^2} \frac{\nu}{2} f^\prime (\nu)   
\nonumber\\
=& - \frac{C_2 (G)g^2}{16\pi^2}  
 \frac{1}{2 \nu^2}[2 \nu^2+2 \nu-\nu^3 \ln \nu 
\nonumber\\&
 +(\nu-2) (\nu+1)^2 \ln (\nu+1)]
 . 
\end{align}
%where we have used
%$\frac{\partial \mu^{-2\epsilon}}{\partial \ln \mu} =-2\epsilon+O(\epsilon^2)$.

Similarly, the anomalous dimension of the gluon  field  is calculated to one-loop order as
\begin{align}
 &  \gamma_{\mathscr{A}}(g^2 ,M^2) 
  :=   \frac{\partial \ln Z_{\mathscr{A}}}{\partial \ln \mu} \Big|_{g_B,M_B}
%\nonumber\\
%=&  \frac{C_2 (G)g^2 }{16 \pi^2} \frac{\nu}{6} 
%\frac{\partial }{\partial \nu} \Biggl[ 
%- \frac{48 }{\nu} + \frac{121}{3}     
%- h(\nu) - \frac{ 3}{\nu} f(\nu)\Biggr] 
\nonumber\\
=&  \frac{C_2 (G)g^2 }{16 \pi^2} \frac{\nu}{6} 
 \Biggl[ 
  \frac{48 }{\nu^2}     
- h^\prime(\nu) + \frac{ 3}{\nu^2} f(\nu) - \frac{ 3}{\nu} f^\prime(\nu) \Biggr]  
\nonumber\\
=& - \frac{C_2 (G)g^2}{16\pi^2} 
\frac{1}{6 \nu^3}
\Biggr[ 
 \left( 17 \nu^2-74 \nu+12\right) \nu
- \nu^5 \ln \nu 
\nonumber\\ &
+(\nu-2)^2 (\nu+1)^2 (2 \nu-3) \ln (\nu+1) 
\nonumber\\ &
+ \nu^{3/2} \sqrt{\nu+4} \left(\nu^3-9 \nu^2+20 \nu-36\right) 
\nonumber\\ & \times
\ln \left(\frac{\sqrt{\nu+4}-\sqrt{\nu}}{\sqrt{\nu+4}+\sqrt{\nu}}\right)
 \Biggr]
 . 
\end{align}
Notice that $\gamma_{\mathscr{C}}$ is always negative ($\gamma_{\mathscr{C}}=0$ at $\nu=0$).
We find that $\gamma_{\mathscr{A}}$ is negative for $\nu>0.28$, becomes zero at $\nu \sim 0.28 \sim 0.53^2$ and  positive for $\nu<0.28$ ($\gamma_{\mathscr{A}}=1/3$ at $\nu=0$).

The $\beta$ function for the gauge coupling constant is obtained from
\begin{align}
 \beta_{g^2}(g^2 ,M^2) 
%= g^2 [ \gamma_{\mathscr{A}}(g^2 ,M^2)+2 \gamma_{\mathscr{C}}(g^2 ,M^2)] 
%\nonumber\\
:=& \frac{\partial g^2}{\partial \ln \mu}  
= g^2 \frac{\partial \ln g^2}{\partial \ln \mu}  
\nonumber\\
=& -2 g^2 \frac{\partial \ln Z_g }{\partial \ln \mu}  
%= -2 g^2 \frac{\partial  Z_g^{(1)}}{\partial \ln \mu}   
= -4 g^2 \nu \frac{\partial  Z_g^{(1)}}{\partial  \nu} + \cdots ,
\end{align}
which is indeed calculated to one-loop  as 
\begin{align}
&  \beta_{g^2}(g^2 ,M^2) 
= g^2 [ \gamma_{\mathscr{A}}(g^2 ,M^2)+2 \gamma_{\mathscr{C}}(g^2 ,M^2)] 
\nonumber\\
%=&  \frac{C_2 (G)g^4 }{16 \pi^2} \frac{\nu}{2} 
%\Biggl[ 
% \frac{16 }{\nu^2}     
%- \frac13 h^\prime(\nu) + \frac{1}{\nu^2} f(\nu) - \frac{1}{\nu} f^\prime(\nu) - 2 f^\prime (\nu)  \Biggr]
=&  - \frac{C_2 (G)g^4 }{16 \pi^2}  2 \nu w(\nu) ,
\nonumber\\
& w(\nu) :=  \frac{1}{4} 
 \Biggl[ 
  - \frac{16 }{\nu^2} + \frac13 h^\prime(\nu) - \frac{1}{\nu^2} f(\nu) + \frac{1}{\nu} f^\prime(\nu) + 2 f^\prime (\nu)  \Biggr]    .
\label{beta-f-safe}
\end{align}
This equation is rewritten into a differential equation with respect to $\nu$
\begin{align}
\frac{\partial (\tilde{g}^2)^{-1}}{\partial \nu}
=  w(\nu) , \ \tilde{g}^2 := \frac{C_2 (G)g^2 }{16 \pi^2} , \ \nu:=\frac{\mu^2}{M^2} .
\end{align}
%This differential equation with respect to $\nu$ reads
%\begin{align}
%\frac{\partial (\tilde{g}^2)^{-1}}{\partial \nu} = w(\nu) .  
%\end{align}
%Therefore, the solution is given by
%\begin{align}
% (\tilde{g}^2(\nu))^{-1} 
%= (\tilde{g}^2(\nu_0))^{-1}  + \int_{\nu_0}^{\nu} d\bar\nu w(\bar\nu) ,  
%\end{align}
%or
%\begin{align}
%  \tilde{g}^2(\nu)  
%= \frac{\tilde{g}^2(\nu_0) }{1+ \tilde{g}^2(\nu_0) \int_{\nu_0}^{\nu} d\bar\nu \ w(\bar\nu) } .  
%\end{align}
Thus, by introducing the indefinite integral $W$ of $w$ which has the closed form
\begin{align}
  W(\nu) :=&  \int^{\nu} d\bar\nu \ w(\bar\nu)
\nonumber\\
 =& 
 \frac{1}{4} 
 \Biggl[ 
   \frac{16 }{\nu } + \frac13 h (\nu) + \frac{1}{\nu } f(\nu)  + 2 f (\nu)  \Biggr] ,
\end{align}
the running gauge coupling constant is given by
%\begin{align}
% (\tilde{g}^2(\nu))^{-1} 
%= (\tilde{g}^2(\nu_0))^{-1}  + [W(\nu)-W(\nu_0)] ,
%\end{align}
%or
\begin{align}
  \tilde{g}^2(\nu)  
= \frac{\tilde{g}^2(\nu_0) }{1+ \tilde{g}^2(\nu_0) [W(\nu)-W(\nu_0)] } .  
\end{align}
Notice that $W$ has the asymptotic expansions for  $\nu \gg 1$  and   $\nu \ll 1$ respectively 
\begin{align}
  W(\nu) =
  \begin{cases}
   \frac{22}{3} \ln \sqrt{ \nu } + \frac{7}{6} + O(\nu)  & (\nu \gg 1) \\
   \frac{1}{3} \ln \sqrt{\frac{1}{\nu}} + \frac{187}{36}  + O(\nu^{-1}) & (\nu \ll 1) 
  \end{cases}   .
\end{align}
Hence, the running gauge coupling constant behaves in the ultraviolet region $\nu \gg 1$  and infrared one $\nu \ll 1$ respectively  as
\begin{align}
  \tilde{g}^2(\nu) =
  \begin{cases}
 \frac{\tilde{g}^2(\nu_0) }{1+ \tilde{g}^2(\nu_0) [\frac{22}{3} \ln \sqrt{ \nu } + \frac{7}{6}+ O(\nu) - W(\nu_0)] } & (\nu \gg 1) \\
 \frac{\tilde{g}^2(\nu_0) }{1+ \tilde{g}^2(\nu_0) [ \frac{1}{3} \ln \sqrt{\frac{1}{\nu}} + \frac{187}{36}+ O(\nu^{-1}) - W(\nu_0)] } & (\nu \ll 1) 
  \end{cases}   .
\end{align}

\begin{figure}%[t]
\begin{center}
\includegraphics[scale=0.9]{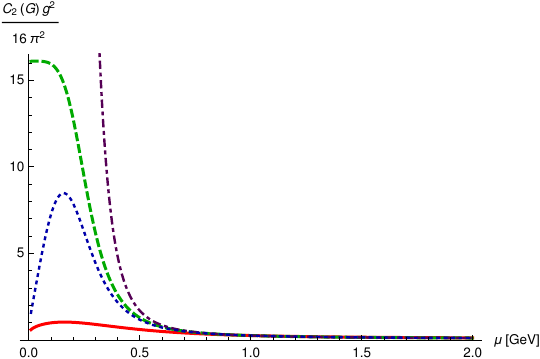}
%\quad\quad\quad\quad
%\includegraphics[scale=0.9]{fig-ep238v3/decoupling_coupling.pdf}
%\quad\quad\quad\quad
%\includegraphics[scale=0.5]{fig-ep238v3/RG_flow3.pdf}
%\includegraphics[scale=0.7]{Kondo-book-fig/CCF/PhaseDiagram.pdf}
%\vskip -0.5cm
\caption{\small
Running coupling constant for the four-dimensional massive Yang-Mills model: Landau pole (purple dot-dash line), scaling solution (green broken  line), decoupling solution (blue dotted line), physical point (red solid line) from top to bottom. 
%One-loop phase diagram and 
%(right)
%RG flow trajectories in the plane 
%($\nu:=\mu^2/M^2$, $\lambda:={\tilde{g}^2}:= \frac{g^2C_2(G)}{16\pi^2}$) 
%($\tilde m^2:=M^2/\mu^2=1/\nu$, $\lambda:={\tilde{g}^2}:= \frac{g^2C_2(G)}{16\pi^2}$) 
%The arrows indicate the flow towards the infrared. Trajectories which connect to the  \textbf{ultraviolet Gaussian fixed point} $(0,0)$ are separated in two classes: those which end at a \textbf{Landau pole} (green) and those which are \textbf{infrared safe} (blue), corresponding to \textbf{decoupling solutions} for the propagators. These are separated by a \textbf{critical trajectory} (red) which relates the Gaussian fixed point to a \textbf{nontrivial infrared fixed point} (red dot) at finite, nonzero values of $\tilde m^2=M^2/\mu^2$ and $\lambda={\tilde{g}^2}$ and corresponds to a \textbf{scaling solution} for the correlators. We also show (orange, lower curve) the trajectory which describes \textbf{lattice results} for the SU($3$) theory.
% (see Appendix~\ref{appsec:lattice}).
}
\label{fig:running-coupling}
\end{center}
\end{figure}

\begin{figure}%[t]
\begin{center}
%\includegraphics[scale=1.0]{fig-ep238v3/decoupling_coupling.pdf}
%\quad\quad\quad\quad
%\includegraphics[scale=0.6]{fig-ep238v3/RSTWfig1_Jul5.pdf}
\includegraphics[scale=0.6]{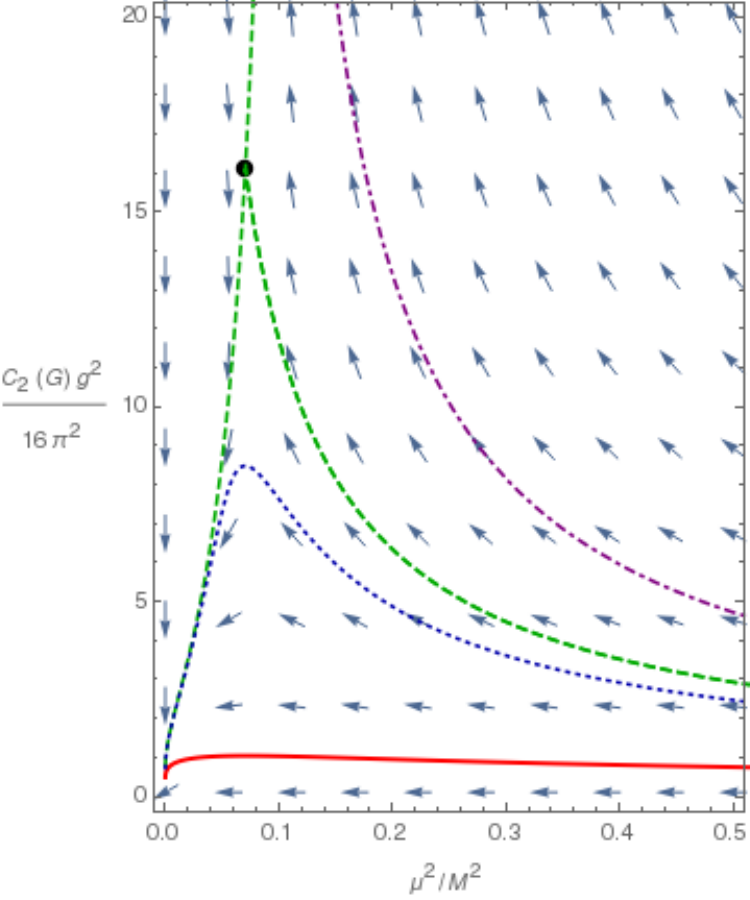}
%\includegraphics[scale=0.6]{fig-ep238v3/RG_flow3.pdf}
%\includegraphics[scale=0.7]{Kondo-book-fig/CCF/PhaseDiagram.pdf}
%\vskip -0.5cm
\caption{\small
RG flows in the parameter space  
($\nu:=\mu^2/M^2$, ${\tilde{g}^2}:= \frac{g^2C_2(G)}{16\pi^2}$) of the four-dimensional massive Yang-Mills model. 
%($\tilde m^2:=M^2/\mu^2=1/\nu$, $\lambda:={\tilde{g}^2}:= \frac{g^2C_2(G)}{16\pi^2}$) 
%One-loop phase diagram and RG flow trajectories in the plane ($\tilde m^2:=M^2/\mu^2$, $\lambda:={\tilde{g}^2}$) in the four-dimensional spacetime. 
The arrows indicate the flow towards the infrared. Trajectories which connect to the  \textit{ultraviolet Gaussian fixed point} $(\infty,0)$ are separated in two classes: those which end at a \textit{Landau pole} (purple dot-dash line) and those which are {infrared safe}, corresponding to \textit{decoupling solutions}  (blue dotted line) for the propagators. These are separated by a \textit{critical trajectory} (green broken line) which relates the Gaussian fixed point to a \textit{nontrivial infrared fixed point} (black dot) at finite, nonzero values of $\nu$ and ${\tilde{g}^2}$ and corresponds to a \textit{scaling solution} for the correlators. We also show (red solid line) the trajectory which describes \textit{lattice results} for the SU($3$) theory.
%The arrows indicate the flow towards the infrared. Trajectories which connect to the  \textbf{ultraviolet Gaussian fixed point} $(0,0)$ are separated in two classes: those which end at a \textbf{Landau pole} (green) and those which are \textbf{infrared safe} (blue), corresponding to \textbf{decoupling solutions} for the propagators. These are separated by a \textbf{critical trajectory} (red) which relates the Gaussian fixed point to a \textbf{nontrivial infrared fixed point} (red dot) at finite, nonzero values of $\tilde m^2=M^2/\mu^2$ and $\lambda={\tilde{g}^2}$ and corresponds to a \textbf{scaling solution} for the correlators. We also show (orange, lower curve) the trajectory which describes \textbf{lattice results} for the SU($3$) theory.
% (see Appendix~\ref{appsec:lattice}).
}
\label{fig:flow}
\end{center}
\end{figure}

In the ultraviolet region $\nu \gg 1$ or $\mu \gg 1$, 
the beta function $\beta_{\tilde{g}^2}$ in the massive Yang-Mills model is negative for $\nu \gg 1$, since  (\ref{beta-f-safe}) has the expansion for $\nu \gg 1$
\begin{align}
&  \beta_{g^2}(g^2 ,M^2) 
\nonumber\\
=&  \frac{C_2 (G)g^4 }{16 \pi^2}  
 \Biggl[ 
-\frac{22}{3}+\frac{\frac{59}{4}-\frac{9}{2} \ln
   \left(\frac{1}{\nu}\right)}{\nu}
%+\frac{\frac{71}{3} \ln \left(\frac{1}{\nu}\right)-\frac{67}{18}}{\nu^2}
+O\left(\nu^{-5/2}\right) \Biggr]  .
\end{align}
This result is in agreement with the standard, universal beta function of the usual Yang-Mills theory reflecting the \textit{ultraviolet asymptotic freedom} 
\begin{align}
  g^2 (\mu) \simeq 
\frac{1}{ \frac{22}{3} \ln \frac{\mu}{M} }
\searrow 0 \ (\mu \nearrow \infty) .  
\end{align}
In the infrared region $\nu \ll 1$ or $\mu \ll 1$, on the other hand, \textit{the beta function $\beta_{\tilde{g}^2}$ of the massive Yang-Mills model becomes positive in the deep infrared regime}, since  (\ref{beta-f-safe}) has the expansion for $\nu \ll 1$
\begin{align}
 & \beta_{g^2}(g^2 ,M^2) 
\nonumber\\
=&   \frac{C_2 (G)g^4 }{16 \pi^2}  
 \Biggl[ 
\frac{1}{3}+\left(\ln \nu-\frac{367}{180}\right) \nu
%+\left(\frac{\ln \nu}{6}-\frac{13}{420}\right) \nu^2
+O\left(\nu^{5/2}\right) \Biggr] .
\end{align}
This implies that the running coupling constant $g^2 (\mu)$ decreases towards the infrared region and vanishes as $\mu \to 0$ 
\begin{align}
  g^2 (\mu) \simeq 
\frac{1}{ \frac{1}{3} \ln \frac{M}{\mu} }
\searrow 0 \ (\mu \searrow 0) .  
\end{align}
Therefore the RG flow drives the system towards a weak coupling region as $\mu$ goes to zero.
This fact justifies the use of the one-loop approximation to study the Yang-Mills theory even in infrared region. 
See Fig.~\ref{fig:running-coupling}.

%Here we have used the expansion of $h$, $f$ and their derivatives  around $s=0$
%\begin{align}
%h^\prime(s) %:=& K_1(s) + K_2(s) + K_3(s) 
%\nonumber\\  
%=& \frac{111}{2 s^2}+\frac{1}{s}+\frac{389}{60}+\left(-\ln  (s)-\frac{88}{35}\right) s
%\nonumber\\ &
%+\frac{269 s^2}{140}+O\left(s^{5/2}\right) , 
%\end{align}
%\begin{align}
%  f^\prime(s)  
%=& \left(\frac{5}{6}-\ln s \right)+\frac{s}{2}-\frac{3s^2}{20}+O\left(s^3\right) .
%\end{align}

We find that the beta function $\beta_{g^2}$ is negative for $\nu>0.07$, becomes zero at $\nu \sim 0.07 \sim 0.26^2$ and positive for $\nu<0.07$.
This implies that the running coupling constant $g^2(\mu)$  of the decoupling solution increases monotonically in decreasing the scale $\mu$ until $\mu$ reaches the value $\mu/M \sim 0.26$, and it turns over at $\mu/M \sim 0.26$ and decreases towards the infrared limit $g^2(\mu) \to 0$ as $\mu \to 0$. 

Finally, we study the RG flow in the two-dimensional parameter space 
($\nu:=\mu^2/M^2$, ${\tilde{g}^2}:= \frac{g^2C_2(G)}{16\pi^2}$) of the four-dimensional massive Yang-Mills model. 
See Fig.~\ref{fig:flow}.
First, we fix the value of $\nu$ to a relatively large value $\nu_0$ (which is equivalent to set $\tilde m^2:=M^2/\mu^2$ to a relatively small value), e.g., $\nu_0=100^2$  and varies the value of $\tilde{g}^2(\nu)$ to see the differences of the resulting trajectories.
Then we find that the running coupling constant $\tilde{g}^2(\nu)$ remains finite for all $\nu$ if the initial value $\tilde{g}^2(\nu_0)$ at $\nu_0$ is smaller than and equal to a certain value $\tilde{g}_*^2(\nu_0)$, while it diverges at a finite $\mu$ if $\tilde{g}^2(\nu_0)$ is greater than the value $\tilde{g}_*^2(\nu_0)$.
Therefore, the decoupling solution exists for $\tilde{g}^2(\nu_0)<\tilde{g}_*^2(\nu_0)$, while the scaling solution is realized at the critical value $\tilde{g}^2(\nu_0)=\tilde{g}_*^2(\nu_0)$ \cite{RSTW17}.
For $\tilde{g}^2(\nu_0)>\tilde{g}_*^2(\nu_0)$, we have an infrared Landau pole. 
Therefore, the coupling constant behaves in decreasing $\nu$ from $\nu_0$, $\nu_0> \nu=\frac{\mu^2}{M^2} \searrow 0$ as
\begin{align}
  \tilde{g}^2(\nu) = 
  \begin{cases}
 \text{Landau~pole} \ \tilde{g}^2(\nu) \nearrow \infty & (\tilde{g}^2(\nu_0)>\tilde{g}_*^2(\nu_0)) \\
 \text{scaling} \ \tilde{g}^2(\nu) \nearrow \tilde{g}_*^2 & (\tilde{g}^2(\nu_0)= \tilde{g}_*^2(\nu_0)) \\
 \text{decoupling} \  \tilde{g}^2(\nu) \nearrow \tilde{g}_{m}^2 \searrow 0 & (\tilde{g}^2(\nu_0)<\tilde{g}_*^2(\nu_0))  
  \end{cases}   .
\end{align}

\subsection{Fitting to the numerical simulations}

%\begin{figure}[t]
%\centering
%\includegraphics[width=7cm]{fig-ep238v3/DrawGluonNRTRaw.pdf}
%\caption{
%The gluon propagator $\mathscr{D}$ as a function of the Euclidean momentum $k_E$ in unit of $\mu$. 
%The numerical data (red points) for the gluon propagator of the SU(3) Yang-Mills theory on the lattice  and the fitted result (blue solid line) to the analytical expression in the one-loop level of the massive Yang-Mills model with two parameters, $g$ and $M$.
%}
%\label{figFitGluonRaw}
%\end{figure}

\begin{figure}[t]
\centering
\includegraphics[width=7cm]{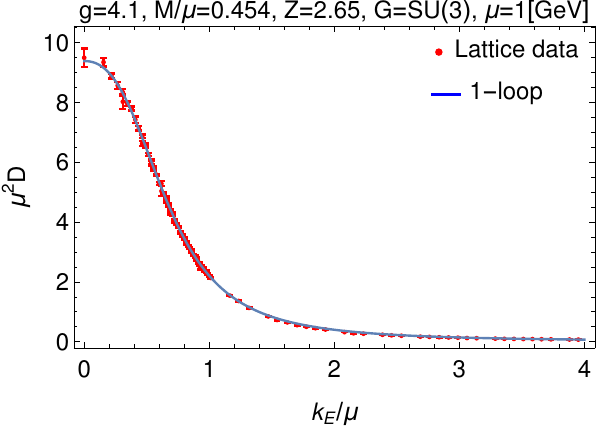}
\caption{
The gluon propagator $\mathscr{D}$ as a function of the Euclidean momentum $k_E$ in unit of $\mu$. 
The  numerical data (red points) for the gluon propagator of the SU(3) Yang-Mills theory on the lattice  and the fitted result (blue solid line) to the scaled analytical expression of the gluon propagator $\mathscr{D}$ in the one-loop level of the massive Yang-Mills model with fitting parameters $g$, $M$ and $Z$ (\ref{exFitParams}).
}
\label{figFitGluon}
\end{figure}

\begin{figure}[t]
\centering
\includegraphics[width=7cm]{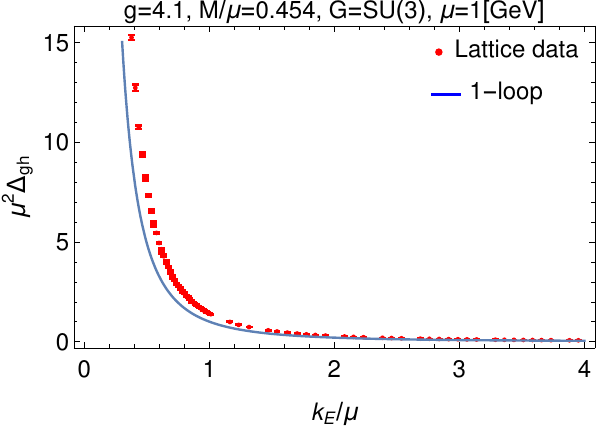}
\caption{
The ghost propagator $\Delta_{gh}$ as a function of the Euclidean momentum $k_E$ in unit of $\mu$. 
(red points) The  numerical data for the ghost propagator of the SU(3) Yang-Mills theory on the lattice  and (blue solid line) the plot of the analytical expression of the ghost propagator to one-loop order   of the massive Yang-Mills model with two parameters $g$, $M$ at the physical point (\ref{exFitParams}).
}
\label{figFitGhost}
\end{figure}

We utilize the data obtained by the numerical simulations on the lattice for the Yang-Mills theory in the covariant Landau gauge  
to determine the parameters, the coupling constant $g$ and the gluon mass parameter $M$, in the massive Yang-Mills model.
%We use our analytical results obtained by the massive Yang-Mills model (to one-loop order) to fit the results of numerical simulations on the lattice  for the $SU(3)$ Yang-Mills theory in the covariant Landau gauge \cite{DOS16}.
%[A.G. Duarte, O. Oliveira, and P.J. Silva, 
%Lattice Gluon and Ghost Propagators, and the Strong Coupling in Pure SU(3) Yang-Mills Theory: Finite Lattice Spacing and Volume Effects,  
%Phys. Rev. D\textbf{94} (2016)  014502.] 
%DOI: 10.1103/PhysRevD.94.014502 

%First, we attempted to fit the data of numerical simulations for the gluon propagator on the lattice \cite{DOS16} to the analytical expression $\mathscr{D}$ for the gluon propagator with one-loop quantum corrections 
%by adjusting the two parameters $g$ and $M$ in the fitting range $0< k_E \le 1$GeV at $\mu = 1$GeV for $G = SU(3)$.
%However, the fitting does not work so well and the appropriate parameters cannot be obtained in this setting. %as shown in Fig.~\ref{figFitGluonRaw}, 
%%%%We choose this shallow fitting range for $k_E$, since we want to see the low-momentum region of the propagator. However, the resulting fitted parameters do not change so much even if we extend the range to $k_E \le 4$.  
%%%%As the primary explanation for this discrepancy, we can raise a limitation of the model in the sense that the model with a one-loop calculation might not be sophisticated enough to capture the intricacies of real Yang-Mills dynamics.
%%%%As the secondary explanation, to be more ambitious, 
%We attribute this failure  to the fact that

In fitting the data of numerical simulations for the gluon propagator on the lattice \cite{DOS16} to the analytical expression $\mathscr{D}$ for the gluon propagator with one-loop quantum corrections, we need to take into account the fact that 
 the renormalization conditions adopted in the lattice simulations \cite{DOS16} are different from those adopted in this paper, leading to the different scale for the gluon propagator.  Otherwise, the fitting does not work so well and the appropriate parameters cannot be obtained. 
For this purpose, we introduce an overall scale factor $Z$ which can scale the  gluon propagator as a whole to absorb the difference of the renormalization conditions. 
In \cite{DOS16}, indeed, such a scaling of data obtained by  numerical simulations for the gluon propagator was adopted to satisfy the renormalization condition $\mathscr{D}_T(k_E^2=\mu^2)=1/\mu^2$ at $\mu=4$~GeV. 
This kind of rescaling was also adopted in \cite{TW10}.
Consequently, the fitting works surprisingly well to give the precise values for the parameters $g$, $M$ and $Z$  as shown in Fig. 5 in the fitting range $0< k_E \le 4$GeV at $\mu = 1$GeV for $G = SU(3)$  where the fitting parameter with errors are given by
%By performing the fitting anew with an additional scale parameter, the fitting result is dramatically improved to give the precise values for the parameters as shown in  Fig.~\ref{figFitGluon} where the fitting parameter with errors are given by 
\begin{align}
& g = 4.1 \pm 0.1 \ \Leftrightarrow \ \lambda := \frac{g^2C_2(G)}{16\pi^2}= 0.32 \pm 0.02  , 
\notag \\
& \frac{M}{\mu} = 0.454 \pm 0.004 \ \Leftrightarrow \ \frac{M^2}{\mu^2} = 0.206 \pm 0.004 , 
\notag \\
& Z = 2.65 \pm 0.02 .
\label{exFitParams}
\end{align}
We use these parameters to plot the ghost propagator using the analytical expression by including quantum corrections to one-loop order in the massive Yang-Mills model, as shown in  Fig.~\ref{figFitGhost}.

Both gluon propagator and ghost propagator in the decoupling solution of the Yang-Mills theory are well reproduced by the values (\ref{exFitParams}) of parameters $g$ and $M$.
In what follows we call these values of the parameters the physical point for the Yang-Mills theory.
%\footnote{ 
%This result should be compared with that of Tissier and  Wschebor \cite{TW10} with the renormalization condition [TW]:
%[M. Tissier and N. Wschebor, 
%Infrared propagators of Yang-Mills theory from perturbation theory, 
%Phys.Rev. D\textbf{82}, 101701 (2010). 
%arXiv:1004.1607 [hep-ph]
%]
%\begin{align}
%\text{[TW]:} \quad g=4.9, \quad M=0.54~\mathrm{GeV} .
%\label{TW-parameter}
%\end{align}
%}

As a side remark, let us add some comments on the validity of the massive Yang-Mills model in reproducing the infrared behaviors of the Yang-Mills theory.
There is  no guarantee in advance that such a specific model  with a ``phenomenological'' mass term for gluons being just included  captures the intricacies of the real Yang-Mills dynamics. 
We acknowledge that the surprising agreement between the numerical lattice data of the Yang-Mills theory and the simple one-loop propagator of the massive Yang-Mills model could be accidental, and that the gluon mass term will, at best, only capture some aspects, not all aspects, of the intricate dynamics of the original Yang-Mills theory or QCD. 
In fact, this type of the massive model for the real QCD is shown to give a poor agreement for the quark sector of QCD with numerical lattice results \cite{PTW14}. 
Nevertheless, we can still claim that this model gives a gluon propagator showing excellent agreement with the lattice data. 
Indeed, it is shown  \cite{GPRT19} that the two-loop calculations for the gluon and ghost propagators considerably improve the one-loop result to show  more excellent agreement with the lattice data.  
In these investigations, it is also confirmed that the pure Yang-Mills sector indicates the infrared-safety, namely, the finiteness of the running gauge coupling constant in all scales, which makes the perturbative method more feasible.  
Incidentally, the one-loop calculation for the three-point gluon vertex functions gives a ``satisfying'' agreement with the available lattice data \cite{PTW13}.
In view of these works,  the massive Yang-Mills model will be valid to capture some aspects of the gluon sector of QCD relevant to our investigation, even though the other important aspects may be missing. 
At least for the gluon, therefore, it will be worthwhile to study the analytic structure of the propagator of this model, which is one of our purposes in this paper.

%\newpage
%\textcolor{green}{\Large\bf $\S$~ $SU(2)$ Yang-Mills from gauge-fundamental scalar  model }
%\setcounter{equation}{0}
%$\bigodot$ Fundamental scalar case: A perturbative approach
%\begin{figure}[!h]
%\centering
%\includegraphics[height=5.0cm]{fig-Higgs/gluon-propagator-Landau.pdf}
%\hspace{5mm}
%\includegraphics[height=5.0cm]{fig-Higgs/ghost-propagator-Landau.pdf}
%\centering\includegraphics[height=4.6cm]{SU3CGgauge12860.png}\hspace{5mm}\includegraphics[height=4.6cm]{SU3CGghost12860.png}
%\caption{
%\small
%(Left) gluon propagator $\tilde D_T(k)$, (Right) ghost propagator $\Delta(k)$.
%}
%\label{SU3CGgauge}
%\end{figure}
%arXiv:1605.00594 [hep-lat]] 

%Both gluon propagator and ghost propagator are well reproduced by a set of parameters $(g,M)$ for the renormalization condition:
%\begin{align}
%\text{[OS]:} \quad %g=2.47 , \quad M=0.329~\mathrm{GeV}.  
%& g  = 4.1 \pm 0.1 , 
%\quad 
%M/\mu  = 0.454 \pm 0.004 ,%~\mathrm{GeV} , 
%\notag \\&
% Z = 2.65 \pm 0.02 ,
%& {\chi}^2_{reduced}  = 1.0009 ,
%\label{OS-parameter}
%\end{align}
%where ${\chi}^2_{\rm reduced}$ is defined as $\chi^2 / (d. o. f.)$, see Appendix~D.

%\newpage
%%%%%%%%%%%%%%%%%%%%%%%%%%%%%%%%%%%%%%%%%%%%%%%%%%%%%%%%%%%%%
%%%%%%%%%%%%%%%%%%%%%%%%%%%%%%%%%%%%%%%%%%%%%%%%%%%%%%%%%%%%%
\section{
 Reflection positivity violation in the massive Yang-Mills model
}
%%%%%%%%%%%%%%%%%%%%%%%%%%%%%%%%%%%%%%%%%%%%%%%%%%%%%%%%%%%%%

In this section, we observe that the Euclidean gluon propagator in the massive Yang-Mills model exhibits violation of reflection positivity.
This result suggests gluon confinement in the Yang-Mills theory.

%Let us explain our standpoint. 
Usually the quantum field theory (QFT) obeying the Wightman axioms \cite{Wightman56,AQFT} is first defined in the Minkowski region and then analytically continued to the Euclidean region to obtain the Euclidean QFT which consequently obeys the \textit{Osterwalder-Schrader (OS) axioms}  \cite{OS73}.  
However, we want to start from the Euclidean QFT obeying the OS axioms (or better axioms if any) and check which kinds of QFT can be defined in the Minkowski spacetime which is to be obtained by analytic continuation from the Euclidean region. 
%[In this sense, our presentation for the Lagrangian should be given in the Euclidean space.] 

In our opinion, only the Euclidean QFT can be rigorously defined as the QFT. Probably, QFT describing only non-confining particles will be defined both in the Euclidean and the Minkowski space in the equivalent way. However, we have no evidences that the QFT describing confining particles can be formulated in the Minkowski spacetime in the same way as QFT for  non-confining particles.  In contrast, we know some examples of Euclidean QFT which exhibit confinement, e.g., the linear potential for the static quark potential is observed in the Euclidean Yang-Mills theory on the lattice. Therefore, the validity of the Euclidean QFT for confining particles is tested everyday on the lattice in the non-perturbative manner.  
   In view of these, we examine the validity of the reflection positivity as an axiom or one of the general properties to be satisfied by the Euclidean QFT. 

%The result is negative for confining case. However, this result is obtained only in the numerical way. To investigate the physical origin of this result and give an analytical derivation, we extend the region of investigation to the complex momentum towards the Minkowski region. Then we have found the existence of a pair of complex poles which can be the origin of the violation of reflection positivity in the Euclidean region. 
%\footnote{
%Notice that the existence of complex poles in the momentum representation of the two-point function does not violate spacelike commutativity. 
%For instance, it was shown by Nakanishi \cite{}[Phys. Rev. D3, 811 (1971)] that the existence of complex pole is compatible with spacelike commutativity in a complex scalar field theory with indefinite metric. 
%}
%A byproduct of this study is to give an explanation why the Euclidean propagator on the lattice shows the Gribov-Stingl form, which has not yet been understood in the other ways to the best of author's knowledge. 

\subsection{Reflection positivity and the Schwinger function}

%Reflection positivity
%In this section, we examine the reflection-positivity violation. 
The \textit{OS axioms} \cite{OS73}  are general  properties to be satisfied for the QFT formulated in the Euclidean space, which are the Euclidean version of the \textit{Wightman axioms} for the relativistic QFT formulated in the Minkowski spacetime. 
%\footnote{See the chapter of Millennium Prize Problems.
%\\ 
%K. Osterwalder and R. Schrader, 
%Commun. Math. Phys \textit{31}, 83  (1973).
%Commun. Math. Phys \textit{42}, 281 (1975). 
%\\
%The lattice version is given in: 
% K. Osterwalder  and E. Seiler,
% Annals Phys. \textit{110}, 440 (1978).
%} 
A relativistic QFT described by a set of the Wightman functions satisfying the Wightman axioms can be constructed from a set of Schwinger functions (Euclidean Green's functions) if they obey the OS axioms.
 In particular, the axiom of \textit{reflection positivity} is the Euclidean counterpart to the positive definiteness of the norm in the Hilbert space of the corresponding Wightman QFT.
If the reflection positivity is violated, a particular Euclidean correlation function  cannot have the  interpretation in terms of stable particle states, which is regarded as a manifestation of confinement. 
To demonstrate the violation of {reflection positivity} in the OS axioms, one counterexample suffices.

For the special case of a single propagator, the reflection positivity reads
\begin{align}
  & \int d^Dx \int d^Dy f^*(\bm{x},-x_D) \mathscr{D}(\bm{x}-\bm{y},x_D-y_D) f(\bm{y},y_D) \ge 0 , 
\nonumber\\ &
 f \in \mathscr{S}_{+}(\mathbb{R}^D) ,
\end{align}
where 
$\mathscr{S}_{+}(\mathbb{R}^D)$ denotes  a complex-valued test (Schwartz) function with support in $\{ (\bm{x},x_D); x_D >0\}$.
The reflection positivity is rewritten as 
\begin{align}
%  & \int d^4x \int d^4y f^*(\bm{x},-x_4)  \int d^3\bm{p} e^{i \bm{p} \cdot (\bm{x}-\bm{y})} \Delta(x_4-y_4, \bm{p})  f(\bm{y},y_4) 
%\nonumber\\
& \int dx_D \int dy_D \int d^{D-1}\bm{p} f^*(\bm{p},-x_D)f(\bm{p},y_D) \Delta( \bm{p}, x_D-y_D) 
\nonumber\\
=& \int_{0}^{\infty} dt \int_{0}^{\infty} dt^\prime \int d^{D-1}\bm{p} f^*(\bm{p},t) f(\bm{p},t^\prime) \Delta(\bm{p}, -(t+t^\prime))
\nonumber\\ &
 \ge 0,
\end{align}
where we defined $\Delta(\bm{p},x_D-y_D)$ by
\begin{align}
\mathscr{D}(x-y) := \int d^{D-1}\bm{p} \ e^{i \bm{p} \cdot (\bm{x}-\bm{y})} \Delta(\bm{p},x_D-y_D)  
 .
\end{align}
In what follows we call $\Delta(\bm{p},x_D-y_D)$ the \textit{Schwinger function}. 
For this inequality to hold for any test function $f \in \mathscr{S}_{+}(\mathbb{R}^D)$, the Schwinger function $\Delta$ must satisfy the positivity
\begin{align}
 \Delta(\bm{p}, -(t+t^\prime)) = \Delta(\bm{p}, t+t^\prime)  \ge 0   
 .
\end{align}

%\newpage
We consider a particular  {Schwinger function} in the $D$-dimensional spacetime defined  by the Fourier transform of the Euclidean propagator $\tilde{\mathscr{D}} (\bm{p},p_{E}^D)$,
%\footnote{
%\begin{align}
%  \Delta(t) = \int d^dx \mathscr{D}( \bm{x},t)
%:=& \int d^dx \int \frac{d^Dp}{(2\pi)^D} e^{ipx} \tilde{\mathscr{D}}(p) 
%\nonumber\\
%=&  \int d^dx \int \frac{d^Dp}{(2\pi)^D} e^{i\bm{p} \cdot \bm{x}}  e^{ip_D t} \tilde{\mathscr{D}}(p)  
%\nonumber\\
%=&   \int \frac{d^Dp}{(2\pi)^D}  e^{ip_D t} \tilde{\mathscr{D}}(p)  \int d^dx e^{i\bm{p} \cdot \bm{x}} 
%\nonumber\\
%=&   \int \frac{d^Dp}{(2\pi)^D}  e^{ip_D t} \tilde{\mathscr{D}}(p)  (2\pi)^d \delta^d(\bm{p}) 
%\nonumber\\
%=&   \int_{-\infty}^{+\infty} \frac{dp_D}{2\pi}  e^{ip_D t} \tilde{\mathscr{D}}(\bm{p}=0,p_D)    ,
%\end{align}
%}
\begin{align}
\Delta(t) :=& \Delta(\bm{p}=\bm{0},t)  
:= \int d^{D-1}x  \ e^{-i\bm{p} \cdot \bm{x}}  \mathscr{D}( \bm{x},t)|_{\bm{p}=\bm{0}}
\nonumber\\  
=& \int_{-\infty}^{+\infty} \frac{dp_{E}^D}{2\pi}  e^{ip_{E}^D t} \tilde{\mathscr{D}} (\bm{p}=\bm{0} ,p_{E}^D) .
\end{align} 
If $\tilde{\mathscr{D}}(\bm{0}, p_{E}^D)$  is even in $p_{E}^D$, namely, $\tilde{\mathscr{D}}(\bm{0},-p_{E}^D) =\tilde{\mathscr{D}}(\bm{0},p_{E}^D)$,
the Schwinger function reduces  to 
\begin{align}
  \Delta(t) 
 = 2 \int_{0}^{\infty} \frac{dp_{E}^D}{2\pi}  \cos (p_{E}^D t) \tilde{\mathscr{D}}(\bm{0} ,p_{E}^D)   
 .
\end{align}
To demonstrate the violation of reflection positivity, one counterexample suffices.
Therefore, non-positivity of the Schwinger function $\Delta(t)$   at some value of $t$ leads to the violation of reflection positivity. 
%\begin{align}
%\Delta(t) \le 0 \  \text{for some} \  t \ge 0
%\Longrightarrow
%\rho(\sigma^2) < 0 \  \text{for some} \ \sigma^2 \ge 0.
%\end{align}
Consequently, the reflection positivity is violated for the gluon propagator.
The corresponding states cannot appear in the physical particle spectrum. 
This is consistent with gluon confinement.

 For the  free massive propagator,
\begin{align}
  \tilde{\mathscr{D}}(p) = \frac{1}{p^2+m^2} \quad (m>0)
 ,
\end{align}
we find $\Delta(t)$ is positive for any $t$: 
\begin{align}
  \Delta(t) 
%=&   \int_{-\infty}^{+\infty} \frac{dp_4}{2\pi}  e^{ip_D t} D(\bm{p}=0,p_D)   
=  \int_{-\infty}^{+\infty} \frac{dp_D}{2\pi}  e^{ip_D t} \frac{1}{p_D^2+m^2} 
= \frac{1}{2m} e^{-m|t|} > 0 %\quad (m>0)
 .
\end{align}
Therefore, there is  no reflection-positivity violation for the free massive propagator, as expected. 
For unconfined particles, the reflection positivity should hold. 
%This case corresponds to the spectral function of
%\begin{equation}
%\rho(\sigma^2)=\delta(\sigma^2-m^2)=\frac{1}{2m}\delta(\sigma-m) > 0 .
%\end{equation}

%--------------------------------------------------
\subsection{Positivity violation for the decoupling solution of the Yang-Mills theory}

\begin{figure}[t]
\centering
\includegraphics[width=7cm]{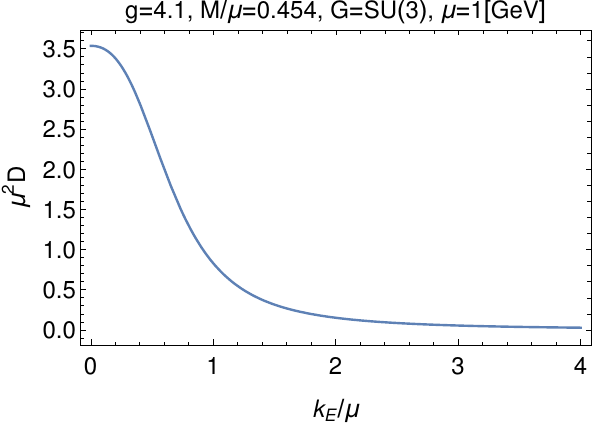}
\includegraphics[width=7cm]{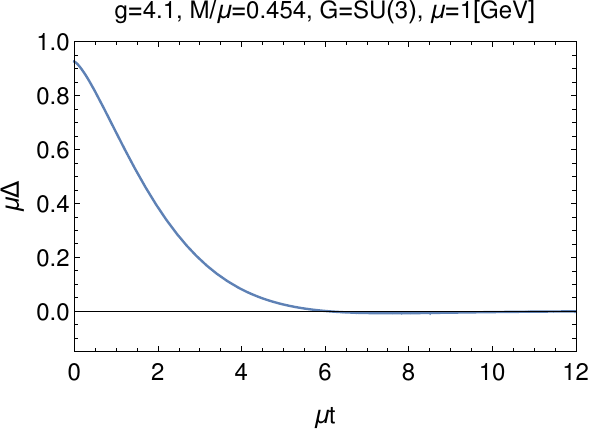}
\caption{
The gluon propagator $\mathscr{D}$ and the Schwinger function $\Delta$  at the physical point of the parameters $g = 4.1$, $M/\mu = 0.454$:
(top) gluon propagator $\mu^2 \mathscr{D}$ as a function of $k_E/\mu$ and (bottom) the  Schwinger function $\mu \Delta$  as a function of $\mu t$, where all quantities are made dimensionless using the rescaling of appropriate powers of $\mu$. 
}
\label{figPropSchPhys}
\end{figure}

In order to examine the violation of the reflection positivity through the behavior of the gluon Schwinger function, we first construct a set of gluon and ghost propagators in such a way that  they are renormalized to satisfy the renormalization conditions [TW2](\ref{TW2}) and (\ref{renor-cond-ghost}) in the massive Yang-Mills model to reproduce the decoupling solution in the Yang-Mills theory to one-loop order. 
The integral in obtaining the Schwinger function as the Fourier transform of the gluon propagator is not so easy to be performed analytically, hence we resort to the numerical calculations for this definite integral.

In Fig.~\ref{figPropSchPhys}, we give the plot for the gluon propagator and the associated Schwinger function in the Landau gauge  $\alpha=0$ for the $SU(3)$ massive Yang-Mills model at the physical point of parameters $g = 4.1$ and $M/\mu = 0.454$.
We observe that the Schwinger function takes  negative values for $\mu t > 6$ and hence the reflection positivity is violated. 
Therefore, this result suggests that the reflection positivity is violated for the decoupling solution in the Yang-Mills theory. 
The more detailed analysis of the reflection positivity will be given in the next section from the viewpoint of the complex structure of the gluon propagator.

\subsection{Positivity violation in the complementary gauge-scalar model}

In what follows, we examine how the gluon propagator and the Schwinger function are modified if the parameters $g$ and $M$ deviate from the physical point.  In this case the massive Yang-Mills model is no longer regarded as a low-energy effective theory of the original Yang-Mills theory. However, the resulting model can be regarded as the gauge-scalar model with the complementarity between Higgs and confinement in the sense that 
the confinement phase in the Yang-Mills theory is analytically  connected with no phase transition to the Higgs phase in the gauge-scalar model through the BEH mechanism, which is called the Fradkin-Shenker continuity.

%-------------------------------
\subsubsection{Smaller coupling constant}

\begin{figure}[t]
\centering
\includegraphics[width=4cm]{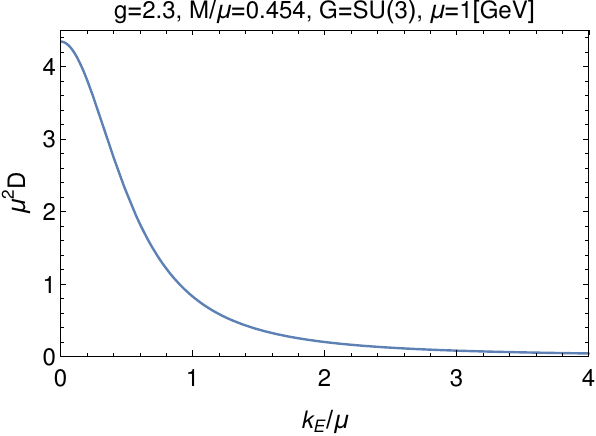}
\includegraphics[width=4cm]{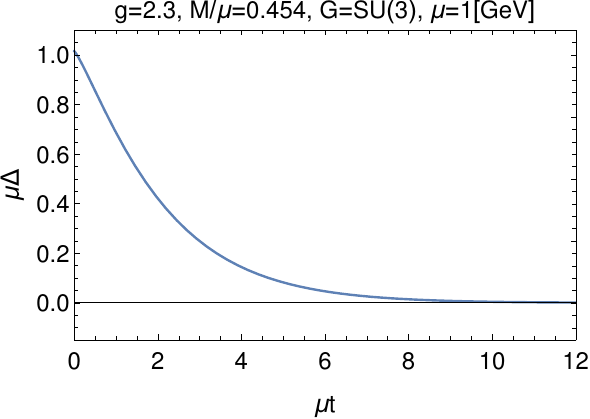}
\caption{
The same plots as those given in Fig.~\ref{figPropSchPhys} for a smaller coupling constant $g = 2.3$ with a physical value $M/\mu=0.454$.
}
\label{figPropSchSC}
\end{figure}

\begin{figure}[t]
\centering
\includegraphics[width=4cm]{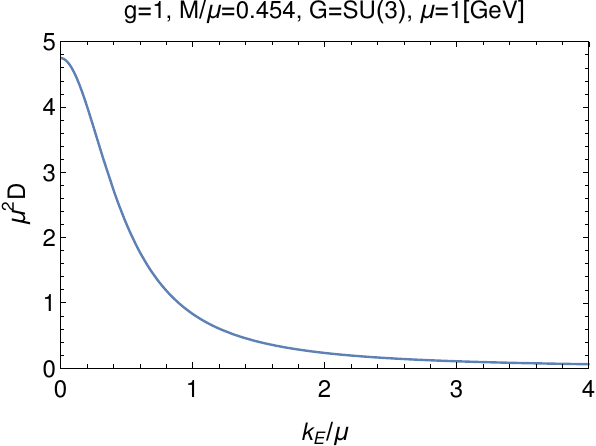}
\includegraphics[width=4cm]{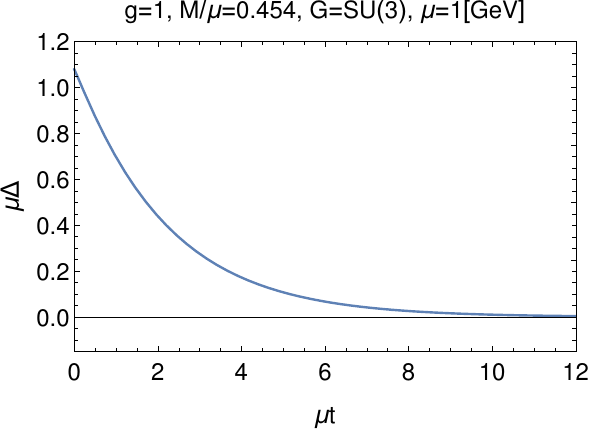}
\caption{
The same plots as those given in Fig.~\ref{figPropSchPhys} for a further smaller coupling constant $g = 1$ with a physical value $M/\mu=0.454$.
}
\label{figPropSchSSC}
\end{figure}

First, we take smaller values for the coupling constant $g$ than the physical value $g = 4.1$ and keep the mass parameter $M$ fixed to the physical value $M/\mu = 0.454$. 
In Fig.~\ref{figPropSchSC}, the gluon propagator and the associated Schwinger functions are given for a smaller value $g = 2.3$. 
For a further smaller value $g = 1$, they are given in  Fig.~\ref{figPropSchSSC}. 

For smaller coupling constant $g$,  the gluon propagator $ {\mathscr{D}} $ seems to be monotonically decreasing in $k_E$.
The Schwinger function falls off very slowly from $t=0$ value and keeps its positivity until very large value of $t$, although it is difficult to see the difference from the graphs.  
Consequently, the smallest value of $t$ giving the negative value of the Schwinger function shifts to larger values of $t$, and eventually goes to infinity as $g \to 0$. 
This result is reasonable, since, in the vanishing coupling limit $g \to 0$, the gluon propagator must reduce to the free massive propagator in the tree level. Therefore, the reflection positivity must be recovered and the Schwinger function keeps positivity everywhere in the  limit $g \to 0$. 
As far as the results of the numerical calculations are concerned, the positivity seems to be not violated and restored for relatively smaller coupling constants. 
%In fact, if the propagator has the ordinary spectral representation, then the positivity $\rho(\sigma^2) \ge 0$ implies that the propagator $\tilde{\mathscr{D}}(k_E^2)$ is monotonically decreasing function of $k_E^2$. 
%The result of Fig.~\ref{pos} seems to be consistent with this statement. 
%However, it should be remarked that the monotonic propagator does not guarantee the positivity, since the converse is not necessarily true, see Appendix~E for the details. 

However, this observation turns out to be wrong. 
In fact, we can prove analytically that the reflection positivity of the gluon  Schwinger function is violated for any value of the parameters  $g $ and $M $  in the massive Yang-Mills  model with one-loop quantum corrections being included.
The proof will be given in the next section.
The Schwinger function $\Delta$ is an oscillating and exponentially fall-off function of $t$ approaching zero finally as $ t \to \infty$.
Therefore, it is difficult to examine the violation of positivity in the large $t$ region in the numerical way due to the restriction on the precision of numerical calculations.
For smaller coupling constant $g$, therefore, the Schwinger function takes a smaller but negative value for larger $t$, until the negativity disappears only in the limit $g \to 0$.

%-------------------------------
\subsubsection{Smaller mass parameter}

\begin{figure}[t]
\centering
\includegraphics[width=4cm]{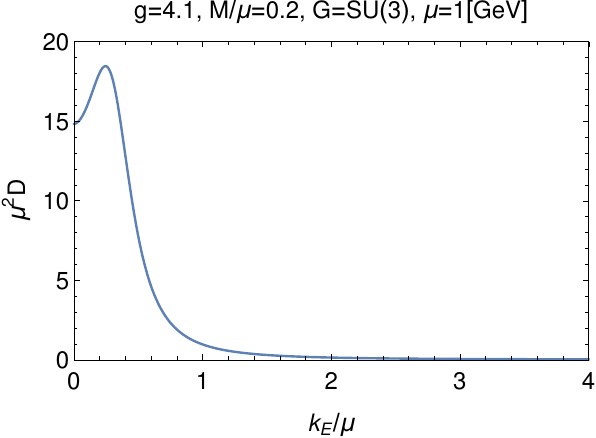}
\includegraphics[width=4cm]{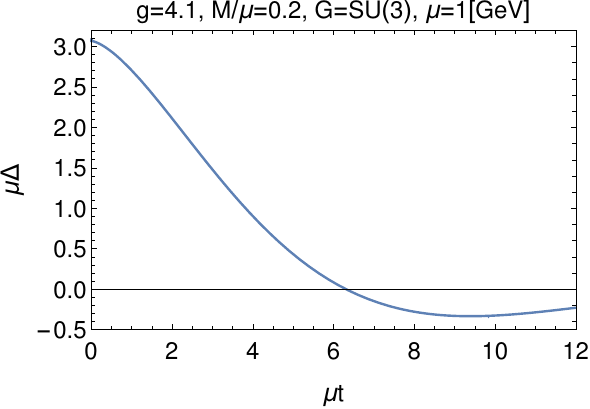}
\caption{
The same plots as those given in Fig.~\ref{figPropSchPhys} for a physical coupling constant $g = 4.1$ and a smaller mass $M/\mu=0.2$.
}
\label{figPropSchSM}
\end{figure}

\begin{figure}[t]
\centering
\includegraphics[width=4cm]{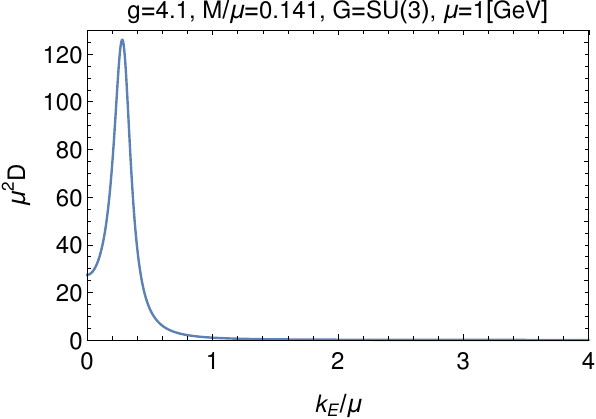}
\includegraphics[width=4cm]{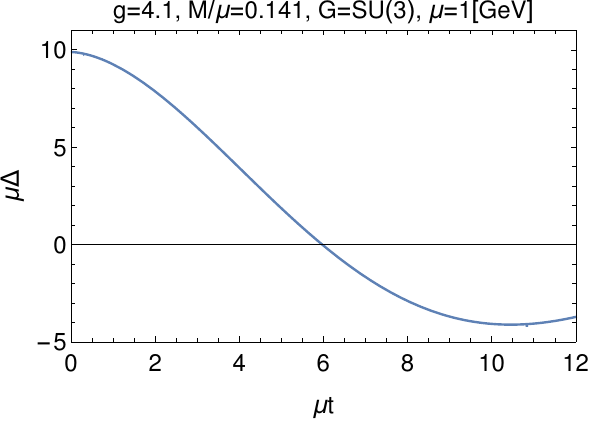}
\caption{
The same plots as those given in Fig.~\ref{figPropSchPhys} for a physical coupling constant $g = 4.1$ and a further smaller mass $M/\mu=0.141$.
}
\label{figPropSchSSM}
\end{figure}

Next, we keep the coupling constant fixed to the physical value $g=4.1$, and take smaller gluon mass parameter $M/\mu$ than the physical value $M/\mu=0.454$.
For a smaller value $M/\mu = 0.2$, the gluon propagator and the associated Schwinger functions are given in 
Fig.~\ref{figPropSchSM}.
For a further smaller value $M/\mu = 0.141$, they are given in 
Fig.~\ref{figPropSchSSM}.

As the value of mass parameter $M/\mu$ is chosen to be smaller and smaller than the physical value for the Yang-Mills theory, the gluon propagator $\tilde{\mathscr{D}}(p)$ exhibits sizable non-monotonic behavior and the Schwinger function exhibits more enhanced negativity, leading to the clearer violation of reflection positivity.

For smaller mass $M$ or larger coupling constant $g$ than the physical value for the Yang-Mills theory, the gluon propagator $\tilde{\mathscr{D}}_{\rm T}(p)$ exhibits stronger non-monotonic behavior.

%In Fig.~\ref{nonPos}, the gluon propagator is given for a different set of parameters $g = 6.0$ and $M = 0.5$ GeV. 
%The gluon propagator clearly shows sizable non-monotonic behavior. 
%Then the Schwinger function $\Delta(t)$ shows the stronger violation of positivity, which is to be compared with Fig.~\ref{lattice} for $g = 4.9$ and $M = 0.54$ GeV. 

%For smaller coupling constant $g = 3$ and $M = 0.5$ than the physical value in the Yang-Mills theory, the gluon propagator $\tilde{\mathscr{D}}_{\rm T}(p)$ is monotonically decreasing in $p$ and the resulting Schwinger function seems to be positive as the numerical calculations demonstrate in Fig.~\ref{pos}. 
%For smaller coupling constant, therefore, this numerical result seems to show that the positivity is not violated and is restored. 
%In fact, if the positivity holds $\rho(\sigma^2) \ge 0$, the propagator $\tilde{\mathscr{D}}(k^2)$ is monotonically decreasing function of $k^2$. 
%The result of Fig.~\ref{pos} seems to be consistent with this statement. 
%However, it should be remarked that the monotonic propagator does not guarantee the positivity, since the converse is not necessarily true, see Appendix~E for the details. 
%In fact, we can prove analytically that the reflection positivity of the gluon \textit{Schwinger function} is violated for any value of the parameters  $g $ and $M $  in the massive Yang-Mills  model with one-loop quantum corrections being included.
%The proof is given in the next subsection.

%-------------------------------
\subsubsection{Presence of Euclidean poles}

\begin{figure}[t]
\centering
\includegraphics[width=4cm]{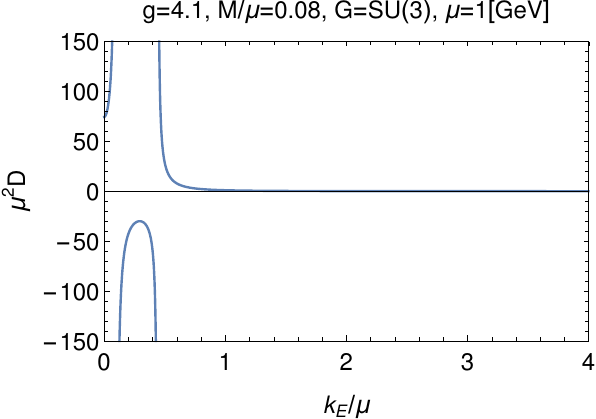}
\includegraphics[width=4cm]{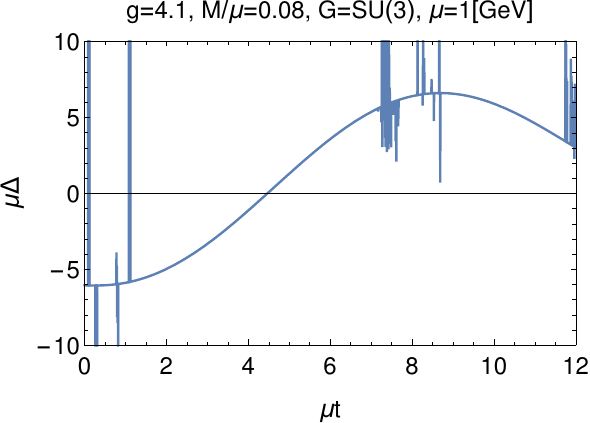}
\caption{
The same plots as those given in Fig.~\ref{figPropSchPhys} for a physical coupling constant $g = 4.1$ and a much smaller mass $M/\mu = 0.08$.
For this choice of the parameters, the Euclidean gluon propagator has poles. 
}
\label{figPropSchTach}
\end{figure}

For quite small mass parameter $M^2/\mu^2$ or large coupling constant $g$, the gluon propagator $\tilde{\mathscr{D}}(k_E^2)$ becomes singular at two values of $k_E^2$ and takes negative values in between.
In Fig.~\ref{figPropSchTach}, the gluon propagator is given for the parameters $g = 4.1$ and $M/\mu = 0.08$.
This result is consistent with the statement \cite{HK18} that the gluon propagator has poles in the Euclidean region (namely, tachyonic poles) with multiplicity two or a pair of complex conjugate poles under some assumptions on the propagator and the spectral function.
The related issue will be discussed in the next section.  

Therefore, this singular behavior affects the associated Schwinger function  $\Delta(t)$. 
This feature will be an artifact due to the limitation of one-loop calculations. 
Therefore, we exclude the relevant region of parameters from the following considerations.

%\newpage
%In Fig.~\ref{lattice}, we give the plot for the gluon propagator and the resulting Schwinger function in the covariant Landau gauge  $\alpha=0$ for the SU(3) Yang-Mills theory under the renormalization condition [TW] for $g = 4.9$ and $M = 0.54$ GeV.

%In Fig.~\ref{latticeOS}, we give the plots under the renormalization condition [OS].
%In both cases, the Schwinger functions clearly show the positivity violation. 

\subsection{Magnitude of positivity violation and the complementary gauge-scalar model}

\begin{figure}[t]
\centering
\includegraphics[width=4cm]{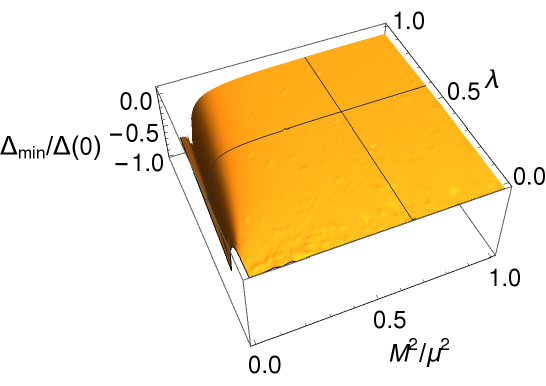}
\includegraphics[width=4cm]{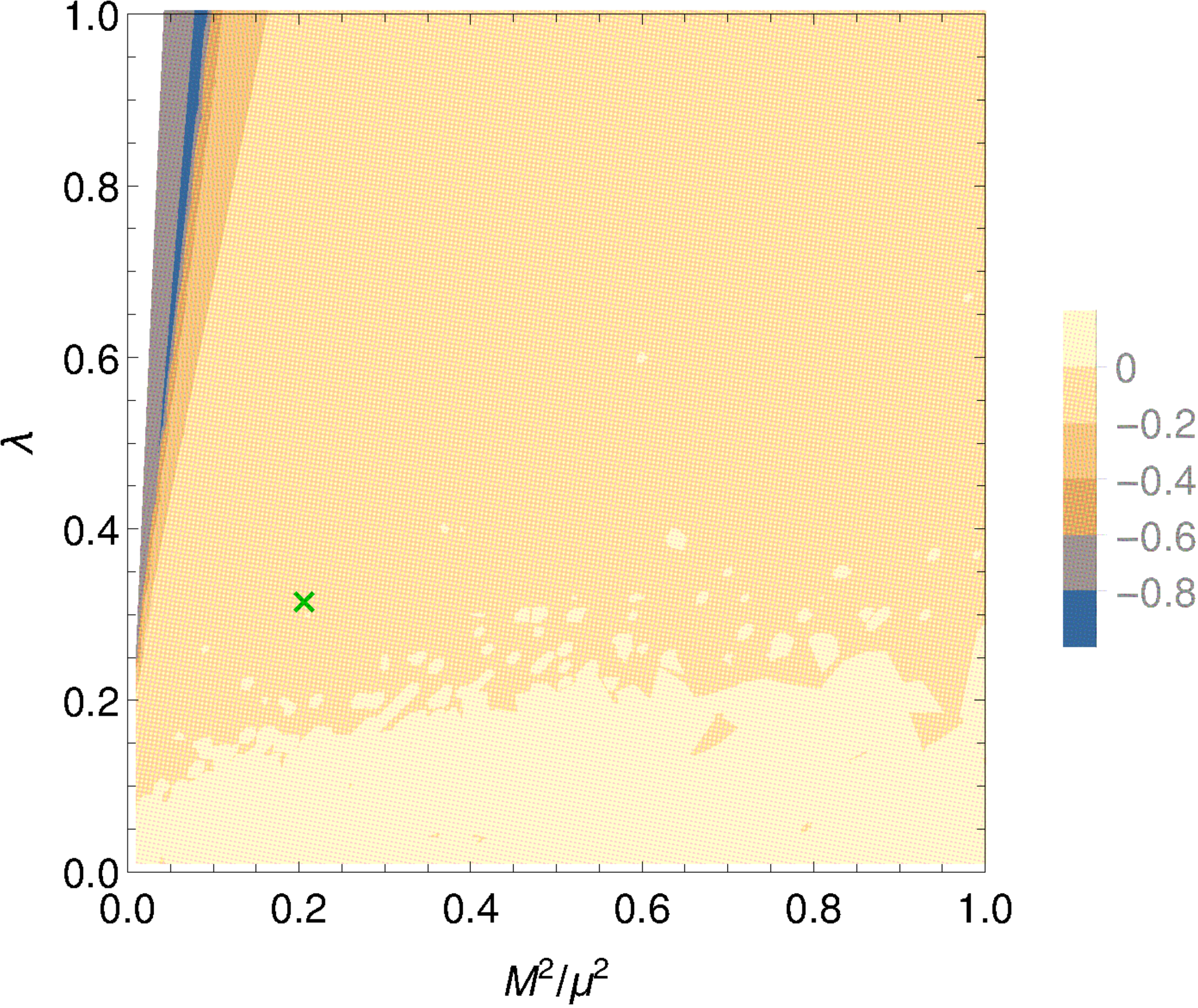}
\caption{
The magnitude of the violation of reflection positivity obtained from the ratio $\min_{0<t<\infty} \Delta(t) / \Delta(t = 0)$ of the Schwinger functions in the smaller range of parameters,    
(left) 3D plot, (right) contour plot.
}
\label{figSchRatio1}
\end{figure}
\begin{figure}[t]
\centering
\includegraphics[width=4cm]{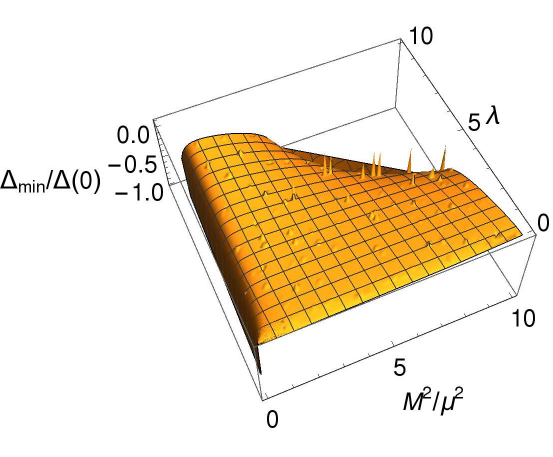}
\includegraphics[width=4cm]{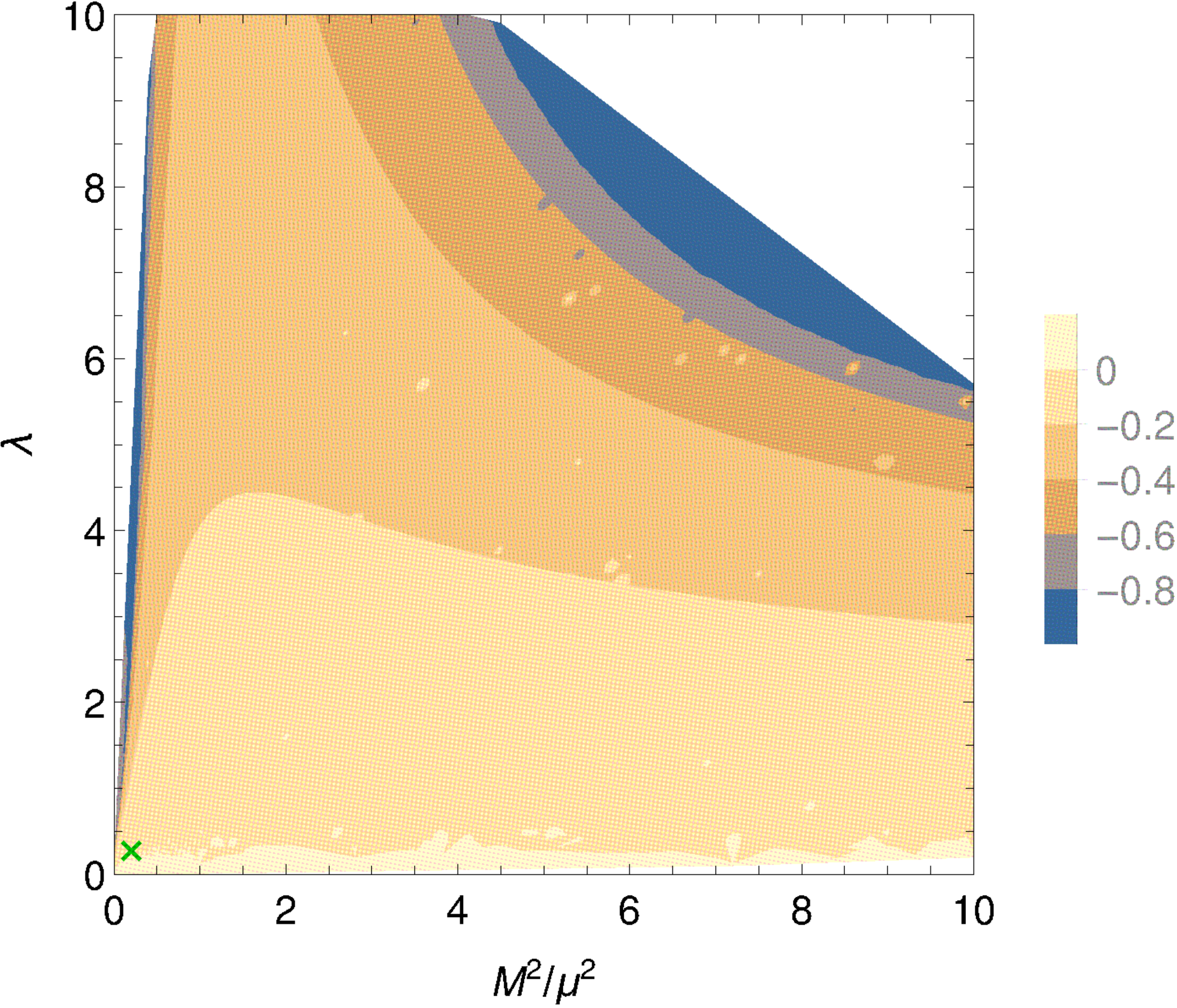}
\caption{
The magnitude of the violation of reflection positivity obtained from the ratio $\min_{0<t<\infty} \Delta(t) / \Delta(t = 0)$ of the Schwinger functions in the larger range of parameters,        
(left) 3D plot, (right) contour plot.
}
\label{figSchRatio10}
\end{figure}

Finally, we investigate to what extent the reflection positivity is violated depending on the choice of the parameters $g$ and $M$, although the reflection positivity is everywhere broken.
We examine the magnitude of positivity violation in the massive Yang-Mills model which could be regarded as the complementary gauge-scalar model.    
To estimate the violation of positivity of the Schwinger function $\Delta(t)$, we adopt the ratio $\min_{0<t<\infty} \Delta(t) / \Delta(t = 0)$ between the smallest value $\min_{0<t<\infty} \Delta(t)$ of  $\Delta(t)$ and the value at the origin $\Delta(t = 0)$. 
Fig.~\ref{figSchRatio1} gives the 3D plot and the contour plot of $\min_{0<t<\infty} \Delta(t) / \Delta(t = 0)$ on the two-dimensional parameter plane ($\frac{M^2}{\mu^2}$,$\lambda$)=($\frac{M^2}{\mu^2}$,$\frac{g^2C_2(G)}{16\pi^2}$). 
Fig.~\ref{figSchRatio10} gives the same plot with larger range of parameters.
Note that the left-upper $(\frac{M^2}{\mu^2} \ll 1,\lambda \gg 1 )$ and right-upper $(\frac{M^2}{\mu^2} \gg 1,\lambda \gg 1)$ regions in the contour plot correspond to the region to be excluded where the Euclidean poles occur. 
Note that the ratio $\min_{0<t<\infty} \Delta(t) / \Delta(t = 0)$ must be negative. 
However, there are spikes showing positive values in Fig.~\ref{figSchRatio10}, which are artifacts of our numerical calculations due to the algorithm used for looking for the very small negative value in the very large $t$ as the minimum. 
These spikes are to be ignored.

If $g^2 \rightarrow 0$, the theory has no interaction and the propagator approaches the free massive propagator $\mathscr{D}(k) = \frac{1}{k^2 + M^2}$.  
In this limit, the Schwinger function is positive for any value of $M$ and there is no violation of reflection positivity. 
%Therefore, the Schwinger function must be positive in the small $g^2$ region for any value of $M$.
%In fact, our results are consistent with this observation.
For small $g^2$ and large $M^2/\mu^2$, namely, for large $1/g^2$ and large $v^2 \simeq (M^2/\mu^2)/g^2$, the Schwinger function exhibits small violation of positivity. This region corresponds to the \textit{Higgs-like region} in the complementary gauge-scalar  model.
% and there is no contribution of the loop calculations (e.g. $\Pi(k)$).
%\noindent
%$\bullet$ 
For large $g^2$ and small $M^2/\mu^2$, namely, for small $1/g^2$ and small $v^2 \simeq (M^2/\mu^2)/g^2$, the Schwinger function exhibits large violation of positivity.  This region corresponds to the \textit{confinement-like region} in the  complementary gauge-scalar  model.

However, there is no phase transition between the positivity violation and restoration. 
There is just a smooth crossover separating large and small violation of positivity.
The massive Yang-Mills model has only one confinement phase. 
This result is interpreted as the Fradkin-Shenker continuity in the  complementary gauge-scalar model from the viewpoint of the gauge-invariant extension from the massive Yang-Mills model to the gauge-invariant complementary gauge-scalar model explained in section II.

\section{
Complex analysis of the gluon propagator
}
%%%%%%%%%%%%%%%%%%%%%%%%%%%%%%%%%%%%%%%%%%%%%%%%%%%%%%%%%%%%%

In the previous section we have investigated the propagator in the Euclidean region. 
We have shown the violation of reflection positivity in the massive Yang-Mills model. However, this result is obtained only in the numerical way. 
In this section, we study the propagator on the complex plane of the squared momentum $k^2$, which follows from the analytic continuation of the propagator from the Euclidean region to the Minkowski region. 
We find that the violation of the reflection positivity in the Euclidean region is understood from the existence of a pair of complex conjugate poles and the discontinuity across the branch cut yielding the negative spectral function represented by the generalized spectral representation of the gluon propagator. 
As a consequence of the complex structure, we give an analytical proof that the reflection positivity is always violated for any choice of the parameters $M$ and $g$ in the massive Yang-Mills model to one-loop order.

\subsection{Spectral representation of a propagator}

It is well-known that a propagator $\mathscr{D}(k^2)$ in the Minkowski region $k^2>0$ (for the time-like momentum $k$) has the \textit{spectral representation} of the K\"all\'en--Lehmann form under assumptions of the general principles of the QFT such as the spectral condition, the Poincar\'e invariance and the completeness of the state space \cite{spectral_repr_UKKL}: The full propagator $\mathscr{D}(k^2)$ of the field $\phi$ is written as the weighted sum of the free propagator,
\begin{align}
 \mathscr{D}(k^2) = \int_0 ^\infty d \sigma^2 \frac{\rho(\sigma^2)}{\sigma^2 - k^2} , \ k^2 \ge 0 , 
\label{eq:KL_spectral_repr-0} 
\end{align} 
with the weight function $\rho(\sigma^2)$ called the \textit{spectral function} being obtained from the state sum
\begin{align}
   \theta(k_0) \rho(k^2) 
%\nonumber\\
:=  (2\pi)^{d} \sum_{n }  |\langle 0 | \phi(0) | P_n \rangle|^2 \delta^D(P_n-k) ,
\end{align}
where $d$ is the space dimension, $D$ is the spacetime dimension,  the sum is over all the intermediate states with the total momentum $P_n$, and $\theta(k_0)$ is a step function ensuring the positivity $k_0 \ge 0$. 
The spectral function $\rho$ has contributions from a \textit{stable single-particle state} with physical mass $m_{P}$ (pole mass) and intermediate many-particle states $| p_1,...,p_n \rangle$ with a continuous spectrum, such as two-particle states, three-particle states, and so on,  
\begin{align}
  \rho(k^2) =& Z \delta(k^2 -m_{P}^2) + \tilde\rho(k^2) , \ k^2 \ge 0 ,
\nonumber\\
\tilde\rho(k^2) =& (2\pi)^{d} \sum_{n=2}^{\infty}  |\langle 0 | \phi(0) | p_1,...,p_n \rangle|^2 \delta^D(p_1+...+p_n-k)
 .
 \label{spectral-f-R}
\end{align}
Then the spectral representation is written as the sum of the contributions from the real pole $k^2=m_P^2$ and the branch cut 
\begin{align}
 \mathscr{D}(k^2) =  \frac{Z}{m_P^2 - k^2} + \int_{0} ^\infty d \sigma^2 \frac{\tilde\rho(\sigma^2)}{\sigma^2 - k^2} 
 , \ k^2 \ge 0 . 
\label{eq:KL_spectral_repr-02} 
\end{align}

%%%%%%%%%%%%%%%%%%%%%%%%%%%%%%%%%%%%%%%%%%%%%%%%%%%%%%%%%%%%
\begin{figure}[tbp]
 \begin{center}
\includegraphics[scale=0.35]{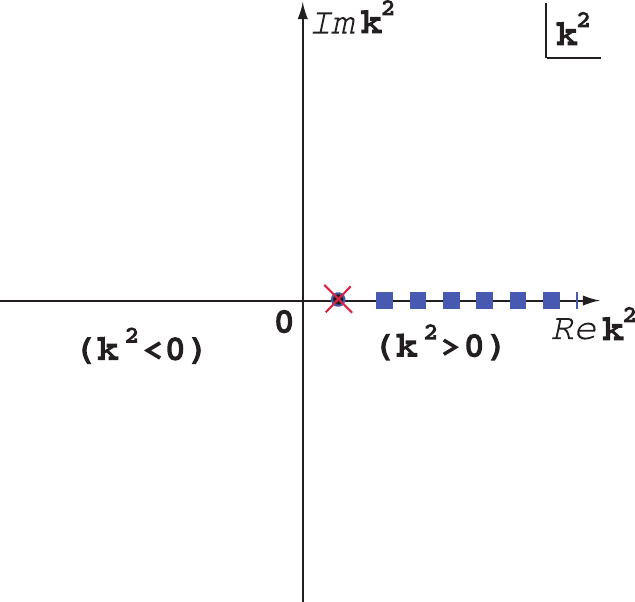}
\quad
\includegraphics[scale=0.35]{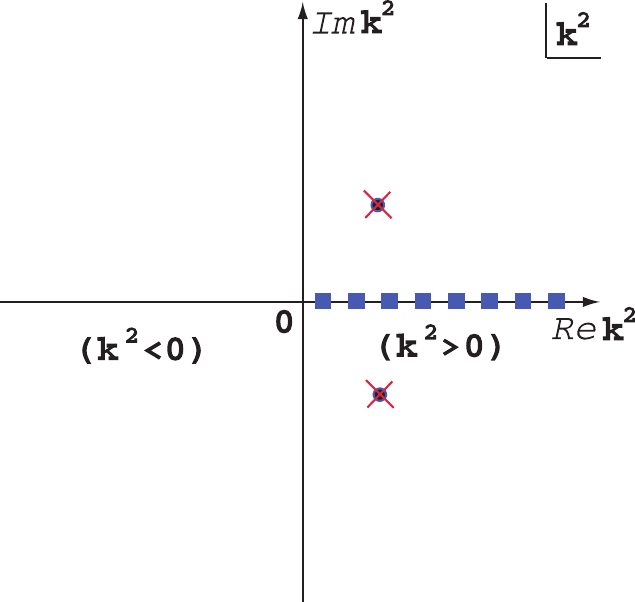}
 \end{center}
 \caption{
 Possible singularities of the propagator on the complex $k^2$ plane,   
 (Left) a real pole and the branch cut on the positive real axis, 
 (Right) a pair of complex conjugate poles and the branch cut  on the positive real axis. 
}
 \label{fig:complex-singularity}
\end{figure}
%%%%%%%%%%%%%%%%%%%%%%%%%%%%%%%%%%%%%%%%%%%%%%%%%%%%%%%%%%%%

This spectral representation can be extended to the complex momentum $k^2 \in \mathbb{C}$. 
See the left panel of Fig.~\ref{fig:complex-singularity}. 
A propagator $\mathscr{D}(k^2)$ as a complex function of the complex variable $z = k^2 \in \mathbb{C}$ has the spectral representation  with the spectral function $\rho$, 
\begin{align}
 \mathscr{D}(k^2) &= \int_0 ^\infty d \sigma^2 \frac{\rho(\sigma^2)}{\sigma^2 - k^2}, \ k^2 \in \mathbb{C}-[s_{\rm min}, \infty) ,
\label{eq:KL_spectral_repr}
\\
 \rho(\sigma^2) &:= \frac{1}{\pi} \operatorname{Im} \mathscr{D}(\sigma^2+i\epsilon).
 \label{spectral-f-C}
\end{align} 
%\begin{align}
%  f(k^2) =   \int_{s_{\rm min}}^{\infty} d\zeta \frac{\rho(\zeta) }{ \zeta-k^2} , 
%  \quad
%  \rho(\zeta) := \frac{1 }{ \pi} \operatorname{Im} f(\zeta+i0) ,
%\label{CCF-dispersionrel}
%\end{align}
This representation (\ref{eq:KL_spectral_repr})  is applied to an arbitrary $k^2$ in the complex plane except for the singularities located on the positive real axis $[s_{\rm min}, \infty)$.
The spectral function $\rho$ (\ref{spectral-f-C})  known as the dispersion relation is obtained from the discontinuity across the branch cut,
$
\mathscr{D}(z+i\epsilon)-\mathscr{D}(z-i\epsilon)
=\mathscr{D}(z+i\epsilon)-\mathscr{D}(z+i\epsilon)^{*}
= 2i \operatorname{Im} \mathscr{D}(z+i\epsilon) .
$
It is explicitly checked that the two definitions of the spectral functions (\ref{spectral-f-R}) and (\ref{spectral-f-C}) agree with each other once the theory is specified. 
% for e.g., the scalar $g\phi^3$ theory.  
The representation (\ref{eq:KL_spectral_repr}) is obtained under the following assumptions \cite{Kondo03}: 
\begin{enumerate}
 \item $\mathscr{D}(z)$ is holomorphic except singularities on the positive real axis.
 \item $\mathscr{D}(z) \rightarrow 0$ as $|z| \rightarrow \infty$.
 \item $\mathscr{D}(z)$ is real on the negative real axis.
\end{enumerate}
%$\mathscr{D}(z)$ vanishes  as $|z| \rightarrow \infty$, 
%Here $\mathscr{D}(z)$ need not to go to zero $\mathscr{D}(z) \rightarrow 0$ in the power-like, i.e., $\mathscr{D}(z) \sim 1/|z|^{\varepsilon} (\varepsilon>0)$ as is usually assumed. 
%In fact, a weaker decay $\mathscr{D}(z) \sim (\ln|z|)^{-\varepsilon} (\varepsilon>0)$ is also allowed.
This is indeed the case of the quantum Yang-Mills theory, see e.g.,
\cite{HK18}. 
% in the \textit{cut complex plane} (complex plane excluding singularities such as poles and a branch cut on the positive real axis) 

The spectral representation has a straightforward generalization in the presence of complex simple poles, see e.g., \cite{Siringo17a,HK18}. 
%The following results are based on \cite{HK18}. 
Suppose that the propagator has simple complex poles at $z = z_\ell $ $(\ell = 1, \cdots, n)$.
See the right panel of Fig.~\ref{fig:complex-singularity}. 
Then the propagator $\mathscr{D}(k^2)$ has the \textit{generalized spectral representation},
\begin{align}
 \mathscr{D}(k^2) 
&=  \mathscr{D}_{p}(k^2) + \mathscr{D}_{c}(k^2) , \ k^2 \in \mathbb{C}-([\sigma^2_{\rm min}, \infty) \cup \{ z_\ell \}_\ell ),
\nonumber\\
 &= \sum_{\ell=1}^n \frac{Z_\ell}{z_\ell - k^2} 
+ \int_0 ^\infty d \sigma^2 \frac{\rho(\sigma^2)}{\sigma^2 - k^2} , 
\label{eq:spec_repr_complex} 
\\
 \rho(\sigma^2) &:= \frac{1}{\pi} \operatorname{Im} \mathscr{D}(\sigma^2+i\epsilon), 
\label{eq:dispersion_complex} 
\\ 
 Z_\ell &:= \oint_{\gamma_\ell} \frac{d k^2}{2 \pi i} \mathscr{D}(k^2), 
\label{eq:dispersion_residue}
\end{align}
where $\gamma_\ell$ is a small contour circulating clockwise around  the pole at $z_\ell$.
Here we have separated the propagator $\mathscr{D}$ into the  contribution from the complex poles $\mathscr{D}_p$ and that from the branch cut $\mathscr{D}_c$. 
This is derived from the following assumptions \cite{HK18}:
\begin{enumerate}
 \item $\mathscr{D}(z)$ is holomorphic except singularities on the positive real axis and a finite number of simple poles.
 \item $\mathscr{D}(z) \rightarrow 0$ as $|z| \rightarrow \infty$.
 \item $\mathscr{D}(z)$ is real on the negative real axis.
\end{enumerate}
%If the poles were not simple, the second term would be modified as
%\begin{align}
%& \sum_{\ell=1}^n \frac{1}{2 \pi i} \oint_{\gamma_\ell} d \zeta \frac{\mathscr{D}(\zeta)}{\zeta - k^2} = \sum_{\ell=1}^n \sum_{n=1}^\infty \frac{a_{-n}^\ell}{(k^2 - z_\ell)^n}, \notag \\
%&a_{-n}^\ell = - \frac{1}{2 \pi i} \oint_{\gamma_\ell} d k^2\ \mathscr{D}(k^2) (k^2 - z_\ell)^{n-1}.
%\end{align}
Note that the poles must appear as real poles or pairs of complex conjugate poles as a consequence of the Schwarz reflection principle $\mathscr{D}(z^*) = [\mathscr{D}(z)]^*$.

From now on, we focus on a propagator with a pair of complex conjugate simple poles.
%For example, the propagator consisting of a pair of complex conjugate poles reproduces well the results of the numerical simulations \cite{Gribov78,Stingl86,Zwanziger90,DGSVV2008,BDGHSVZ10}. 
This is indeed the case for the gluon propagator of the massive Yang-Mills model as will be shown in the next subsection.
For a propagator with one pair of complex conjugate simple poles at $k^2 = v \pm i w$,  
%where we can set $w > 0$ without loss of generality, 
the generalized spectral representation (\ref{eq:spec_repr_complex}) reduces to 
\begin{align}
 \mathscr{D}(k^2) &= \mathscr{D}_p(k^2) + \mathscr{D}_c(k^2)  ,
 \nonumber\\
\mathscr{D}_p(k^2)  &:=  
   \frac{Z}{(v+iw) - k^2} + \frac{Z^*}{(v-iw) - k^2} , 
 \nonumber\\
 \mathscr{D}_c(k^2) & := \int_0 ^\infty d \sigma^2 \frac{\rho(\sigma^2)}{\sigma^2 - k^2} .
\label{eq:one_pair_complex}
\end{align}

\subsection{Gluon propagator on the complex momentum plane}

\begin{figure}[t]
\centering
\includegraphics[width=8cm]{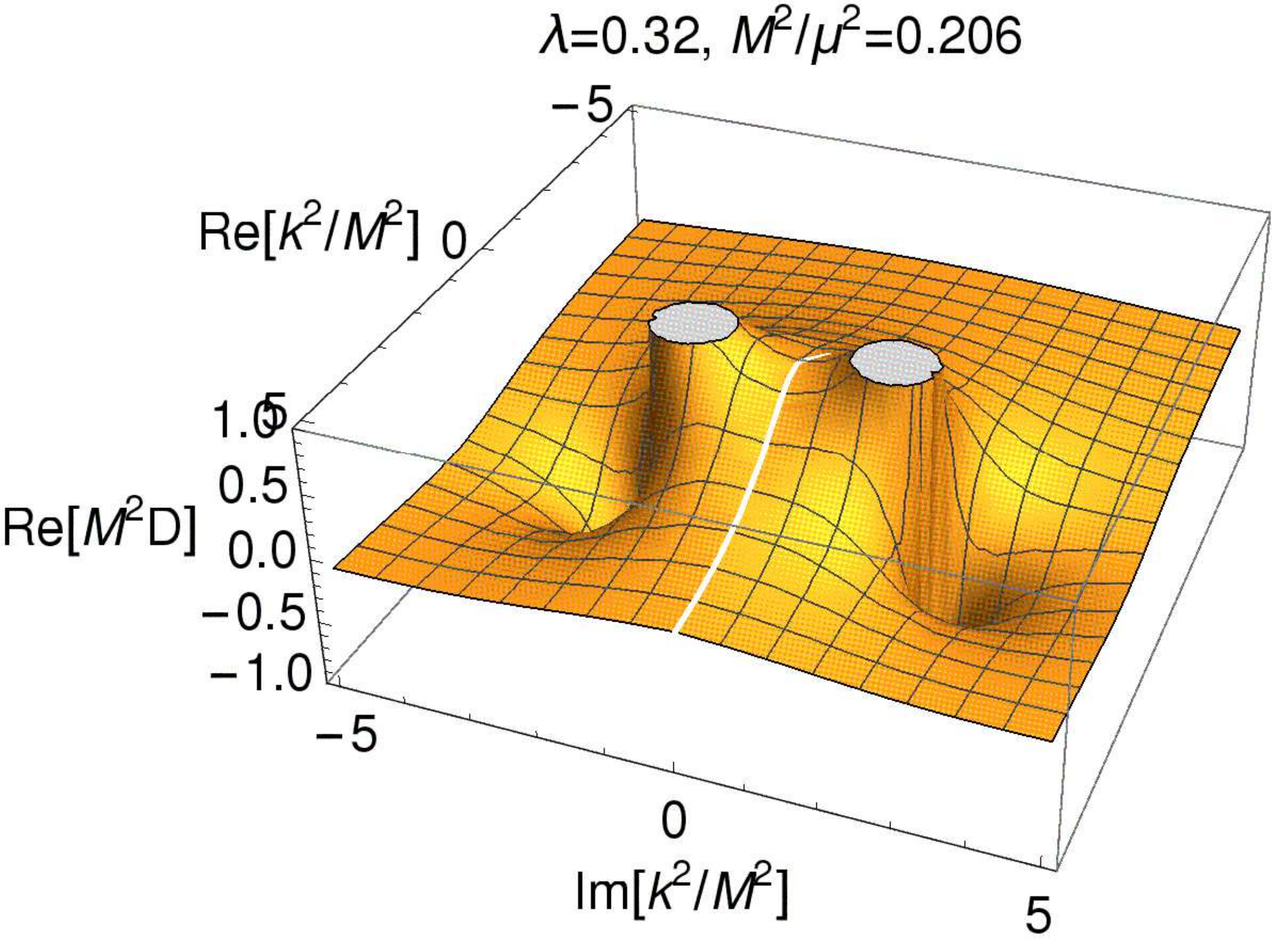}
\includegraphics[width=8cm]{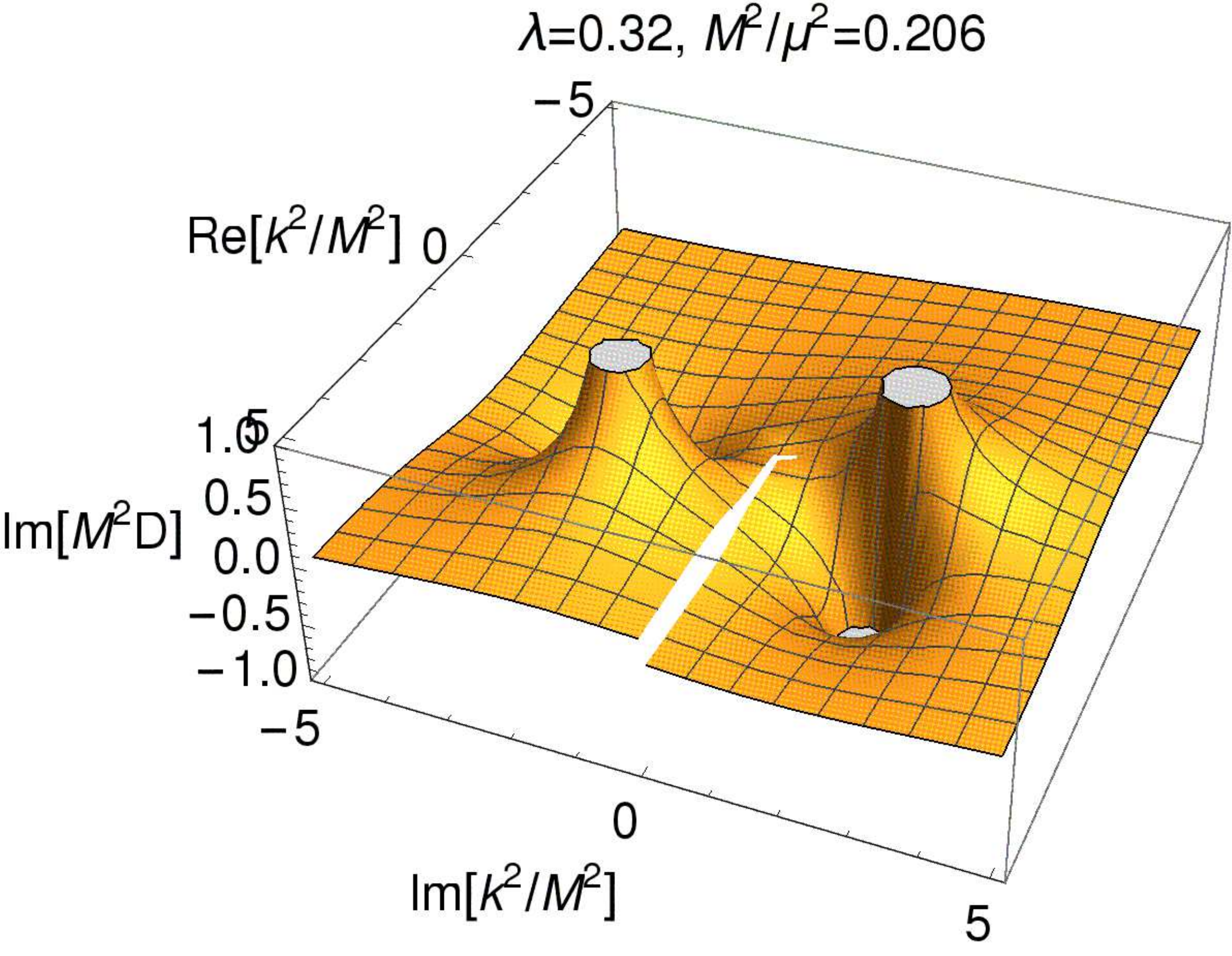}
\caption{
The gluon propagator $\mathscr{D}(k^2)$ as a complex function of  the complex squared momentum $k^2 \in \mathbb{C}$,  
(top) the real part $\operatorname{Re}\mathscr{D}(k^2)$, (bottom) the imaginary part $\operatorname{Im}\mathscr{D}(k^2)$, 
at the physical point of the parameters $\lambda := Ng^2/(4\pi)^2= 0.32$, $M^2/\mu^2 = 0.206$.
}
\label{figPropAnaConPhys}
\end{figure}

\begin{figure}[t]
\centering
\includegraphics[width=8cm]{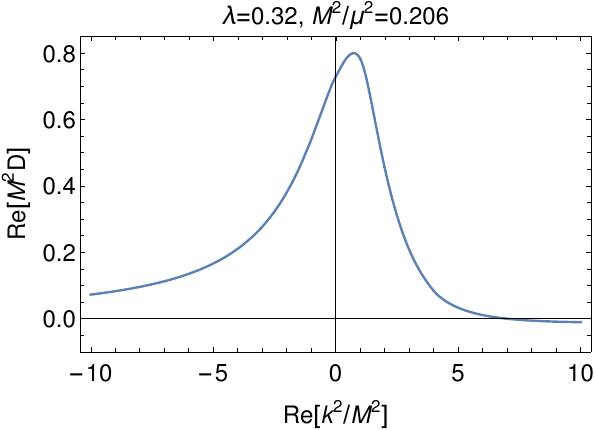}
\includegraphics[width=8cm]{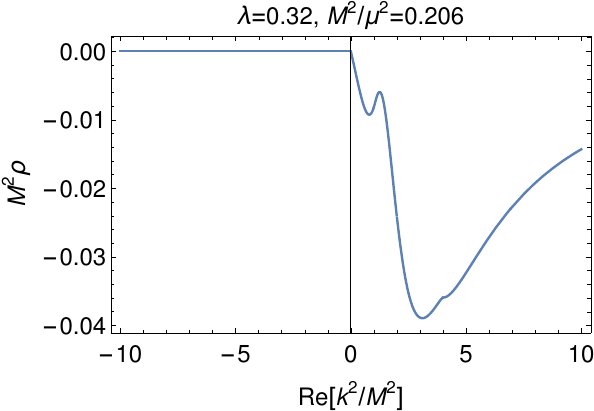}
\caption{
The gluon propagator $\mathscr{D}(k^2)$ as a function of $k^2 $ restricted on the real axis $k^2 \in \mathbb{R}$, 
(top) the real part $\operatorname{Re}\mathscr{D}(k^2)$, (bottom) the scaled imaginary part $\operatorname{Im}\mathscr{D}(k^2+i\epsilon)/\pi$ which is equal to the spectral function $\rho(k^2)$, 
at the physical point of the parameters $\lambda := Ng^2/(4\pi)^2= 0.32$, $M^2/\mu^2 = 0.206$.
}
\label{figPropRealAxisPhys}
\end{figure}

We first perform the analytic continuation of the propagator $\mathscr{D}$ in the Euclidean region $k^2=-k_E^2<0$ to the entire complex plane $k^2 \in \mathbb{C}$.  
Fig.~\ref{figPropAnaConPhys} is the plot of the real and imaginary parts of the complex-valued gluon propagator $\mathscr{D}(k^2)$ on the complex momentum plane $k^2 \in \mathbb{C}$, at the physical point of the parameters (\ref{exFitParams}) in the massive Yang-Mills model. 
Note that the gluon propagator $\mathscr{D}(k^2)$ is real-valued on the negative real axis (space-like momentum) $k^2=-k_E^2<0$, since  the imaginary part $\operatorname{Im}\mathscr{D}(k^2)$ is zero on the negative real axis (space-like momentum).
The real part $\operatorname{Re}\mathscr{D}(k^2)$ on the negative real axis $k^2=-k_E^2<0$ is identical to the Euclidean propagator.
We observe that the gluon propagator has a pair of complex conjugate poles and the imaginary part has discontinuities across the branch cut on the positive real axis $\mathscr{D}(k^2 + i\epsilon) \not= \mathscr{D}(k^2- i\epsilon)$ ($k^2>0$, $\epsilon \downarrow 0$), while there are no discontinuities on the negative real axis  $\mathscr{D}(k^2 + i\epsilon) = \mathscr{D}(k^2- i\epsilon)$ ($k^2<0$, $\epsilon \downarrow 0$). 
Therefore, in discussing the behavior of the propagator on the positive real axis, we must specify which side is used.  In what follows we use the limit $\mathscr{D}(k^2+i\epsilon)$ ($\epsilon \downarrow 0$).

\begin{figure}[t]
\centering
\includegraphics[width=8cm]{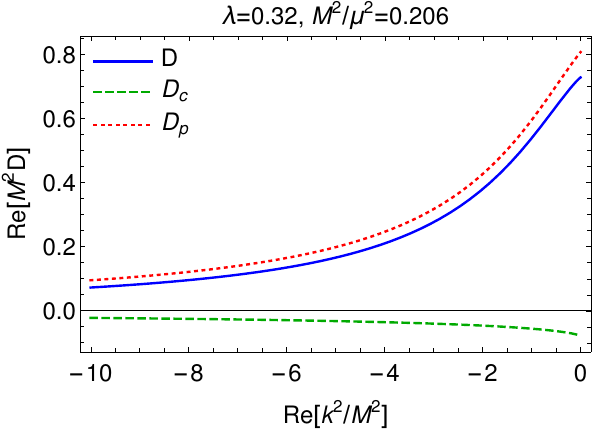}
\includegraphics[width=8cm]{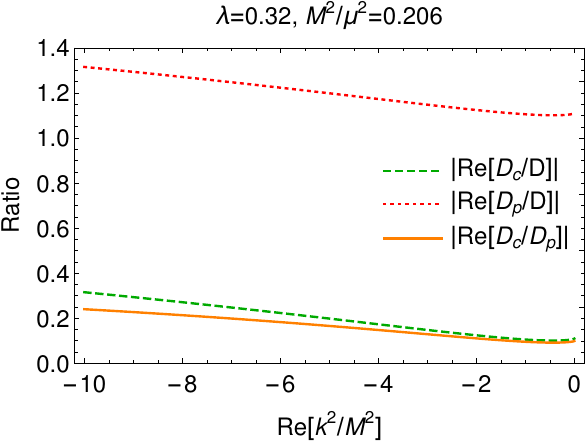}
\caption{
The comparison of the pole  and cut parts with the original gluon propagator in the Euclidean region $\mathscr{D}(k^2)=\mathscr{D}_{p}(k^2)+\mathscr{D}_{c}(k^2)$ for $k^2<0$:  
(top) the pole part $\mathscr{D}_p(k^2)$ (red dotted line) and the cut part $\mathscr{D}_c(k^2)$ (green broken line) in the gluon propagator $\mathscr{D}(k^2)$ (blue solid line), 
(bottom) 
the absolute values of the real part of the ratio of the pole and  cut parts to the total gluon propagator,  
$|\operatorname{Re}[\mathscr{D}_{p}(t)/\mathscr{D}(t)]|$ (red dotted line), $|\operatorname{Re}[\mathscr{D}_{c}(t)/\mathscr{D}(t)]|$ (green broken line), and $|\operatorname{Re}[\mathscr{D}_{c}(t)/\mathscr{D}_{p}(t)]|$ (orange solid line), 
at the physical point of the parameters $\lambda := Ng^2/(4\pi)^2= 0.32$, $M^2/\mu^2 = 0.206$.
}
\label{figPropBreakPhys}
\end{figure}

\begin{figure}[t]
\centering
\includegraphics[width=8cm]{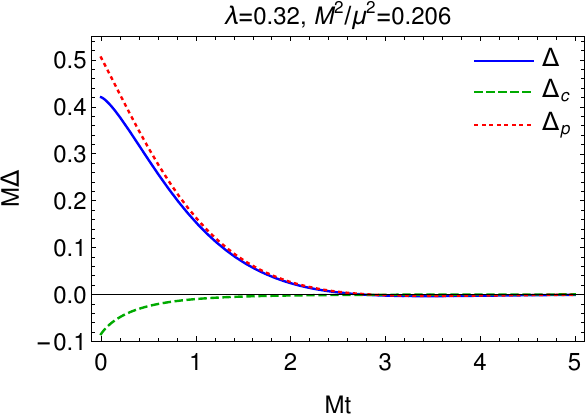}
\includegraphics[width=8cm]{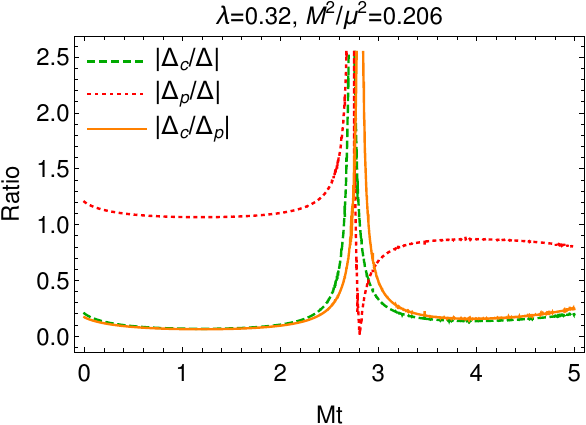}
\caption{
The comparison of the pole  and cut  parts with the original gluon Schwinger function  in the Euclidean region $\Delta(t)=\Delta_p(t)+\Delta_c(t)$ for $k^2<0$:  
(top) the pole part $\Delta_{p}(t)$ (red dotted line) and the cut part $\Delta_{c}(t)$ (green broken line) in the gluon Schwinger function $\Delta (t)$ (blue solid line), 
(bottom) 
the absolute values of the ratios $\Delta_{c,p}(t)/\Delta(t)$ of the pole $\Delta_{p}(t)$ and cut $\Delta_{c}(t)$ parts to the total gluon Schwinger function,  
$|\Delta_{p}(t)/\Delta(t)|$ (red dotted line), $|\Delta_{c}(t)/\Delta(t)|$ (green broken line), and $|\Delta_{c}(t)/\Delta_{p}(t)|$ (orange solid line),
at the physical point of the parameters $\lambda := Ng^2/(4\pi)^2= 0.32$, $M^2/\mu^2 = 0.206$.
Notice that the ratio blows up at a zero of the Schwinger function.
}
\label{figSchBreakPhys}
\end{figure}

Next, we focus on the real axis $k^2 \in \mathbb{R}$ to see the behavior of the complex-valued gluon propagator $\mathscr{D}(k^2)$ as a function of a real-valued momentum $k^2 \in \mathbb{R}$.  
 Fig.~\ref{figPropRealAxisPhys} is the plot of the real and imaginary parts of the complex-valued gluon propagator on the real axis $k^2 \in \mathbb{R}$ at the physical point of the parameters (\ref{exFitParams})  in the massive Yang-Mills model. 
On the negative real axis $k^2=-k_E^2<0$ (the Euclidean region), we find that the real part $\operatorname{Re}\mathscr{D}(k^2)$ is always positive, % $\operatorname{Re}\mathscr{D}(k^2)>0$, 
and the imaginary part $\operatorname{Im}\mathscr{D}(k^2)$ is identically zero.
On the positive real axis $k^2>0$ (the Minkowski region),  $\operatorname{Re}\mathscr{D}(k^2)$ changes the sign such that it is positive for small $k^2$, %$\operatorname{Re}\mathscr{D}(k^2)>0$ for $0<k^2 \ll 1$ 
and negative for large $k^2$, 
%$\operatorname{Re}\mathscr{D}(k^2)<0$ for   $k^2 \gg 1$, 
which implies the existence of (at least one) zeros of $\operatorname{Re}\mathscr{D}(k^2)$ in the Minkowski region $k^2>0$. 
The scaled imaginary part $\operatorname{Im}\mathscr{D}(k^2+i\epsilon)/\pi$ is identical to the spectral function $\rho(k^2)$. Therefore, the spectral function is identically zero in the Euclidean region, 
\begin{align}
 \rho(k^2) \equiv 0 \ \text{for} \  k^2=-k_E^2<0 . 
\end{align}
However, it is non-trivial in the Minkowski region. 
It is remarkable that the spectral function is always negative, %$\rho(k^2) < 0$ for $k^2>0$
\begin{align}
\rho(k^2) := \frac{1}{\pi} \operatorname{Im}\mathscr{D}(k^2+i\epsilon)  < 0 \ \text{for} \ k^2>0 ,
\label{def-spectral}
\end{align}
in the massive Yang-Mills model to one-loop order.

For a given propagator $\mathscr{D}(k^2)$, we can decompose it into  the contribution from the branch cut $\mathscr{D}_{c}(k^2)$ and that from the poles $\mathscr{D}_{p}(k^2)$.
Fig.~\ref{figPropBreakPhys} gives this decomposition of the gluon  propagator for the Euclidean momentum $\mathscr{D}(k^2)=\mathscr{D}_{p}(k^2)+\mathscr{D}_{c}(k^2)$ for $k^2<0$.

According to the separation of the propagator, the Schwinger function is also separated into the two parts:
the continuous cut part $\Delta_{c}(t)$ coming from the spectral function and the pole part $\Delta_{p}(t)$ coming from the pole part $\mathscr{D}_{p}$ of the propagator $\mathscr{D}$,
\begin{align}
\Delta(t)  =&  \Delta_{p}(t)  +  \Delta_{c}(t) , 
\nonumber\\
\Delta_{p}(t) :=& \int_{-\infty}^{+\infty} \frac{dk_E}{2\pi}  e^{ik_E t}  {\mathscr{D}}_{p}(k_E^2) ,
\nonumber\\
\Delta_{c}(t) :=& \int_{-\infty}^{+\infty} \frac{dk_E}{2\pi}  e^{ik_E t}  {\mathscr{D}}_{c}(k_E^2) .
\label{CCF-spec-Sch2m}
\end{align}
Especially, the cut part $\Delta_{c}(t)$ is directly written as an integral of the spectral function as follows 
\begin{align}
 \Delta_{c}(t)  
=& \int_{-\infty}^{+\infty} \frac{dk_E}{2\pi}  e^{ik_E t}  \int_0 ^\infty d \sigma^2 \frac{\rho(\sigma^2)}{\sigma^2 + k_E^2}
\nonumber\\
=& \int_0 ^\infty d \sigma^2 \rho(\sigma^2) \int_{-\infty}^{+\infty} \frac{dk_E}{2\pi}  e^{ik_E t}   \frac{1}{\sigma^2 + k_E^2}  
\nonumber\\
=& \int_{0}^{\infty} d\sigma^2    \rho(\sigma^2) \frac{1}{2\sqrt{\sigma^2}} e^{-\sqrt{\sigma^2} t}  .
%\nonumber\\
%=& \int_{0}^{\infty} d\sigma   \rho(\sigma^2) e^{-\sigma t} .
\label{Schwinger-f-cut}
\end{align}

The same procedure is also applied to the Schwinger function.
Fig.~\ref{figSchBreakPhys} shows the respective ratio $\Delta_{c,p}(t)/\Delta(t)$ of the pole or cut part $\Delta_{c,p}(t)$ to the total Schwinger function $\Delta (t)$. 
Using the already known spectral function $\rho(k^2)$ calculated according to $\rho(k^2)=\operatorname{Im}\mathscr{D}(k^2)/\pi$, the cut part $\Delta_c(t)$ of the Schwinger function is obtained by integrating $\rho(k^2)$ according to (\ref{Schwinger-f-cut}).  Then the pole part $\Delta_p(t)$ of the Schwinger function is obtained as the difference $\Delta_p(t)=\Delta(t)-\Delta_c(t)$ from the total Schwinger function $\Delta (t)$. 
Note that the ratio can become divergent at a zero $t_0$ of the Schwinger function $\Delta (t_0)=0$, which should be ignored as an artifact of this procedure, see also the caption of Fig.~\ref{figSchBreakPhys}.

%----------------------------------------------------
\subsection{A pair of complex conjugate poles and Gribov-Stingl form}

If the propagator has no complex poles besides the singularities on the real positive axis, the complex  pole part vanishes $\mathscr{D}_p(k^2) = 0$ and the Euclidean gluon propagator obeys the usual spectral representation 
\begin{align}
\mathscr{D}(-k_E^2) = \mathscr{D}_c(-k_E^2) = \int_0^{\infty} d \sigma^2 \rho(\sigma^2) \frac{ 1 }{ \sigma^2 + k_E^2 } .
\end{align}
Then the Schwinger function is calculated from the cut part alone
\begin{align}
\Delta(t) = \Delta_c(t) = \int_0^{\infty} d \sigma^2  \rho(\sigma^2) \frac{1}{2 \sqrt{\sigma^2} }  e^{ - \sqrt{\sigma^2} |t| } .
\end{align}
In this case, we find that the positivity of the spectral function $\rho$ implies the positivity of the Schwinger function $\Delta$
\begin{align}
\rho(\sigma^2) \ge 0 \ \text{for} \ {}^{\forall} \sigma^2 \ge 0 
\ \Rightarrow \ 
\Delta(t) \ge 0 \ \text{for} \  {}^{\forall} t \ge 0 ,
\end{align}
which implies that non-positivity of the Schwinger function $\Delta$ yields non-positivity of the spectral function $\rho$, namely, violation of reflection positivity,
\begin{align}
\Delta(t) < 0 \ \text{for} \ {}^{\exists} t \ge 0
\ \Rightarrow \ 
\rho(\sigma^2) < 0 \ \text{for} \ {}^{\exists} \sigma^2 \ge 0 .
\end{align}
Thus, when the propagator has no singularities other than the positive real axis, the positivity of the spectral function is directly related to the positivity of the Schwinger function, or the reflection positivity.
The violation of reflection positivity can be seen as the non-positivity of the spectral function. 
However, this is not the case for the Yang-Mills theory, as demonstrated in the massive Yang-Mills model shortly. 

Suppose that the propagator has a pair of complex conjugate poles at   
 $k^2 = v \pm i w$ $(v,w \in \mathbb{R}, w > 0)$ with the  respective residues $Z, Z^* \in \mathbb{C}$.
Then the pole part of the propagator in the Euclidean region is represented  as 
\begin{align}
\mathscr{D}_p(k^2=-k_E^2) 
%&= - \frac{Z}{k^2 - (v + i w)} - \frac{Z^*}{k^2 - (v - i w)} \notag \\
%&= - \frac{Z}{-k_E^2 - (v + i w)} - \frac{Z^*}{-k_E^2 - (v - i w)} \notag \\
&= \frac{Z}{k_E^2 + (v + i w)} + \frac{Z^*}{k_E^2 + (v - i w)} \notag \\
&= 2 \frac{\operatorname{Re}[Z] k_E^2 + (v \operatorname{Re}[Z] + w \operatorname{Im}[Z])}{k_E^4 + 2 v k_E^2 + (v^2 + w^2)} .
\label{pp}
\end{align}
This pole part of the propagator agrees with the \textit{Gribov-Stingl form}% 
\footnote{
If we apply the definition of the spectral function $\rho$ given in eq.(\ref{eq:dispersion_complex}) to the  Gribov-Stingl  propagator (\ref{GS-propagator}), we obtain the identically vanishing spectral function, since the Gribov-Stingl  propagator does not have the branch cut on the real $k^2$ axis across which there is a discontinuity: 
$\rho(k^2) := \frac{1}{\pi} \operatorname{Im}\mathscr{D}(k^2+i\epsilon)
= \frac{1}{2i \pi}[\mathscr{D}(k^2+i\epsilon)-\mathscr{D}(k^2-i\epsilon)]$ for $k^2 \in \mathbb{R}$. 
Therefore, the Gribov-Stingl  propagator has only the pole part and does not have the continuous cut part.
%has its meaning only in the Euclidean region, since the Gribov-type argument can be formulated only in the Euclidean region. 
}
\cite{Stingl86} with real parameters $c_0, c_1, c_2, d_0, d_1 \in \mathbb{R}$,
\begin{align}
 {\mathscr{D}}_{\rm GS}(k_{E}^2)
=& \frac{d_0 + d_1 k_E^2}{c_0 + c_1 k_E^2 + c_2 k_E^4} 
=  \frac{ \frac{d_0}{c_2} + \frac{d_1}{c_2} k_E^2 }{ \frac{c_0}{c_2} + \frac{c_1}{c_2} k_E^2 + k_E^4} \ , 
\nonumber\\ &
\ c_0, c_1, c_2, d_0, d_1 \in \mathbb{R} .
\label{GS-propagator}
\end{align}
Note that all the coefficients $c_0,c_1,c_2,d_0,d_1$ are not independent. 
%One of them can take any value by the rescaling, e.g., we choose an overall rescaling by $c_2=1$. 
The Gribov-Stingl form actually has four independent parameters, since one of them is eliminated by the rescaling.
This number of independent parameters agrees with that of the pole part of the propagator with a pair of complex conjugate poles characterized by the four parameters $v,w,\operatorname{Re}(Z),\operatorname{Im}(Z)$.

For instance, the correspondence between two sets of parameters is given as 
\begin{align}
\begin{cases}
\ \frac{c_0}{c_2} = v^2 + w^2 ,  \\
\ \frac{c_1}{c_2} = 2 v ,  \\
\ \frac{d_0}{c_2} = 2 (v \operatorname{Re}[Z] + w \operatorname{Im}[Z]) ,  \\
\ \frac{d_1}{c_2} = 2 \operatorname{Re}(Z) ,
\end{cases}  
\end{align}
which has the inverse relation  
\begin{align}
\begin{cases}
\ v = \frac{1}{2} \frac{c_1}{c_2} ,  \\
\ w = \sqrt{\frac{c_0}{c_2} - v^2} = \sqrt{ \frac{c_0}{c_2} - \left( \frac{1}{2} \frac{c_1}{c_2} \right)^2 } ,  \\
\ \operatorname{Re}[Z] = \frac{1}{2} \frac{d_1}{c_2} ,  \\
\ \operatorname{Im}[Z] = \frac{1}{w} \left( \frac{1}{2} \frac{d_0}{c_2} - \frac{1}{2} \frac{d_1}{c_2} v  \right) = \frac{  \frac{1}{2} \frac{d_0}{c_2} - \frac{1}{4} \frac{d_1}{c_2} \frac{c_1}{c_2} }{ \sqrt{ \frac{c_0}{c_2} - \left( \frac{1}{2} \frac{c_1}{c_2} \right)^2 } } .
\end{cases}  
\label{g2c}
\end{align}

For $v+iw$ to be a complex number (namely, $w$ to be a real number), the parameters of the Gribov-Stingl  form must satisfy the restriction
\begin{align}
\label{exGSRestrict}
\frac{c_0}{c_2} - \left( \frac{1}{2} \frac{c_1}{c_2} \right)^2 > 0 
\ \Leftrightarrow \ 
\frac{c_1^2}{4 c_0 c_2} < 1.
\end{align}
Assuming this condition, we can obtain the closed form for the pole part of the Schwinger function 
\begin{align}
\Delta_p(t)
 =  \int_{- \infty}^{\infty} \frac{dk_E}{2 \pi} e^{i k_E t} \left[ \frac{Z}{k_E^2 + (v + i w)} + \frac{Z^*}{k_E^2 + (v - i w)} \right] .
\end{align}
Indeed, the contribution from one of the poles is exactly evaluated as 
\begin{align}
&\int_{- \infty}^{\infty} \frac{dk_E}{2 \pi} e^{i k_E t} \frac{Z}{k_E^2 + (v + i w)} 
\notag \\
&= i Res \left. \left[ e^{i k_E t} \frac{Z}{(k_E - \alpha)(k_E + \alpha)} \right] \right|_{k_E = \pm \alpha} \notag \\
&= i e^{i \alpha |t|} \frac{Z}{2 \alpha} \notag \\
&= i \exp \left[ - t r^{1/2} e^{i \varphi} - i \varphi \right] \frac{Z}{i 2 r^{1/2}} \notag \\
&= \frac{Z}{2 r^{1/2}} \exp \left[ - t r^{1/2} \cos \varphi \right] 
\nonumber\\ & \times 
\exp \left[ - i t r^{1/2} \sin \varphi - i \varphi \right]  ,
\end{align}
where we have defined 
\begin{align}
& \alpha^2 = - ( v + i w ) = - \sqrt{v^2 + w^2} e^{i \theta} = - r e^{i \theta} \notag \\
& \Rightarrow \alpha =  (v^2 + w^2)^{1/4} e^{i \theta / 2 + i \pi / 2} = i r^{1/2} e^{i \varphi} , \notag \\
& r = \sqrt{v^2 + w^2} \ , \  \theta = \arctan \frac{w}{v} \ , \ \varphi = \frac{\theta}{2}  = \frac{1}{2}  \arctan \frac{w}{v},
\end{align}
where $\alpha$ must be located on the upper half plane of the complex $k_E$ plane.
Therefore, the pole part of the Schwinger function coming from a pair of complex conjugate poles is exactly obtained as \cite{Kondo11}
\begin{align}
\Delta_p(t)
%=& \int_{- \infty}^{\infty} \frac{dk_E}{2 \pi} e^{i k_E t} \left[ \frac{Z}{k_E^2 + (v + i w)} + \frac{Z^*}{k_E^2 + (v - i w)} \right]
%\notag \\
%=& 2 \operatorname{Re} \left[ \frac{Z}{2 r^{1/2}} \exp \left[ - t r^{1/2} \cos \varphi \right] \exp \left[ - i t r^{1/2} \sin \varphi - i \varphi \right]  \right] 
%\notag \\
%=& \frac{ 1 }{r^{1/2}} e^{ - t r^{1/2} \cos \varphi }  \Big\{ \operatorname{Re}[Z] \cos \left( t r^{1/2} \sin \varphi + \varphi \right) 
%\nonumber\\&
%+ \operatorname{Im}[Z] \sin \left( t r^{1/2} \sin \varphi + \varphi \right)  \Big\} 
%\nonumber\\
 =& \frac{\sqrt{\operatorname{Re}(Z)^2+\operatorname{Im}(Z)^2}}{ (v^2+w^2)^{1/4}  } \exp [ - t(v^2+w^2)^{1/4} \cos \varphi]  
\nonumber\\ & \times
 \cos  [ t(v^2+w^2)^{1/4} \sin \varphi + \varphi - \delta ]
, 
\nonumber\\ &
\ \varphi := \frac{1}{2} \arctan \frac{w}{v} ,
\ \delta :=  \arctan \frac{\operatorname{Im}(Z)}{\operatorname{Re}(Z)}  .
\label{exSchPole}
\end{align}
%where $c.c.$ denotes the complex conjugate. 
At $t=0$, $\Delta_{p} (0)$ has the value,
\begin{align}
\Delta_{p} (0) = \frac{\sqrt{\operatorname{Re}(Z)^2+\operatorname{Im}(Z)^2}}{ (v^2+w^2)^{1/4}  }  
%\nonumber\\& \times 
%[ \operatorname{Re}(Z) \cos \varphi +  \operatorname{Im}(Z) \sin \varphi ]
\cos  ( \varphi - \delta ) .
\end{align}
We find that $\Delta_{p} (t)$ is oscillating between positive and negative values, although the absolute value $|\Delta_{p} (t)|$ becomes smaller for larger $t>0$.
\footnote{
The pole part of the Schwinger function becomes positive only when the poles become real ones $w=0$ (or $\varphi=0$)
\begin{align}
  \Delta_{p}^{\rm GS}(t) 
 = \frac{|\operatorname{Re}(Z)|}{ \sqrt{|v|} } \exp [ - t\sqrt{|v|}  ]  .
\end{align}
}

%The spectral function $\rho(\sigma^2)$ always takes the negative value and the resulting cut part $\Delta_c(t)$ of the Schwinger function always  takes the negative value according to (\ref{Schwinger-f-cut}):
%\begin{align}
%\Delta_c(t) < 0 \ \text{for} \ {}^{\forall} \ t \ge 0 .
%\end{align}

On the other hand, the cut part $\Delta_{c} (t)$ of the Schwinger function is estimated using the integral representation (\ref{Schwinger-f-cut}):
\begin{align}
\Delta_c(t) 
= \int_0^{\infty} d \sigma^2  \rho(\sigma^2) \frac{1}{2 \sqrt{\sigma^2} }  e^{ - \sqrt{\sigma^2} |t| } .
%= \int_0^{\infty} d \zeta  \rho(\zeta) \frac{1}{2 \sqrt{\zeta} }  e^{ - \sqrt{\zeta} |t| } .
%\label{exSchCont}
\label{Schwinger-c-partf}
\end{align}
This representation is an exact relation between the spectral function and the cut part of the Schwinger function, which holds irrespective of the existence or non-existence of complex poles. 

At least to one-loop order in the massive Yang-Mills model, 
 the spectral function $\rho(\sigma^2)$ takes the negative value $\rho(\sigma^2)<0$ for all $\sigma^2>0$: 
\begin{align}
  & \rho(\sigma^2)<0 \ \text{for} \ {}^{\forall} \  \sigma^2 > 0 ,
\label{negative-spec-f} 
\end{align} 
as demonstrated numerically in Fig.~\ref{figPropRealAxisPhys} (at the physical point) in this paper and shown analytically for any value of the parameters $g$ and $M$ in \cite{HK18}.
According to (\ref{Schwinger-c-partf}), therefore, the cut part $\Delta_{c} (t)$ of the Schwinger function takes  negative value for any value of $t$, 
%if the spectral function is negative $\rho(\sigma^2)<0$ for all $\sigma^2$: 
\begin{align}
%& \rho(\sigma^2)<0 \ \text{for} \ {}^{\forall} \  \sigma^2 > 0  
%\nonumber\\
%\Longrightarrow   
\Delta_c(t) < 0 \ \text{for} \ {}^{\forall} \ t \ge 0 ,
%= \int_{\sigma_{\rm min}}^{\infty} d\sigma  \rho(\sigma^2)   e^{- \sigma t} < 0  \text{any $t >0$}  ,
\end{align} 
although $\Delta_{c} (t)$ takes smaller and smaller negative value for larger and larger $t>0$.

Moreover, \textit{the negative spectral function (\ref{negative-spec-f}) yields the existence of one pair of complex conjugate poles or two real poles in the Euclidean region} as shown in \cite{HK18}.  
According to (\ref{exSchPole}), therefore, the pole part $\Delta_p(t)$ of the Schwinger function due to a pair of complex conjugate poles takes negative value for a certain value of $t$:
\begin{align}
\Delta_p(t) < 0 \ \text{for} \ {}^{\exists} \ t \ge 0 .
\end{align}

Thus, the Schwinger function $\Delta (t)$ obtained as a sum of two parts, $\Delta (t) = \Delta_{p} (t)  + \Delta_{c} (t)$
%\begin{align}
% \Delta (t) = \Delta_{p} (t)  + \Delta_{c} (t) ,
% \ \Delta_{c}^{\rm GS}(t) = \int_{\sigma_{\rm min}}^{\infty} d\sigma  \rho(\sigma^2)   e^{- \sigma t} .
%\end{align} 
has necessarily negative value at a certain value of $t$,
%, if the spectral function is negative $\rho(\sigma^2)<0$ for all $\sigma^2>0$.
\begin{align}
\Delta(t) = \Delta_c(t) + \Delta_p(t) < 0 \ \text{for} \ {}^{\exists} \ t \ge 0 .
\end{align}
Thus we complete the analytical proof that \textit{the reflection positivity is always violated irrespective of the choice of the parameters $g$ and $M$ in the massive Yang-Mills model to one-loop order}. 

In particular, the propagator of the Gribov type is a special case corresponding to $c_1=0$ and $d_0=0$
\begin{equation}
% \tilde{\mathscr{D}}_{\rm cc}(p) 
 \tilde{\mathscr{D}}_{\rm G}(p)=   \frac{ d_1}{ c_2 } \frac{ p_{E}^2}{ \frac{c_0}{c_2} + p_{E}^4} , 
\label{CCF-Gribov}
\end{equation}
which has a pair of pure imaginary poles  
\begin{align}
 v=0, \ \pm  iw ,  \quad  w = \sqrt{\frac{c_0}{c_2} },   
\end{align}
with the real-valued residue
\begin{align}
  \operatorname{Re}(Z) = \frac12 \frac{ d_1}{ c_2 }, \ \operatorname{Im}(Z) = 0  .
\end{align}
The pole part of the Schwinger function for the propagator of the Gribov type (\ref{CCF-Gribov}) is given by 
\begin{align}
 \Delta_{p}^{\rm G}(t)  
=&  \frac{d_1}{2c_2r^{1/2}} e^{-\frac{r^{1/2}}{\sqrt{2}}t}  \cos \left( \frac{r^{1/2}}{\sqrt{2}}t + \frac{\pi}{4} \right)   
\nonumber\\
 =&  \frac{Z}{ \sqrt{|w|}} e^{-\frac{\sqrt{|w|}}{\sqrt{2}}t }    \cos \left( \frac{\sqrt{|w|}}{\sqrt{2}}t + \frac{\pi}{4} \right)  , 
\end{align}
where we have used $\varphi = \frac{\pi}{4}$ and $r^{1/2}=\left( c_0/c_2  \right)^{1/4}=\sqrt{|w|}$.

%====================================================
\subsection{Fitting of the pole part at the physical point}

\begin{figure}[t]
\centering
\includegraphics[width=8cm]{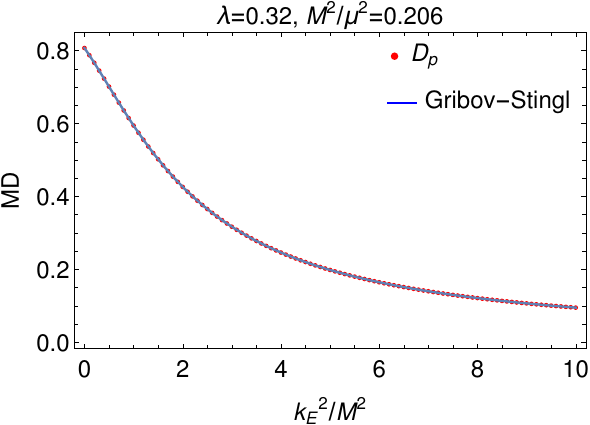}
\caption{
The comparison between the pole part of the gluon propagator (red dotted line) and the fit to the Gribov-Stingl form (blue solid line), 
at the physical values of the parameters $\lambda = 0.32$, $M^2/\mu^2 = 0.206$. 
}
\label{figPropFitPhys}
\end{figure}

\begin{figure}[t]
\centering
\includegraphics[width=8cm]{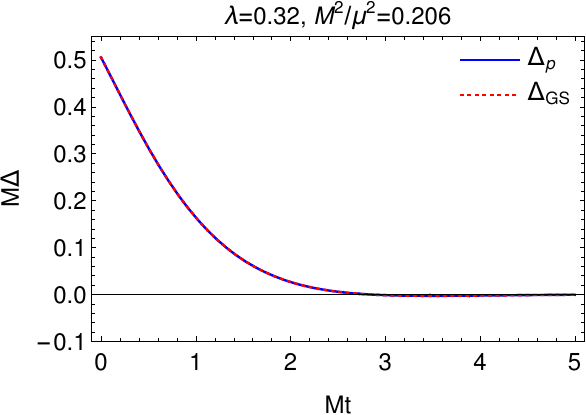}
\caption{
The comparison between the pole part of the Schwinger function calculated from the spectral function (blue solid line) and the Schwinger function calculated from the Gribov-Stingl fit (red dotted line), at the physical values of the parameters $\lambda = 0.32$, $M^2/\mu^2 = 0.206$. 
}
\label{figRepSchPhys}
\end{figure}
The general analysis given in the above can be substantiated by choosing the physical point for the parameters of the massive Yang-Mills model. 
We have obtained the pole part of the gluon propagator, as demonstrated in Fig.~\ref{figPropBreakPhys}.
This pole part $\mathscr{D}_{p}(k^2)$ is obtained by the difference $\mathscr{D}_{p}(k^2)= \mathscr{D} (k^2) - \mathscr{D}_{c}(k^2)$ from the propagator $\mathscr{D} (k^2)$ once the cut part $\mathscr{D}_{c}(k^2)$ of the propagator is specified according to (\ref{eq:one_pair_complex}) from the imaginary part of the propagator on the positive real axis through the spectral function $\rho(k^2)$. 
%By choosing an appropriate number of sample points from it, 
We have fitted the resulting pole part $\mathscr{D}_{p}(k^2)$ of the gluon propagator to the Gribov-Stingl form. 
Fig.~\ref{figPropFitPhys} is the result of fitting of the pole part of the gluon propagator to the Gribov-Stingl form at the physical point of the parameters.  
%can be used in fitting to determine the parameters $c_0,c_1,c_2,d_0,d_1$ of the Gribov-Stingl form. 
Here we have introduced dimensionless versions of the parameters $\hat{c_0},\hat{c_1},\hat{c_2},\hat{d_0},\hat{d_1}$ for  $c_0,c_1,c_2,d_0,d_1$ and squared momentum $s$ for $k^2$, which are scaled by appropriate powers of the gluon mass $M$ to make the dimensionless Gribov-Stingl form
\begin{align}
\frac{\hat{d}_0  + \hat{d}_1 s}{\hat{c}_0 + \hat{c}_1 s + \hat{c}_2 s^2} .
\label{GS-form-dimensionless}
\end{align}
The fitting parameters are determined as \cite{Watanabe19}
\begin{align}
\begin{cases}
\ \hat{c}_0 = 1.7678 \pm 3 \times 10^{-5} ,  \\
\ \hat{c}_1 = 0.73006 \pm 5 \times 10^{-5} ,  \\
\ \hat{c}_2 = 0.32505 \pm 8 \times 10^{-5} ,  \\
\ \hat{d}_0 = 1.4268 \pm 3 \times 10^{-5} ,  \\
\ \hat{d}_1 = 0.2512 \pm 1 \times 10^{-4} ,  \\
\end{cases} 
%\ \frac{\hat{c}_1^2}{4 \hat{c}_0 \hat{c}_2} = 0.2319 \pm 1 \times 10^{-4} 
\end{align}
which is subject to the restriction (\ref{exGSRestrict})
\begin{align}
 \ \frac{\hat{c}_1^2}{4 \hat{c}_0 \hat{c}_2} = 0.2319 \pm 1 \times 10^{-4} .
\end{align}
This result is translated into the complex pole and the residue of the gluon propagator
\begin{align}
\begin{cases}
\ \hat{v} = 1.123 \pm 4 \times 10^{-4} ,  \\
\ \hat{w} = 2.044 \pm 2 \times 10^{-4} ,  \\
\ \operatorname{Re}[Z] = 0.3863 \pm 2 \times 10^{-4} ,  \\
\ \operatorname{Im}[Z] = 0.8615 \pm 2 \times 10^{-4} 
\end{cases} .
\end{align}
It is ensured that this data reproduces the location of the poles given in Fig.~\ref{figPropAnaConPhys}.
We find that the fitting errors are very small and to good accuracy  the pole part of the gluon propagator is identical to the Gribov-Stingl form. 
This result strongly suggests that the pole part of the gluon propagator indeed stems from a pair of complex conjugate poles. 
For the other argument for understanding the Gribov-Stingl form, see \cite{Kondo11}.

The pole part of the Schwinger function can be calculated according to (\ref{exSchPole}) once all the parameters of the Gribov-Stingl form are determined.
If our analysis of the complex structure of the propagator is correct, the result should agree with the pole part of the Schwinger function given in Fig.~\ref{figSchBreakPhys}.
In fact, 
Fig.~\ref{figRepSchPhys} shows excellent agreement between pole part of the Schwinger function obtained from the spectral function and the expression (\ref{exSchPole}) with the parameters obtained through the fit of the gluon  propagator to the Gribov-Stingl form.
This result supports the validity of our arguments. 

Thus we have shown that the gluon propagator consists of the pole part due to a pair of complex conjugate poles and the cut part due to the branch cut on the positive real axis, in agreement with the generalized spectral representation (\ref{eq:one_pair_complex}). 
This is also the case for the associated Schwinger function.
%and that each contribution can be extracted to be analyzed. 
In this way we can conclude that the reflection positivity is violated in the massive Yang-Mills model at the physical point.
% for any choice of the parameters to one-loop order.
\footnote{
Notice that the existence of complex poles in the momentum representation of the two-point function does not necessarily violate spacelike commutativity. 
For instance, it was shown by Nakanishi \cite{Nakanishi71} that the existence of complex pole is compatible with spacelike commutativity in a complex scalar field theory with indefinite metric.
Notice that this theory is manifestly Lorentz covariant in a finite duration of time, against the title of the paper.  
}

%====================================================
\subsection{Parameter dependence other than the physical point}

We investigate the gluon propagator and the associated Schwinger function at choices of the parameters other than the physical point.

%-------------------------------------------
\subsubsection{Smaller gauge coupling}

\begin{figure}[t]
\centering
\includegraphics[width=7cm]{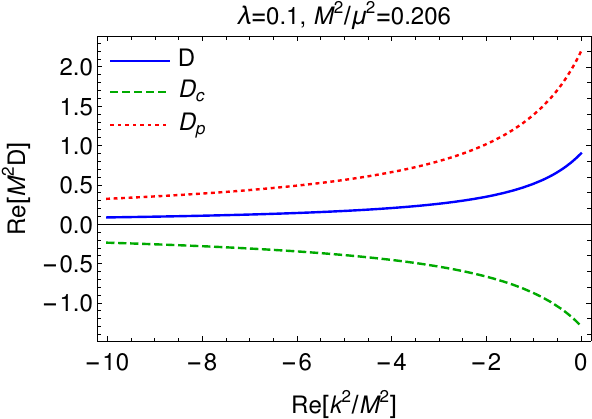}
\includegraphics[width=7cm]{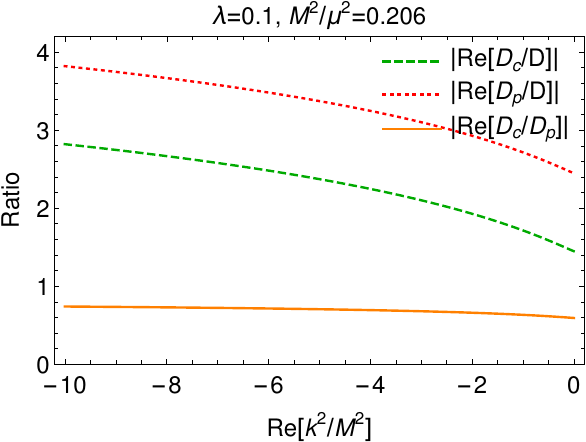}
\caption{
The same plots for the gluon propagator in the Euclidean region as those in Fig.~\ref{figPropBreakPhys} 
for the choice of parameters  with a smaller coupling constant, $\lambda = 0.1$, $M^2/\mu^2 = 0.206$.
%, gluon propagator (blue solid line), the cut part (green broken line), the pole part (red dotted line),
%(bottom) the absolute value of the real part of the ratio.
}
\label{figPropBreakSmallLambda}
\end{figure}

\begin{figure}[t]
\centering
\includegraphics[width=7cm]{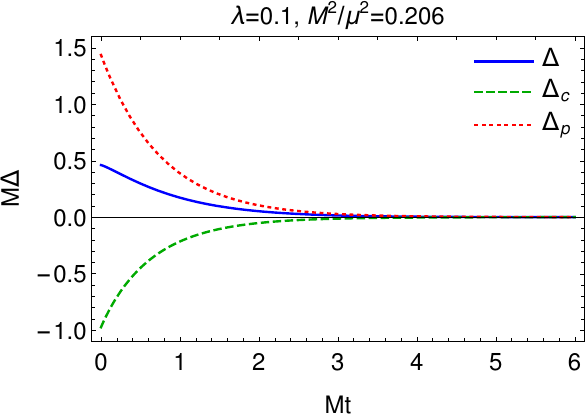}
\includegraphics[width=7cm]{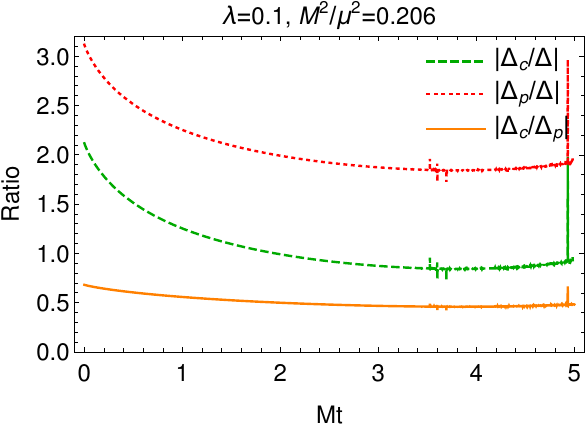}
\caption{
The same plots for the gluon Schwinger function  in the Euclidean region as those in Fig.~\ref{figSchBreakPhys}
for the choice of parameters  with a smaller coupling constant, $\lambda = 0.1$, $M^2/\mu^2 = 0.206$.
}
\label{figSchBreakSmallLambda}
\end{figure}

 For smaller value of the coupling constant $\lambda := Ng^2/(4\pi)^2 = 0.1$ with the physical value for $M$, we obtain the gluon propagator in Fig.~\ref{figPropBreakSmallLambda}
and the Schwinger function in Fig.~\ref{figSchBreakSmallLambda}.
The cut part is relatively large and has the opposite sign to the pole part to cause cancellation. 
The fall-off of both parts of the Schwinger function is slow for smaller value of the coupling constant.  
Therefore, the large $t$ behavior of the Schwinger function must be investigated to see the violation of positivity due to the cancellation between two parts for this choice of parameters other than the physical point which can be identified with an effective model of the pure Yang-Mills theory.

%-------------------------------------------
\subsubsection{Smaller gluon mass}

\begin{figure}[t]
\centering
\includegraphics[width=7cm]{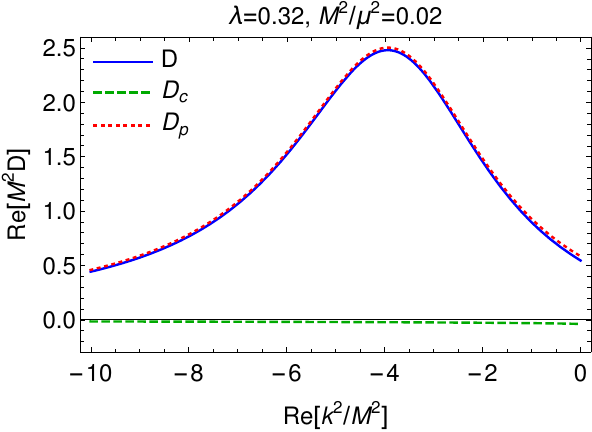}
\includegraphics[width=7cm]{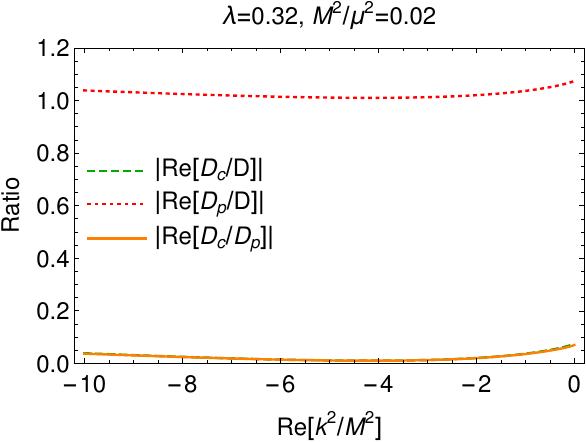}
\caption{
The same plots for the gluon propagator  in the Euclidean region as those in Fig.~\ref{figPropBreakPhys} 
for the choice of parameters  with a smaller mass, $\lambda = 0.32$, $M^2/\mu^2 = 0.02$.
}
\label{figPropBreakSmallMass}
\end{figure}

\begin{figure}[t]
\centering
\includegraphics[width=7cm]{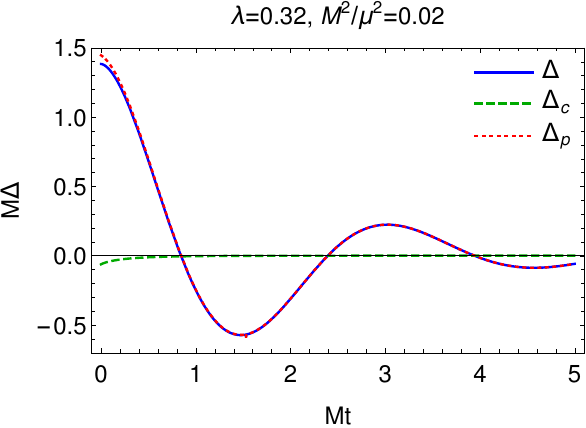}
\includegraphics[width=7cm]{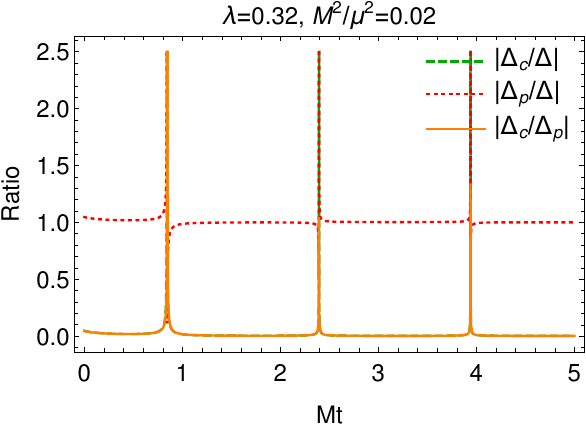}
\caption{
The same plots for the gluon Schwinger function  in the Euclidean region as those in Fig.~\ref{figPropBreakPhys} 
for the choice of parameters with a smaller mass, $\lambda = 0.32$, $M^2/\mu^2 = 0.02$.
}
\label{figSchBreakSmallMass}
\end{figure}

For smaller value of the gluon mass at $M^2/\mu^2 = 0.02$, we obtain the gluon propagator in Fig.~\ref{figPropBreakSmallMass}
and the Schwinger function in Fig.~\ref{figSchBreakSmallMass}.
The cut part is relatively rather small.
Therefore, violation of reflection positivity largely comes from the pole part.

We find that the cut parts $\mathscr{D}_c$ and $\Delta_c$ of the gluon propagator and the Schwinger function are always negative. 
This result reconfirms in a numerical way that the spectral function is negative irrespective of the choice of the parameters. 
Thus, the reflection positivity is always violated for any choice of the parameters.

%%%%%%%%%%%%%%%%%%%%%%%%%%%%%%%%%%%%%%%%%%%%%%%%%%%%%%%%%%%%%
%%%%%%%%%%%%%%%%%%%%%%%%%%%%%%%%%%%%%%%%%%%%%%%%%%%%%%%%%%%%%
\section{
Conclusion and discussion
}
%%%%%%%%%%%%%%%%%%%%%%%%%%%%%%%%%%%%%%%%%%%%%%%%%%%%%%%%%%%%%
%%%%%%%%%%%%%%%%%%%%%%%%%%%%%%%%%%%%%%%%%%%%%%%%%%%%%%%%%%%%%

\noindent
%$\bigodot$
We have examined the mass-deformed Yang-Mills theory or the massive Yang-Mills model in the covariant Landau gauge with two parameters, the coupling constant $g$ and the mass parameter $M$, in order to reproduce the confining decoupling solution of the pure Yang-Mills theory. 
By choosing appropriate values for a set of two parameters $g$ and $M$, we have shown that the massive Yang-Mills model well reproduces simultaneously the gluon and ghost propagators of the decoupling solution obtained by the numerical simulations on the lattice (at least) in the low-momentum region. Such a choice of the parameters is called the physical point for the Yang-Mills theory. 

Then we have shown that the reflection positivity is violated in the massive Yang-Mills model at the physical point of the parameters by observing the negativity of the Schwinger function which is obtained in a numerical way as the Fourier transform of the gluon propagator. 

The violation of reflection positivity was also confirmed by examining the complex structure of the complex-valued gluon propagator obtained by performing the analytic continuation of the Euclidean propagator to the entire complex squared momentum plane.  
We have verified that the violation of reflection positivity in the Euclidean region detected by the Schwinger function associated with  the Euclidean gluon propagator is a consequence of the complex structure of the complex-valued gluon propagator: (i) the negativity of the spectral function obtained from the discontinuity of the gluon propagator across the branch cut on the positive real axis on the complex squared momentum plane, (ii) the existence of a pair of complex conjugate poles in the gluon propagator. 
At the physical point, the contribution from the cut part to the gluon propagator in the Euclidean region is relatively small compared with that from the pole part.
Therefore, the propagator in the Euclidean region is well approximated by the contribution from a pair of complex conjugate poles in the complex region, which implies that the propagator  in the Euclidean region is well described by the Gribov-Stingl form, in agreement with the lattice result \cite{DOS18}. 
The violation of reflection positivity is regarded as a necessary condition for gluon confinement. 
Therefore, our results of reflection positivity violation at the physical point of the massive Yang-Mills model support strongly  gluon confinement in the Yang-Mills theory. 

We have regarded the massive Yang-Mills model at the physical point  as the low-energy effective model of the pure Yang-Mills theory.  
However, the massive Yang-Mills model with the parameters $g$ and $M$ other than the physical point has another meaning. 
We have discussed that the massive Yang-Mills model in the covariant Landau gauge has the gauge-invariant extension, which is identified with the complementary gauge-scalar model with a radially fixed fundamental scalar field which is subject to an appropriate reduction condition. 
In other words, the gauge-scalar model with a radially fixed fundamental scalar field subject to the reduction condition can be gauge-fixed to becomes the massive Yang-Mills model in the covariant Landau gauge. 
The gauge-invariant extension of a non-gauge theory is performed through the gauge-independent description of the BEH mechanism  \cite{Kondo18} without relying on the spontaneous symmetry breaking which was first proposed for the adjoint scalar field \cite{Kondo16}. 

Therefore, the Yang-Mills theory in the confinement phase characterized by the decoupling solution is identified with the massive Yang-Mills model with the physical point of the parameters $g$ and $M$.
This physical point of the massive Yang-Mills model represents a point in the parameter space of the complementary gauge-scalar model obtained as a gauge-invariant extension of the massive Yang-Mills model.
%A physical quantity is regarded as a smooth function of the parameters in the analyticity domain which includes the physical point in the confinement-like region and the points in the Higgs-like region to the whole parameter space including confinement-like and Higgs-like regions without discontinuities in light of the Fradkin-Shenker continuity.
%region in the phase diagram of  which could have a single confinement phase as suggested from the Fradkin-Shenker continuity.
%Therefore, the positivity violation/restoration could be gluon confinement/deconfinement criterion in the presence of matter fields
Thus, the violation of reflection positivity in the massive Yang-Mills model for any value of the parameters $g$ and $M$ is consistent with the Fradkin-Shenker continuity in the sense that the massive Yang-Mills model describes both confinement-like and Higgs-like regions in the single confinement phase of the complementary gauge-scalar model. 
Our result seems to be consistent with the other approaches \cite{CPRW18,RSTW17}. 

Let us make comments on the sum rule for the spectral function  called the superconvergence relation \cite{OZ80},
\begin{align}
\int_0 ^\infty d \sigma^2 \rho (\sigma^2) = 0 .
\label{eq:the_superconvergence}
\end{align}
It is obvious that this sum rule cannot be satisfied for the negative spectral function. 
In \cite{HK18}, remarkably, the generalized sum rule for the spectral function has been derived in the presence of a pair of complex conjugate poles
\begin{align}
2 \operatorname{Re} Z + \int_0 ^\infty d \sigma^2 \rho (\sigma^2) = 0, 
\label{eq:modified_superconvergence}
\end{align}
provided that the propagator has the asymptotic behavior
\begin{align}
\lim_{|k^2| \rightarrow \infty} k^2 \mathscr{D}(k^2) = 0 , 
\label{UVpropbehav}
\end{align}
 in the region far from the origin of the complex $k^2$ plane.
In fact, it is shown that the gluon propagator fulfills this condition in Yang-Mills theories in the Landau gauge  due to  the asymptotic freedom and the negativity of the anomalous dimension \cite{OZ80}. 
%Indeed, the renormalization group (RG) equation for the propagator and the ultraviolet asymptotic freedom yield the asymptotic form for $|k^2| \rightarrow \infty$
%\begin{align}
%D(k^2) \sim - \frac{Z_{UV}}{k^2 (\ln |k^2|)^\gamma}, \label{eq:section2_UV_asymptotic}
%\end{align}
%where $Z_{UV}$ is a positive constant and $\gamma$ is the ratio between the coefficient $\gamma_0$ of the leading term of the anomalous dimension and the first coefficient $\beta_0$ of the beta function. 
%For the gluon propagator of the pure Yang-Mills theory in the Landau gauge, $\gamma$ has the value,
%\begin{align}
%\gamma_0 = - \frac{1}{16 \pi^2} \frac{13}{6} C_2 (G) &< 0, ~~ \beta_0 = - \frac{1}{16 \pi^2} \frac{11}{3} C_2 (G) < 0, \notag \\
%\gamma &= \frac{\gamma_0}{\beta_0} = \frac{13}{22} > 0, \label{eq:gamma_0_and_beta_0}
%\end{align}
%where $C_2(G)$ is the quadratic Casimir invariant of the gauge group $G$. The condition (\ref{UVpropbehav}) holds due to (\ref{eq:section2_UV_asymptotic}) and (\ref{eq:gamma_0_and_beta_0}).
%The residue $Z$ in (\ref{eq:one_pair_complex}) as a special case of (\ref{eq:spec_repr_complex}) can be measured according to (\ref{eq:dispersion_residue}). 
The assumption (\ref{UVpropbehav}) is enough to obtain
\begin{align}
Z &= -\frac{1}{2 \pi i} \int_{- \infty}^{\infty} dx~ \mathscr{D}(x + i \epsilon), ~~ x = \operatorname{Re}k^2. 
\label{eq:section2_residue_formula}
\end{align}
The real part of (\ref{eq:section2_residue_formula}) leads to  a generalized sum rule for the spectral function (\ref{eq:modified_superconvergence})  
%as a consequence of the assumption (\ref{UVpropbehav}) for a propagator of the form (\ref{eq:one_pair_complex}), 
by taking into account the relation  (\ref{eq:dispersion_complex}): 
\begin{align}
\operatorname{Im} \ \mathscr{D}(x + i \epsilon) = \begin{cases}
0 & \ (x < 0) \\
\pi \rho(x) & \ (x > 0) 
\end{cases}
,
\end{align}
while the imaginary part leads to another relation,
\begin{align}
2 \operatorname{Im} Z = \frac{1}{ \pi} \int_{- \infty}^{\infty} dk^2~ \operatorname{Re}\mathscr{D}(k^2 + i \epsilon). 
\end{align}
Therefore, the usual superconvergence relation (\ref{eq:the_superconvergence}) does not hold unless the residue of the complex pole is pure imaginary $\operatorname{Re}Z = 0$.
% (\ref{eq:gamma_0_and_beta_0}).
The preliminary results for the massive Yang-Mills model to one-loop order at the physical point are \cite{Watanabe19}
\begin{align}
 & \operatorname{Re}Z =  0.386322, \
 \operatorname{Im}Z =  0.861514 ,
\nonumber\\
 & \int_0 ^\infty d \sigma^2 \rho (\sigma^2) =  -0.694533 <0 , 
\nonumber\\
 & \frac{1}{\pi} \int_{- \infty}^{\infty} dk^2~ \operatorname{Re} \mathscr{D}(k^2 + i \epsilon) =  1.74006 > 0 
  ,
\label{eq:modified_superconvergence2}
\end{align}
which leads to 
\begin{align}
2 \operatorname{Re} Z + \int_0 ^\infty d \sigma^2 \rho (\sigma^2) = 0.0781108 , 
\nonumber\\
 2 \operatorname{Im} Z - \frac{1}{\pi} \int_{- \infty}^{\infty} dk^2~ \operatorname{Re} \mathscr{D}(k^2 + i \epsilon) =& - 0.0170369 
  .
\label{eq:modified_superconvergence3}
\end{align}
It will be interesting to examine whether the generalized sum rule holds or not, and to what extent it is satisfied beyond one-loop level when the gluon propagator has a pair of complex conjugate poles.

Moreover, it is desirable to extend the results obtained in this paper  to a finite temperature to see whether or not the reflection positivity violated in the low-temperature confinement phase is recovered in the high-temperature deconfinement phase even in the pure Yang-Mills theory. 
Then we can ask whether or not the transition detected by the positivity violation/restoration agrees with the confinement/deconfinement transition detected by the Polyakov loop average.
It is also interesting to examine how the relevant complex structure changes depending on the temperature. 
These issues will be discussed in subsequent papers.

Finally, we give some comments on the obstructions stemming from the  presence of complex poles in the gluon propagator to the formal field theoretic issues such as loss of locality.  
In local QFT, it is recognized that any 2-point correlation function is an analytic function in the cut complex  $p^2$(squared momentum)-plane with singularities along the time-like (positive) real axis only.  The assumptions to establish this analytic  property are \cite{AQFT}: Lorentz covariance (covariance under space-time translations),   the spectrum condition, local (or space-like) commutativity,  and uniqueness and cyclicity of the vacuum.  For any other singularity structure of 2-point correlations, at least one of these assumptions must be violated. 

In order to consider this issue in the Yang-Mills theory, we can take into account the observation  \cite{OZ80,Oehme95} that the correlation functions of the Yang-Mills field vanish in the limit $p^2 \to \infty$ in all directions of the complex $p^2$-plane due to the ultraviolet asymptotic freedom of the Yang-Mills theory.  However, non-trivial entire functions with that property do not exist.  Therefore, they must have singularities somewhere.  
Usually, the singularities are supposed to exist on the positive real $p^2$ axis.  However, this does not deny the  existence of complex conjugate poles discussed in this paper.  
See also \cite{Oehme95} and section 2.5 and 5.4 of \cite{AS01} for the review. 

Indeed, the existence of complex poles does not only play a crucial role in the violation of the reflection positivity but also breaks the spectral representation, which is a fundamental implication of local field theories. In the standard point of view, e.g., from the Jost-Lehmann-Dyson (JLD) representation \cite{,JL57,Dyson58}, complex poles must violate the local spacelike commutativity.

 One might claim that the non-locality of the Yang-Mills theory in a gauge-fixed picture is rather ``natural'' due to the Gribov-Singer obstruction, see \cite{Gribov78,Singer78,Maas13} and \cite{Zwanziger89,Baulieu-etal}. The problem of locality is discussed in \cite{Stingl96,HKRSW90}, in which they assert that complex poles describe short-lived excitations, and the locality is broken in the level of propagators, but the corresponding $S$-matrix remains causal.
However,  their way of reconstructing the Minkowskian propagator from the Euclidean propagator is questionable. By a straightforward reconstruction of the analytic continuation in the complex time plane, the resulted Minkowskian propagator differs from the previous one \cite{Stingl96} and predicts the Lee-Wick type propagator \cite{LW69}.  In fact, without the positive definiteness of the state space and the spectral condition, which are not guaranteed for confined degrees of freedom, complex spectra can appear. The Lee-Wick type theories can yield complex poles without the loss of the spacelike commutativity, see, e.g. \cite{Nakanishi71b} as the simplest example for the propagator with complex poles. Notice that the complex spectra enable a theory to evade the restriction of the axiomatic or analytic theorems like the JLD representation because it deviates from the framework of tempered distributions. In this scenario, complex poles are not an indication of the non-locality but just a reflection of unphysical degrees of freedom, such as timelike photons. 
%In the quantum theory with complex conjugate poles, it is claimed \cite{Nakanishi06} that the Lorentz invariance is spontaneously violated, as far as the locality is maintained. 
This issue will be further discussed elsewhere.

Another important issue to be addressed in the presence of the complex poles in the gluon propagator is to answer the question how the correlation function of color singlet composite operators can have the real poles, since such composite states must be observed.  This issue was investigated in \cite{Zwanziger89} and \cite{,Baulieu-etal}.  It is further argued in \cite{Stingl96} that complex singularities with time-like real part might be acceptable for the propagators of unphysical colored fields, by the reason that such singularities might conspire to cancel with singularities or zeros in other unphysical correlation functions so as to be absent from physical amplitudes. This will give rise to an infinite hierarchy of constraints on such unphysical singularities in arbitrary high $n$-point functions.  An example of such compensating singularities are those in the non-perturbative expansion scheme, see \cite{Stingl96,HKRSW90} and sections 2.5 and 5.4 of \cite{AS01}. 
More serious discussion on the related issues will be given in future works.

\section*{Acknowledgements}

The authors would like to thank Professor Oliveira for providing us with the data of numerical simulation for gluon and ghost propagators presented in this paper. 
K.-I. K.  thanks Matthieu Tissier, Nicolas Wschebor, Julien Serreau, and Urko Reinosa for the discussion on positivity violation. 
This work was  supported by Grant-in-Aid for Scientific Research, JSPS KAKENHI Grant Number (C) No.19K03840 and No.15K05042.
R. M. was supported by Grant-in-Aid for JSPS Research Fellow Grant Number 17J04780.

\appendix
\section{Recursive construction of the gauge-invariant transverse field}

\crefname{equation}{}{}
\crefrangelabelformat{equation}{(#3#1#4--#5#2#6)}
\crefname{figure}{Fig.}{Figs.}

%\section{How to obtain the solution of reduction condition in the expansion of the gauge field}

In this section, we derive the recurrence relation for obtaining the power-series solution of the reduction condition which is equivalent to the transverse condition,
\begin{align}
  \partial_\mu \mathscr A_\mu^{h[\mathscr A]} = 0. \label{red}
\end{align}

First, we expand $h[\mathscr A]$ into the power series in the gauge field $\mathscr A$,
\begin{align}
  h[\mathscr A] =& \bm 1 + h^{(1)} + h^{(2)} + \cdots, 
\\
  h^{-1}[\mathscr A] =& \bm 1 + (h^{-1})^{(1)} + (h^{-1})^{(2)} +  \cdots,
\end{align}
where the superscript $(k)$ denotes $k$-th term in the power series.
Note that $(h^{-1})^{(k)} = (h^{(k)})^\dag$ because $h^{-1}=h^\dag$.
Then the $k$-th term of $\mathscr A_\mu^h = h \mathscr{A}_{\mu} h^\dagger - ig^{-1} \partial_\mu h  h^\dagger$ is given by
%\begin{widetext}
\begin{align}
  (\mathscr A_\mu^h)^{(k)} 
 =& \sum_{l=0}^{k-1} h^{(l)} \mathscr A_\mu  h^{(k-l-1)}{}^\dag
%\notag \\&
  -ig^{-1}\sum_{l=0}^k \partial_\mu h^{(l)} h^{(l-k)}{}^\dag 
\notag \\
  =& -ig^{-1}\partial_\mu h^{(k)}
\notag \\&
  + \sum_{l=0}^{k-1} \left[ h^{(l)} \mathscr A_\mu h^{(k-l-1)}{}^\dag
  - ig^{-1} \partial_\mu h^{(l)} h^{(k-l)}{}^\dag \right] .
\label{Ah}
\end{align}
%\end{widetext}
By substituting (\ref{Ah}) into \cref{red} we obtain the recurrence relation as
\begin{align}
  ig^{-1} h^{(k)} = \frac{\partial_\mu}{\partial^2} \sum_{l=0}^{k-1} 
   \left[ h^{(l)}\mathscr A_\mu  h^{(k-l-1)}{}^\dag 
   -ig^{-1}\partial_\mu h^{(l)} h^{(k-l)}{}^\dag \right]. \label{zen}
\end{align}

By using this recurrence relation, we can derive several features of $\mathscr A^h_\mu$.
First, we show that $\mathscr A_\mu^h$ can be written as
\begin{align}
  \mathscr A_\mu^h = \left(\delta_{\mu\nu}-\frac{\partial_\mu\partial_\nu}{\partial^2}\right)\Psi_\nu. \label{phi}
\end{align}
This is shown as follows.
Indeed, by substituting the recurrence relation \cref{zen} into \cref{Ah}, we obtain
\begin{align}
  \mathscr A_\mu^{h(k)} = \left(\delta_{\mu\nu} - \frac{\partial_\mu\partial_\nu}{\partial^2}\right) F_\nu^{(k)} ,
\end{align}
where we have defined 
\begin{align}
  F_\mu^{(k)} := \sum_{l=0}^{k-1}\left[
  h^{(l)}\mathscr{A}_\mu  h^{(k-l-1)}{}^\dagger - ig^{-1}\partial_\mu h^{(l)} h^{(k-l)}{}^\dagger \right].
\end{align}
By taking into account the fact that $F^{(k)}_\mu$ is not Hermitian, the $k$-th term of $\Psi_\mu$ in \cref{phi} can be written as
\begin{align}
  \Psi_\mu^{(k)} = \frac12\left( F_\mu^{(k)} + F_\mu^{\dag(k)}\right) .
\end{align}

Next, we show that $A_\mu^h$ is gauge invariant by using \cref{zen}.
Now let the gauge transformation of $\mathscr A$ by $V$ be
\begin{align}
  \mathscr A_\mu \quad\rightarrow\quad V\mathscr A_\mu V^{-1}+ig^{-1}V\partial_\mu V^{-1}.
\end{align}
We observe that $\mathscr A_\mu^h$ is indeed gauge invariant 
if the gauge transformation of $h$ obeys
\begin{align}
  h \quad\rightarrow\quad hV^{-1} \label{transf} ,
\end{align}
whose infinitesimal form for  $V=e^{-ig\Lambda}$ is given by
\begin{align}
  \delta h = -igh\Lambda .  
\label{inf_transf}
\end{align}
In the following, we show that \cref{inf_transf} holds order by order.

Note that because $\delta\mathscr A_\mu =  \partial_\mu\Lambda -ig [\mathscr A_\mu,\Lambda]$, $\delta h^{(k)}$ contains $k$-th term and $(k-1)$-th term of power series of $\delta h$.
For this reason, we separate $\delta h^{(k)}$ into two parts as
\begin{align}
  \delta h^{(k)} &= \delta h^{(k)}_- + \delta h^{(k)}_= 
\end{align}
where $\delta h^{(k)}_-$ has the order of $k-1$ and $\delta h^{(k)}_=$ has the order of $k$.
Therefore, the $k$-th order term of \cref{inf_transf} is written as
\begin{align}
  ig^{-1}\delta h^{(k)}_= + ig^{-1}\delta h^{(k+1)}_- &= h^{(k)}\Lambda. 
\label{transf_hk}
\end{align}
This relation is shown to hold with mathematical induction as follows. 
For $k=0$, it follows from \cref{zen} that
\begin{align}
  ig^{-1}h^{(1)} &= \frac{\partial_\mu}{\partial^2}\mathscr{A}_\mu. 
\end{align}
By taking the variation under the gauge transformation, we obtain
\begin{align}
  ig^{-1}\delta h^{(1)}_- &= \Lambda .
\end{align}
Thus we have shown \cref{transf_hk} holds for $k=0$.
Next, suppose that \cref{transf_hk} holds for $k-1$.
Then we proceed to show that it holds for $k$.
By taking variation of \cref{zen} for $k+1$ under the gauge transformation we obtain
\begin{widetext}
\begin{align}
  ig^{-1}\delta h^{(k+1)}_- &= \frac{\partial_\mu}{\partial^2} \sum_{l=0}^k
  \left(\delta h^{(l)}_- \mathscr A_\mu (h^{(k-l)})^\dag
  + h^{(l)} \mathscr A_\mu (\delta h^{(k-l)}_-)^\dag
  + h^{(l)} \partial_\mu\Lambda  (h^{(k-l)})^\dag\right. \notag\\
  &\quad \left. -ig^{-1}\partial_\mu \delta h^{(l)}_- (h^{(k-l+1)})^\dag
  -ig^{-1}\partial_\mu h^{(l)} (\delta h^{(k-l+1)}_-)^\dag\right) \notag\\
  &= \frac{\partial_\mu}{\partial^2}\biggl[ h^{(k)}\partial_\mu\Lambda -ig^{-1} \partial_\mu h^{(k)}(\delta h^{(1)}_-)^\dag  
\notag\\
  &\quad + \sum_{l=0}^{k-1}\left( h^{(l)} \partial_\mu\Lambda  (h^{(k-l)})^\dag
  -ig^{-1}\partial_\mu h^{(l)} (\delta h^{(k-l+1)}_-)^\dag\right) \notag\\
  &\quad+ \sum_{l=0}^k \left(
  \delta h^{(l)}_- \mathscr A_\mu (h^{(k-l)})^\dag
  + h^{(l)} \mathscr A_\mu (\delta h^{(k-l)}_-)^\dag
  -ig^{-1}\partial_\mu \delta h^{(l)}_- (h^{(k-l+1)})^\dag \right) \biggr] 
\notag\\
  &= h^{(k)}\Lambda + \frac{\partial_\mu}{\partial^2}
  \sum_{l=0}^{k-1}\left( h^{(l)} \partial_\mu\Lambda  (h^{(k-l)})^\dag
  -\dashuline{ig^{-1}\partial_\mu h^{(l)} (\delta h^{(k-l+1)}_-)^\dag} \right.\notag\\
  &\quad \left. +\uline{\delta h^{(l+1)}_- \mathscr A_\mu (h^{(k-l-1)})^\dag}
  + \uuline{h^{(l)}\mathscr A_\mu (\delta h^{(k-l)}_-)^\dag}
  -\uwave{ig^{-1}\partial_\mu \delta h^{(l+1)}_- (h^{(k-l)})^\dag}\right). \label{k+1-}
\end{align}
By taking the variation of \cref{zen} for $k$ we obtain
\begin{align}
  ig^{-1}\delta h^{(k)}_= &= \frac{\partial_\mu}{\partial^2}\sum_{l=0}^{k-1}\left(
  \uline{\delta h^{(l)}_=\mathscr A_\mu(h^{(k-l-1)})^\dag} + \uuline{h^{(l)}\mathscr A_\mu(\delta h^{(k-l-1)}_=)^\dag}
  -igh^{(l)}[\mathscr A_\mu,\Lambda](h^{(k-l-1)})^\dag \right.\notag\\
  &\quad \left. -\uwave{ig^{-1}\partial_\mu\delta h^{(l)}_=(h^{(k-l)})^\dag} -\dashuline{ ig^{-1}\partial_\mu h^{(l)}(\delta h^{(k-l)}_=)^\dag} \right). \label{k=}
\end{align}
\end{widetext}
By summing up the underlined part, double-underlined part, wavy-lined part and broken-lined part respectively, we obtain
\begin{align}
  \uline{\qquad} &= -igh^{(l)}\Lambda \mathscr A_\mu(h^{(k-l-1)})^\dag, \label{ul}\\
  \uuline{\qquad} &= igh^{(l)}\mathscr A_\mu\Lambda (h^{(k-l-1)})^\dag, \label{uul}\\
  \uwave{\qquad} &= \partial_\mu(h^{(l)}\Lambda)(h^{(k-l)})^\dag, \label{uw}\\
  \dashuline{\qquad} &= -\partial_\mu h^{(l)}\Lambda (h^{(k-l)})^\dag, \label{ud} 
\end{align}
where we have used the assumption of induction.
The sum of \cref{ul} and \cref{uul} cancels the third term in the parentheses of \cref{k=}.
The sum of \cref{uw} and \cref{ud} cancels the first term in the parentheses of \cref{k+1-}.
Therefore \cref{transf_hk} is satisfied for $k$.
Thus we have shown that $\mathscr A^h$ is invariant under a gauge transformation.

\section{
Solving the reduction condition in another way
}

By using the massive vector field mode $\mathscr{W}_\mu$ (\ref{W1b-SU2}), the reduction condition reads
\begin{align}
 \chi(x) = \chi^A(x) T_A 
=& \partial_\mu \mathscr{W}_\mu (x) - ig[\mathscr{A}_\mu (x) , \mathscr{W}_\mu (x)] 
\nonumber\\
=&  
\partial_\mu \mathscr{A}_{\mu}(x) -ig^{-1} \partial_\mu (\hat{\Theta}(x) \partial_{\mu} \hat{\Theta}(x)^\dagger )
\nonumber\\ &
- [\mathscr{A}_{\mu}(x) , \hat{\Theta}(x) \partial_{\mu} \hat{\Theta}(x)^\dagger]  .
\end{align}  
For the scalar field $\hat{\Theta}(x)$, we introduce the Lie algebra $\mathscr{G}$-valued field $\theta(x)$ as   
\begin{align}
  \hat{\Theta}(x) =& e^{-ig \theta (x) } \in G  , \ \theta (x)  := \theta^A(x)T_A \in \mathscr{G} .
\end{align}
In the following, we solve the reduction condition by expressing the scalar field $\hat{\Theta}(x)$ as a power series in the gauge field $\mathscr{A}_{\mu}$. 
The Lie algebra form of the pure gauge reads 
%\begin{widetext}  
%\begin{align}
%\hat{\Theta} \partial_{\mu} \hat{\Theta}^\dagger  =& \left( 1 -ig \theta^AT_A - \frac12 g^2 \theta^AT_A \theta^BT_B \right) \partial_{\mu} \left( 1 +ig \theta^AT_A - \frac12 g^2 \theta^AT_A \theta^BT_B \right) + O(\theta^3) 
%\nonumber\\
%=& \left( 1 -ig \theta^AT_A - \frac12 g^2 \theta^A \theta^B T_AT_B \right)  \left(  ig \partial_{\mu} \theta^AT_A - \frac12 g^2 \partial_{\mu} \theta^A \theta^B T_AT_B - \frac12 g^2  \theta^A \partial_{\mu} \theta^B T_AT_B  \right) + O(\theta^3) 
%\nonumber\\
%=& ig \partial_{\mu} \theta^A T_A - \frac12 g^2 \partial_{\mu} \theta^A \theta^B T_AT_B - \frac12 g^2  \theta^A \partial_{\mu} \theta^B T_AT_B  + g^2 \theta^A  \partial_{\mu} \theta^B T_AT_B + O(\theta^3)
%\nonumber\\
%=& ig \partial_{\mu} \theta^A T_A + \frac12 g^2  \theta^A \partial_{\mu} \theta^B [T_A,T_B]  + O(\theta^3) ,
%\end{align}
%or
\begin{align}
 & \hat{\Theta} \partial_{\mu} \hat{\Theta}^\dagger  \nonumber\\
=& \left( 1 -ig \theta  - \frac12 g^2 \theta \theta   \right) \partial_{\mu} \left( 1 +ig \theta  - \frac12 g^2 \theta \theta  \right) + \mathcal{O}(\theta^3) 
\nonumber\\
=& \left( 1 -ig \theta  - \frac12 g^2 \theta \theta  \right)  \left(  ig \partial_{\mu} \theta   - \frac12 g^2 \partial_{\mu} \theta \theta  - \frac12 g^2  \theta  \partial_{\mu} \theta   \right) 
\nonumber\\&
 + \mathcal{O}(\theta^3) 
\nonumber\\
=& ig \partial_{\mu} \theta  - \frac12 g^2 \partial_{\mu} \theta \theta  - \frac12 g^2  \theta  \partial_{\mu} \theta   + g^2 \theta   \partial_{\mu} \theta  + \mathcal{O}(\theta^3)
\nonumber\\
=& ig \partial_{\mu} \theta  - \frac12 g^2 \partial_{\mu} \theta \theta  + \frac12 g^2  \theta  \partial_{\mu} \theta    + \mathcal{O}(\theta^3) ,
\nonumber\\
=&  ig \partial_{\mu} \theta  + \frac12 g^2 [\theta ,\partial_{\mu} \theta]  + \mathcal{O}(\theta^3) .
\end{align}
%\end{widetext}  
The more general expression is given as
%\footnote{
%M. Esole, 
%The Non-local massive Yang-Mills action as a gauged sigma model, 
%e-Print: hep-th/0407069
%\\
%R. Delbourgo and G. Thompson,
%Massive, Unitary, Renormalizable Yang-mills Theory Without Higgs Mesons, 
%Phys.Rev.Lett. 57 (1986) 2610 
%Print-86-1161 (TASMANIA) 
%DOI: 10.1103/PhysRevLett.57.2610 
%}
\begin{align}
 & \hat{\Theta}(x) \partial_{\mu} \hat{\Theta}(x)^\dagger  =   - \sum_{n=0}^{\infty}   \frac{(-ig)^{n+1}}{(n+1)!} [Ad~\theta(x)]^n \partial_\mu \theta(x) ,
\nonumber\\&
  Ad~X(Y) := [X,Y] .
\end{align}
By substituting this result into the reduction condition $\chi=0$, we have 
\begin{align}
0 =&  
\partial_\mu \mathscr{A}_{\mu} -ig^{-1} \partial_\mu (\hat{\Theta} \partial_{\mu} \hat{\Theta}^\dagger ) 
- [\mathscr{A}_{\mu} , \hat{\Theta} \partial_{\mu} \hat{\Theta}^\dagger]  
\nonumber\\
=& \partial_\mu \mathscr{A}_{\mu} -ig^{-1} \partial_\mu (ig \partial_{\mu} \theta  + \frac12 g^2 [\theta ,\partial_{\mu} \theta] ) 
\nonumber\\ &
- \left[ \mathscr{A}_{\mu} , ig \partial_{\mu} \theta  + \frac12 g^2 [\theta ,\partial_{\mu} \theta] \right]  + \mathcal{O}(\theta^3)
\nonumber\\
=& \partial_\mu \mathscr{A}_{\mu} + \partial_\mu \partial_{\mu} \theta  - \frac12 i g  [\theta ,\partial_\mu \partial_{\mu} \theta]  
%\nonumber\\ &
- ig [\mathscr{A}_{\mu} ,  \partial_{\mu} \theta ] 
%- \frac12 g^2 [\mathscr{A}_{\mu} ,  [\theta ,\partial_{\mu} \theta] ] 
 + \mathcal{O}(\theta^3)
 .
\end{align}  
This is recast into 
\begin{align}
 \partial^2 \theta   
=& -\partial \cdot \mathscr{A}  + \frac12 i g  [\theta ,\partial^2 \theta]  
%\nonumber\\ &
+ ig [\mathscr{A}_{\mu} ,  \partial_{\mu} \theta ] 
%+ \frac12 g^2 [\mathscr{A}_{\mu} ,  [\theta ,\partial_{\mu} \theta] ] 
 + \mathcal{O} (\theta^3)
 ,
\end{align} 
which yields 
\begin{align}
  \theta   
=& -\frac{1}{\partial^2} \partial \cdot \mathscr{A}  + \frac12 i g \frac{1}{\partial^2}  [\theta ,\partial^2 \theta]  
%\nonumber\\ &
+ ig \frac{1}{\partial^2}[\mathscr{A}_{\mu} ,  \partial_{\mu} \theta ] 
%+ \frac12 g^2 [\mathscr{A}_{\mu} ,  [\theta ,\partial_{\mu} \theta] ] 
 + \mathcal{O}(\theta^3)
 .
\end{align} 
Substituting recursively for $\theta$, we obtain a power series, 
\begin{align}
  \theta(x)   
=& -\frac{1}{\partial^2} \partial \cdot \mathscr{A}(x)  + \frac12 i g \frac{1}{\partial^2}  \left[ \frac{1}{\partial^2} \partial \cdot \mathscr{A}(x) ,   \partial \cdot \mathscr{A}(x) \right]  
\nonumber\\ &
- ig \frac{1}{\partial^2} \left[\mathscr{A}_{\mu}(x) ,  \partial_{\mu}  \frac{1}{\partial^2} \partial \cdot \mathscr{A}(x) \right] 
%+ \frac12 g^2 [\mathscr{A}_{\mu} ,  [\theta ,\partial_{\mu} \theta] ] 
 + \mathcal{O}(\mathscr{A}^3)
 .
\end{align}

The massive vector field mode $\mathscr{W}_\mu$ (\ref{W1b-SU2}) is written as
\begin{align}
 \mathscr{W}_\mu 
 =&  \mathscr{A}_{\mu}(x)  -ig^{-1} \hat{\Theta}(x) \partial_{\mu} \hat{\Theta}(x)^\dagger .
\nonumber\\  
=&  \mathscr{A}_{\mu}  + \partial_{\mu} \theta  -i \frac12 g  [\theta ,\partial_{\mu} \theta]  + \mathcal{O}(\theta^3)
\nonumber\\  
=&  \mathscr{A}_{\mu}^T  + \frac12 i g  \frac{1}{\partial^2} \partial_{\mu} \left[ \frac{\partial \cdot \mathscr{A}}{\partial^2} , \partial \cdot \mathscr{A} \right]  
\nonumber\\ &
- ig  \frac{1}{\partial^2} \partial_{\mu} \left[ \mathscr{A}_{\lambda} ,  \partial_{\lambda}  \frac{\partial \cdot \mathscr{A}}{\partial^2}  \right] 
\nonumber\\ &
 -i \frac12 g  \left[\frac{\partial \cdot \mathscr{A}}{\partial^2} ,\partial_{\mu} \frac{\partial \cdot \mathscr{A}}{\partial^2} \right]  + \mathcal{O}(\mathscr{A}^3)  ,
 \label{W-exp}
\end{align}
where we have defined the transverse  field $\mathscr{A}_{\mu}^T$ in the lowest order term linear in $\mathscr{A}$ as 
\begin{align}
\mathscr{A}_{\mu}^T := \mathscr{A}_{\mu} -\partial_{\mu} \frac{\partial \cdot \mathscr{A}}{\partial^2} .
\end{align}
Notice that $\mathscr{W}_\mu$ agrees with $\mathscr{A}_{\mu}$ in the Landau gauge $\partial \cdot \mathscr{A}=0$.   
  
Thus, by substituting $\mathscr{W}$ of (\ref{W-exp}) into (\ref{kin=mass}),  the term $S_{\rm kin}$ reads 
%the mass term of $W$ is  
\begin{align}
 S_{\rm kin}^*[\mathscr{A}]  
=& \int d^Dx \ M^2{\rm tr}(\mathscr{W}_\mu \mathscr{W}_\mu )
\nonumber\\  
=&   \int d^Dx \ M^2{\rm tr}\Big\{  \mathscr{A}_{\mu}^T  \mathscr{A}_{\mu}^T     
\nonumber\\ &
 +   i g  \mathscr{A}_{\mu}^T \partial_{\mu} \frac{1}{\partial^2}  \left[\frac{\partial \cdot \mathscr{A}}{\partial^2}  ,   \partial \cdot \mathscr{A} \right]  
\nonumber\\ &
- 2ig \mathscr{A}_{\mu}^T \partial_{\mu} \frac{1}{\partial^2}  \left[ \mathscr{A}_{\lambda} ,  \partial_{\lambda}  \frac{\partial \cdot \mathscr{A}}{\partial^2}  \right] 
\nonumber\\ &
 -i   g \mathscr{A}_{\mu}^T \left[ \frac{\partial \cdot \mathscr{A}}{\partial^2} ,\partial_{\mu} \frac{\partial \cdot \mathscr{A}}{\partial^2} \right] 
 \Big\} 
 + \mathcal{O}(\mathscr{A}^4) .
\end{align}
By performing integration by parts and taking into account the transversality $\partial_{\mu} \mathscr{A}_{\mu}^T=0$, the action $S_{\rm kin}^*$ takes the form, 
\begin{align}
 S_{\rm kin}^*[\mathscr{A}]  =&   \int d^Dx \ M^2  {\rm tr}\Big\{  \mathscr{A}_{\mu}^T  \mathscr{A}_{\mu}^T     
 -i   g \mathscr{A}_{\mu}^T \left[\frac{\partial \cdot \mathscr{A}}{\partial^2} ,\partial_{\mu} \frac{\partial \cdot \mathscr{A}}{\partial^2} \right] 
 \Big\} 
\nonumber\\ &
 + \mathcal{O}(\mathscr{A}^4) 
 .
\end{align}
This indeed agrees with the expression (\ref{Sh}).

%In the minimum reduction condition, if we choose the unitary gauge $=1$, the gauge-scalar model with the reduction condition reduces to the massive Yang-Mills model, 

%\newpage
%%%%%%%%%%   REFERENCES   %%%%%%%%%%

\end{document}